\documentclass{aa} % for a paper on 2 columns  
\usepackage{graphicx}
\usepackage{txfonts}
\usepackage{hyperref}
\begin{document}

   \title{Surveys of clumps, cores, and condensations in Cygnus-X: \\ SMA observations of SiO (5$-$4)}
   
   \titlerunning{SiO (5$-$4) in Cygnus-X}

   \author{Kai Yang\inst{1,2}          
          \and
          Keping Qiu\inst{1,2} 
          \and
          Xing Pan\inst{1,2}   
          }

   \institute{School of Astronomy and Space Science, Nanjing University, 163 Xianlin Avenue, Nanjing 210023, P.R.China\\
             \email{kpqiu@nju.edu.cn}
             \and            
                 Key Laboratory of Modern Astronomy and Astrophysics (Nanjing University), Ministry of Education, Nanjing 210023, P.R.China  
             }

   \date{Received XXX; Accepted XXX}
        
% \abstract{}{}{}{}{}
% 5 {} token are mandatory
 
  \abstract
   {The SiO emissions are usually used to trace high-velocity outflow shocks in star-forming regions. However, several studies have found low-velocity and widespread SiO emissions not associated with outflows in molecular clouds.}
   {We aim to detect and characterize the SiO emissions in massive dense cores (MDCs), and explore the properties of the  central sources of SiO emission.}
   {We present high-angular-resolution ($\sim$1.5$^{\prime\prime}$) observations of the SiO (5$-$4) line made with the Submillimeter Array towards a sample of 48 MDCs in the Cygnus-X star-forming complex. We studied the SiO emission structures, including their morphologies, kinematics, and energetics, and investigated their relationship with the evolution of the central sources.}
   {The SiO (5$-$4) emission is detected in 16 out of 48 MDCs. We identify 14 bipolar and 18 unipolar SiO (5$-$4) outflows associated with 29 dust condensations. Most outflows (24 out of 32) are associated with excess Spitzer 4.5 $\mu$m emissions.
We also find diffuse low-velocity ($\Delta{v}$ $\le$ 1.2 km s$^{-1}$) SiO (5$-$4) emission closely surrounding the dust condensations in two MDCs, and suggest that it may originate from decelerated outflow shocks or large-scale shocks from global cloud collapse.}
   {We find that the SMA SiO (5$-$4) emission in MDCs is mostly associated with outflows. Probably due to the relatively high excitation of SiO (5$-$4) compared to SiO (2$-$1) and due to the spatial filtering effect, we do not detect large-scale low-velocity SiO (5$-$4) emission, but detect more compact low-velocity emission in close proximity to the dust condensations. We group the sources into different evolutionary
stages based on the infrared emission, radio continuum emission, and gas temperature properties of the outflow central sources,  and find that the 24 $\mu$m luminosity tends to increase with evolution.}


   \maketitle
%
%________________________________________________________________

\section{Introduction} \label{sec:intro}

Massive stars dominate the energy input in galaxies and play a crucial role in shaping their host galaxies through feedback in the form of outflows, wind, UV radiation, and supernovae \citep{2013Natur.499..450B, 2015EAS....75..227S}. Over the last few decades, the formation of high-mass stars has been studied both observationally and theoretically (e.g., \citealt{1979ARA&A..17..345H}; \citealt{1997MNRAS.285..201B}; \citealt{1998MNRAS.298...93B}; \citealt{2003ApJ...585..850M}; \citealt{2007ARA&A..45..339B}; \citealt{2008ApJ...685.1005Q}; \citealt{2015ApJ...804..141Z}; \citealt{2020ApJ...900...82P}). However, high-mass star-forming regions are difficult to observe due to their large distance, high extinction, small population, and short timescales \citep{2007ARA&A..45..481Z, 2018ARA&A..56...41M}. By removing angular momentum from the accreting gas, outflows and jets are thought to be unique forms of feedback generated from massive star formation and feeding back to their parent molecular clouds and the surrounding interstellar medium (ISM; e.g., \citealt{2007prpl.conf..245A}; \citealt{2015ApJ...804..141Z}; \citealt{2016ARA&A..54..491B}; \citealt{2016A&ARv..24....6B}). These phenomena can also provide vital information about the massive star formation processes.

Outflows and jets are found in massive star-forming regions at different evolutionary stages (e.g., \citealt{2002ARA&A..40...27C}; \citealt{2007prpl.conf..245A}; \citealt{2007ARA&A..45..339B}; \citealt{2007ApJ...668..348B}; \citealt{2007ApJ...654..361Q}; \citealt{2012A&A...548L...2C}; \citealt{2016A&A...595A.122L}; \citealt{2021ApJ...909..177L}). CO rotational lines are widely employed to study outflows; they typically trace the low-velocity, extended component of the outflow and are usually contaminated by surrounding material. The rotational transitions of silicon monoxide (SiO) exclusively trace shocks in outflows and jets because Si atoms are released from dust grains by sputtering or vaporization due to shock activity \citep{2008A&A...482..809G, 2008A&A...490..695G, 2009A&A...497..145G}, and SiO is mainly formed by the reaction between Si atoms and O$_{2}$ or OH in the gas phase \citep{1989A&A...222..205H, 1997A&A...322..296C, 1997A&A...321..293S, 2014CPL...610..335R}. 
Additionally, SiO can also be formed through a less prominent scenario involving Si$^{+}$, SiO$^{+}$, and HSiO$^{+}$ \citep{2013A&A...554A..35L}. Therefore, the abundance of SiO in molecular outflows can be highly enhanced by several orders of magnitude compared to the surroundings (e.g., \citealt{1992A&A...254..315M}; \citealt{2005MNRAS.361..244C}; \citealt{2007A&A...462..163N}; \citealt{2011A&A...526L...2L}; \citealt{2012ApJ...756...60S}; \citealt{2016A&A...586A.149C}), which makes SiO emission a good tracer for studying shocks in star-forming regions. 

The SiO emission in star-forming regions is usually used to trace high-velocity outflow shocks (e.g., \citealt{2007ApJ...654..361Q}; \citealt{2011A&A...526L...2L}; \citealt{2021ApJ...909..177L}). However, low-velocity and widespread SiO emissions not associated with outflows in molecular clouds have been found in several studies (e.g., \citealt{2007A&A...476.1243M}; \citealt{2013ApJ...775...88N}; \citealt{2013ApJ...773..123S}; \citealt{2014A&A...570A...1D}; \citealt{2016A&A...595A.122L}). For example, \citet{2013A&A...557A..94S} carried out CARMA (Combined Array for Research in Millimeter-wave Astronomy) SiO observations towards IRDC G028.23-00.19 and speculated that the narrow ($\sigma_{v}$ $\le$ 2 km s$^{-1}$) SiO emission is generated from a ``subcloud--subcloud'' interaction or by unresolved low-mass stars. \citet{2014A&A...570A...1D} found narrow ($\sigma_{v}$ $\le$ 1.5 km s$^{-1}$) and dispersed SiO emission in six Cygnus-X MDCs with Plateau de Bure Interferometer (PdBI) observations and concluded that this emission arises from collisions between large-scale flows. Until now, for most of the SiO observations toward star-forming regions, the angular resolution or sensitivity has not been good enough to explore the origin of the SiO emission.

In this work, we report a SiO (5$-$4) line survey towards 48 MDCs in Cygnus-X with the Submillimeter Array\footnote{The Submillimeter Array is a joint project between the Smithsonian Astrophysical Observatory and the Academia Sinica Institute of Astronomy and Astrophysics, and is funded by the Smithsonian Institution and the Academia Sinica.}  (SMA, \citealt{2004ApJ...616L...1H}). With a distance of 1.4 kpc to the Sun \citep{2012A&A...539A..79R}, the Cygnus-X star-forming complex is one of the richest and most active massive star-forming regions in our Galaxy. This complex contains $\sim$4$\times$10$^{6}$ M$_{\odot}$ of molecular gas, and harbors H {\scriptsize II} regions, OB associations, and a large sample of MDCs \citep{1977A&A....58..197H, 1991A&A...241..551W, 1992ApJS...81..267L, 2001A&A...371..675U, 2004ApJ...601..952S, 2007A&A...476.1243M, 2007A&A...474..873S, 2010A&A...524A..18B, 2013A&A...556A..16G,2019ApJS..241....1C}. The aim of our project, Surveys of \textbf{C}lumps, Cor\textbf{E}s, and Co\textbf{N}den\textbf{S}ations in Cygn\textbf{US}-X (CENSUS, PI: Keping Qiu), is to study the hierarchical cloud structures  and high-mass star formation processes taking place  in Cygnus-X. 

Thanks to the high resolution of the SMA and the relatively short distance to Cygnus-X, we can achieve a spatial resolution of $\sim$0.01 pc and can therefore resolve the shock-associated SiO (5$-$4) structures. First, we describe the sample and observations in Section \ref{sec:obs}. We then present the results and data analysis in Section \ref{sec:results}. We discuss the observational results in Section \ref{sec:dis} and summarize the main findings in Section \ref{sec:sum}.

%-----------------------------------------------------------------------

\begin{table*}
\begin{center}
\caption{The MDC sample for the SMA observations.} \label{tab:ob}
\begin{tabular}{l ll cc lc}
\hline\hline
\noalign{\smallskip}
Field & Motte07$^{~\rm b}$ & Cao21$^{~\rm c}$ & R.A. & Decl. & SiO (5$-$4) & r.m.s noise \\
 &  &  & (J2000) & (J2000) &  & (Jy beam$^{-1}$) \\
\hline
1 & N02/N03 & 220 & 20:35:34.630 & +42:20:08.787 & Y & 0.067 \\
2 & N05/N06 & 274 & 20:36:07.300 & +41:39:57.994 & N & 0.073 \\
3 & N10 & 725 & 20:36:52.199 & +41:36:22.991 & N & 0.052 \\
4 &N12/N13 & 248 &  20:36:57.397 & +42:11:27.997 & Y & 0.053 \\
5 & N14 & 714 & 20:37:00.900 & +41:34:57.002 & N & 0.058 \\
6 & N22/N24 & 1267 & 20:38:04.599 & +42:39:53.997 & N & 0.068 \\
7 & N30/N31/N32 & 1112/1231 & 20:38:37.401 & +42:37:32.994 & Y/N & 0.075 \\
8 & N56 & 698/1179 & 20:39:16.897 & +42:16:07.003 & Y/N & 0.056 \\
9 & N63 & 341 & 20:40:05.202 & +41:32:13.003 & Y & 0.056 \\
10 & N64/N65 & 801 & 20:40:28.397 & +41:57:10.997 & Y & 0.063 \\
11 & N68 & 684 & 20:40:33.499 & +41:59:02.995 & Y & 0.047 \\
12 & N69 & 4315/4797 & 20:40:33.698 & +41:50:59.000 & N & 0.057 \\
13 & NW01/NW02 & 327/742 & 20:19:38.998 & +40:56:45.006 & N & 0.076 \\
14 & NW04/NW05/NW07 & 640/675 & 20:20:30.503 & +41:21:39.998 & N & 0.069 \\
15 & NW12 & 839 & 20:24:14.301 & +42:11:43.001 & N & 0.059 \\
16 & NW14 & 310 & 20:24:31.701 & +42:04:22.999 & Y & 0.044 \\
17 & S06/S07/S08/S09 & 507/753 & 20:20:38.599 & +39:38:00.006 & Y & 0.061 \\
18 & S10 & 798 & 20:20:44.400 & +39:25:19.999 & N & 0.046 \\
19 & S15 & 874 & 20:27:13.996 & +37:22:57.997 & N & 0.085 \\
20 & S29 & 723 & 20:29:58.297 & +40:15:57.994 & N & 0.080 \\
21 & S30/S31 & 509 & 20:31:12.599 & +40:03:15.996 & Y & 0.088 \\
22 & S32 & 351 & 20:31:20.303 & +38:57:16.002 & Y & 0.065 \\
23 & S34 & 1225 & 20:31:57.801 & +40:18:30.004 & N & 0.054 \\
24 & S41 & 892 & 20:32:33.397 & +40:16:42.997 & N & 0.058 \\
25 & S42/S43 & 540 & 20:32:40.799 & +38:46:30.994 & N & 0.043 \\
26 & - & 214/247 & 20:30:28.503 & +41:15:55.000 & N & 0.179 \\
27 & - & 302/520 & 20:35:09.499 & +41:13:29.995 & N & 0.254 \\
28 & - & 340 & 20:32:22.500 & +41:07:55.996 & N & 0.239 \\
29 & - & 370 & 20:28:09.402 & +40:52:49.995 & N & 0.293 \\
30 & - & 608 & 20:34:00.003 & +41:22:25.001 & N & 0.184 \\
31 & - & 1460/2320 & 20:35:00.002 & +41:34:57.002 & N & 0.213 \\
32$^{~\rm a}$  & S36/S37 & 1454 & 20:32:21.850 & +40:20:00.708 & Y & 0.125 \\
     & S38 & 2210 & 20:32:22.302 & +40:19:19.524 & N & 0.125 \\
33$^{~\rm a}$ & N36/N40/N41 & 1018 & 20:38:59.333 & +42:23:37.183 & N & 0.167 \\
     & N37/N43 & 5417 & 20:38:58.299 & +42:24:35.896 & N & 0.172 \\
     & N38/N48 & 699 & 20:39:00.023 & +42:22:16.036 & Y & 0.150 \\
     & N44 & 1467 & 20:38:59.642 & +42:23:06.864 & Y & 0.170 \\
     & N51 & 1243 & 20:39:02.409 & +42:25:09.124 & N & 0.166 \\
     & N52/N53 & 1599 & 20:39:03.131 & +42:26:00.013 & Y & 0.176 \\
\hline\hline
\end{tabular}
\end{center}
Notes. $^{~\rm a} $ SMA mosaic mapping fields.
$^{~\rm b} $ MDCs from \citet{2007A&A...476.1243M}.
$^{~\rm c} $ MDCs from \citet{2021ApJ...918L...4C} covered by the fields.
\end{table*}

%-----------------------------------------------------------------------

\section{Observations and data reduction} \label{sec:obs}

The MDC sample observed in this work is summarized in Table \ref{tab:ob}, which covers 48 MDCs identified by \citet{2021ApJ...918L...4C}. We first made SMA observations towards MDCs identified by \citet{2007A&A...476.1243M}. We then added eight MDCs with 850 $\mu$m continuum peaks in the JCMT (James Clerk Maxwell Telescope) maps of the Cygnus OB2 region \citep{2021ApJ...918L...4C}, which is not covered by \citet{2007A&A...476.1243M}. As a result, the presented SMA observations consist of 31 single-pointing and two mosaic observations.

The SMA observations were conducted with the Subcompact, Compact, and Extended configurations. The raw data were calibrated with the IDL (Interactive Data Language) superset MIR\footnote{https://www.cfa.harvard.edu/$\sim$cqi/mircook.html}. The calibrated visibilities were exported into CASA (the Common Astronomy Software Applications, \citealt{2007ASPC..376..127M}) for joint imaging, and the final maps have a synthesized beam of $\sim$1.8$^{\prime\prime}$. Considering the upgrade of the SMA correlator from ASIC to SWARM during our observation and the archival observations, the total bandwidths of the data in this study range from 4 GHz to 16 GHz, and the spectral resolution is also not uniform. Nevertheless, each observation covered the SiO (5$-$4) line, and we smoothed all the data with a uniform spectral resolution of 812.5 kHz ($\sim$1.12 km s$^{-1}$ at 217.105 GHz) during imaging. The simultaneously observed CO (2$-$1) line emission serves as an auxiliary tool in the identification of SiO outflows, and detailed information on CO (2$-$1) can be found in \citet{Pan2023}.

%-----------------------------------------------------------------------

\section{Results} \label{sec:results}

 \subsection{SiO detection} \label{sec:detec}
We identify the SiO (5$-$4) emission using the following two steps: (1) We check the position-position-velocity (PPV) data cubes of the SiO emission. If more than three consecutive pixels have a peak flux of higher than 3$\sigma$ noise level, we mark this MDC as a detection. (2) Furthermore, we check the integrated intensity of SiO (5$-$4) within $\pm$15 km s$^{-1}$ and $\pm$5 km s$^{-1}$ relative to the MDC systemic velocity, and pick those with a detection with a signal-to-noise ratio of greater than 3. The systemic velocity for each MDC is determined by the centroid velocity, $v_{\rm lsr}$, of the DCN (3$-$2) line, while the C$^{18}$O (2$-$1) line is used when the DCN (3$-$2) line is not detected. Table \ref{tab:obs} summarizes the SiO detections and the systemic velocities.

%-----------------------------------------------------------------------
\begin{sidewaystable*}
\caption{Physical properties of the SiO (5$-$4)-detected MDCs and observational parameters.} \label{tab:obs}
\begin{center}
\footnotesize
\begin{tabular}{c cc ccr rrrrcc rrcc}
\hline\hline
MDCs & R.A. & Decl. & \multicolumn{3}{c}{Fitted FWHM for MDCS$^{~\rm a}$} & $M$$^{~\rm a}$ & $T_{\rm dust}$$^{~\rm a}$ & $L_{\rm FIR}$$^{~\rm a}$ & $F_{\nu, 24 \mu m}$$^{~\rm a}$ & $N_{\rm H_{2}}$$^{~\rm a}$ & $n_{\rm H_{2}}$$^{~\rm a}$ & \multicolumn{2}{c}{$v_{\rm lsr}$} & $\theta_{\rm cont}$$^{~\rm b}$ & $\theta_{\rm SiO}$$^{~\rm c}$ \\ 
\cline{4-6} \cline{13-14} 
  & (J2000) & (J2000) & D$_{\rm maj}$ & D$_{\rm min}$ & P. A. &  &  &  &  &  &  &  & tracer &  &  \\
  & (h m s) & ($^{\circ}$ $^{\prime}$ $^{\prime\prime}$) & (pc) & (pc) & (deg) & (M$_{\odot}$) & (K) & (L$_{\odot}$) & (Jy) & (cm$^{-2}$) & (cm$^{-3}$) & (km s$^{-1}$) &  & ($^{\prime\prime}\times$ $^{\prime\prime}$) & ($^{\prime\prime}\times$ $^{\prime\prime}$) \\
\hline
 220 & 20:35:34.17 & +42:20:10.8 & 0.287 & 0.247 &  31.0 & 383.10 & 17.75 & 1151.8 &   9.23 & 3.1E+23 & 5.6E+05 &  14.5  & C$^{18}$O (2$-$1) & 1.75 $\times$ 1.50 & 1.76 $\times$ 1.41 \\
 248 & 20:36:57.55 & +42:11:35.2 & 0.267 & 0.204 & 145.5 & 202.05 & 18.17 &  698.9 &   1.04 & 2.1E+23 & 4.4E+05 &  14.4  &    DCN (3$-$2)    & 1.73 $\times$ 1.46 & 1.80 $\times$ 1.42 \\
 310 & 20:24:31.73 & +42:04:20.4 & 0.240 & 0.195 & 152.2 & 152.78 & 20.48 & 1084.3 &  33.06 & 1.9E+23 & 4.2E+05 &  6.2   &    DCN (3$-$2)    & 2.07 $\times$ 1.85 & 2.61 $\times$ 2.16 \\
 341 & 20:40:05.40 & +41:32:13.3 & 0.245 & 0.235 &   4.4 & 160.91 & 17.44 &  435.2 &   0.10 & 1.6E+23 & 3.2E+05 & $-$4.2 &    DCN (3$-$2)    & 1.51 $\times$ 1.21 & 1.78 $\times$ 1.39 \\
 351 & 20:31:20.72 & +38:57:15.4 & 0.258 & 0.190 & 141.2 & 78.63  & 18.29 &  283.2 &   0.38 & 9.1E+22 & 2.0E+05 & $-$0.6 &    DCN (3$-$2)    & 2.69 $\times$ 2.44 & 2.38 $\times$ 2.02 \\
 507 & 20:20:38.43 & +39:37:45.4 & 0.322 & 0.231 &   6.6 & 372.46 & 21.40 & 3439.6 & 189.06 & 2.8E+23 & 5.1E+05 &  1.6   &    DCN (3$-$2)    & 2.38 $\times$ 2.17 & 3.18 $\times$ 2.67 \\
 509 & 20:31:13.29 & +40:03:12.6 & 0.341 & 0.287 & 134.6 & 300.64 & 18.18 & 1045.0 &   1.20 & 1.7E+23 & 2.7E+05 &  7.4   & C$^{18}$O (2$-$1) & 2.77 $\times$ 2.51 & 2.38 $\times$ 1.92 \\
 684 & 20:40:33.83 & +41:59:03.5 & 0.257 & 0.203 &  38.8 & 109.49 & 17.62 &  315.5 &   0.18 & 1.2E+23 & 2.5E+05 & $-$6.5 &    DCN (3$-$2)    & 1.92 $\times$ 1.71 & 2.12 $\times$ 1.77 \\
 698 & 20:39:17.42 & +42:16:10.4 & 0.225 & 0.190 & 109.8 & 81.14  & 16.98 &  186.8 &   5.89 & 1.1E+23 & 2.5E+05 &  19.6  &    DCN (3$-$2)    & 1.67 $\times$ 1.41 & 1.81 $\times$ 1.41 \\
 699 & 20:39:00.02 & +42:22:16.0 & 0.397 & 0.282 & 128.8 &  1283  & 20.92 &10332.1 &  31.03 & 6.5E+23 & 9.5E+05 & $-$1.5 &    DCN (3$-$2)    & 2.02 $\times$ 1.74 & 1.79 $\times$ 1.45 \\
 753 & 20:20:37.57 & +39:38:25.8 & 0.302 & 0.213 &  92.2 & 189.53 & 17.82 &  582.8 &   0.11 & 1.7E+23 & 3.2E+05 &  1.6   &    DCN (3$-$2)    & 2.38 $\times$ 2.17 & 3.18 $\times$ 2.67 \\
 801 & 20:40:28.06 & +41:57:05.7 & 0.331 & 0.236 &  26.9 & 143.46 & 17.84 &  444.4 &   2.76 & 1.0E+23 & 1.8E+05 & $-$9.8 &    DCN (3$-$2)    & 1.69 $\times$ 1.48 & 1.75 $\times$ 1.39 \\
1112 & 20:38:36.70 & +42:37:48.6 & 0.420 & 0.369 & 109.6 & 499.31 & 28.08 &  23522 & 925.99 & 1.8E+23 & 2.3E+05 &   9.0  & C$^{18}$O (2$-$1) & 1.62 $\times$ 1.37 & 1.91 $\times$ 1.54 \\
1454 & 20:32:21.85 & +40:20:00.7 & 0.508 & 0.414 & 127.8 & 420.81 & 16.17 &  252.7 &   0.22 & 1.1E+23 & 1.8E+05 &   1.9  & C$^{18}$O (2$-$1) & 1.69 $\times$ 1.33 & 2.93 $\times$ 2.39 \\
1467 & 20:38:59.64 & +42:23:06.9 & 0.341 & 0.233 &  27.6 & 257.92 & 23.4 &  2216.3 &  75.95 & 1.8E+23 & 4.9E+04 & $-$3.7 &    DCN (3$-$2)    & 2.02 $\times$ 1.74 & 1.79 $\times$ 1.45 \\
1599 & 20:39:03.13 & +42:26:00.0 & 0.232 & 0.186 & 168.5 & 164.34 & 17.19 &15187.4 &    ... & 2.2E+23 & 2.5E+05 & $-$4.4 &    DCN (3$-$2)    & 2.02 $\times$ 1.74 & 1.79 $\times$ 1.45 \\
\hline\hline
\end{tabular}
\end{center}
Notes.\\
$^{~\rm a} $ The coordinates and physical parameters of the MDCs obtained from \citet{2021ApJ...918L...4C}.\\
$^{~\rm b} $ Synthesized beam sizes of the 1.37 mm continuum.\\
$^{~\rm c} $ synthesized beam sizes of the the SiO (5$-$4) line.
\end{sidewaystable*}
%-----------------------------------------------------------------------

 \subsection{SiO outflow morphology and kinematics}

High-velocity SiO emissions detected in star-forming regions are typically indicative of outflows (e.g., \citealt{2007ApJ...654..361Q}; \citealt{2021ApJ...909..177L}). Several studies have explored the narrow SiO emissions using interferometer arrays. \citet{2013ApJ...773..123S} reported a SiO component with narrow line widths of approximately 2 km s$^{-1}$ from CARMA observations toward the infrared-dark cloud G028.23-00.19. \citet{2014A&A...570A...1D} conducted PdBI observations and identified narrow-line SiO (2$-$1) emissions ($\sigma_{v}$ $\leq$ 1.5 km s$^{-1}$) in six MDCs located within Cygnus-X, four of which are included in the  sample of the present study. Furthermore, \citet{2022MNRAS.512.5214D} detected SiO (2$-$1) emission with narrow-line profiles of around 1.5 km s$^{-1}$ towards Class 0 protostars in NGC 1333 through NOrthern Extended Millimeter Array (NOEMA) observations. Given the limited velocity resolution ($\sim$1.12 km s$^{-1}$) of our data, we are forced to work with only one velocity channel to check the presence of this kind of emission within our sample and, if it exists, to derive its spatial distribution. We derive high-velocity SiO emission ---which most likely arises from outflows--- with $\Delta{v}$ $\ge$ two channel widths ($\sim$ 2.3 km s$^{-1}$) with respect to the systemic velocity.

\begin{figure*}
\centering
   \includegraphics[width=240pt]{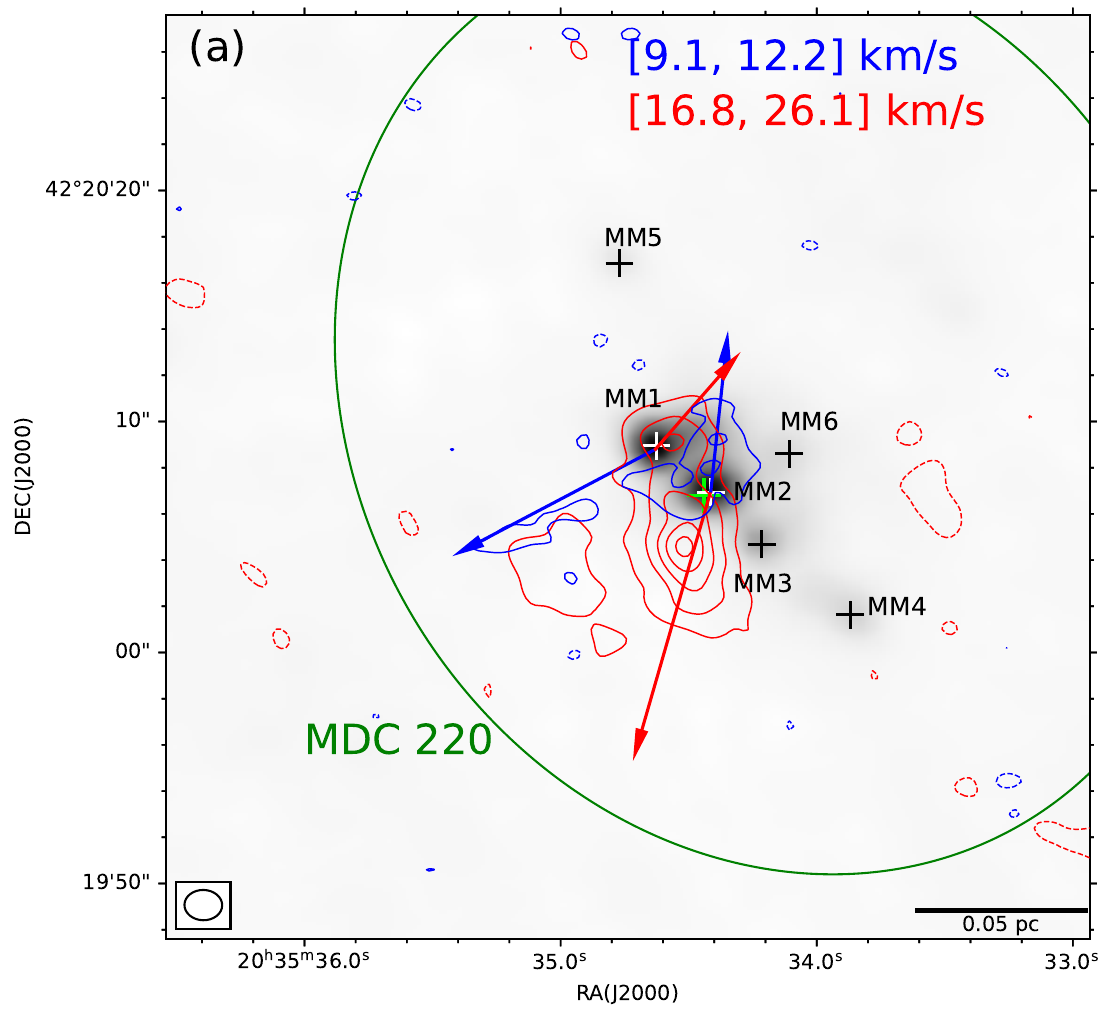} 
   \includegraphics[width=240pt]{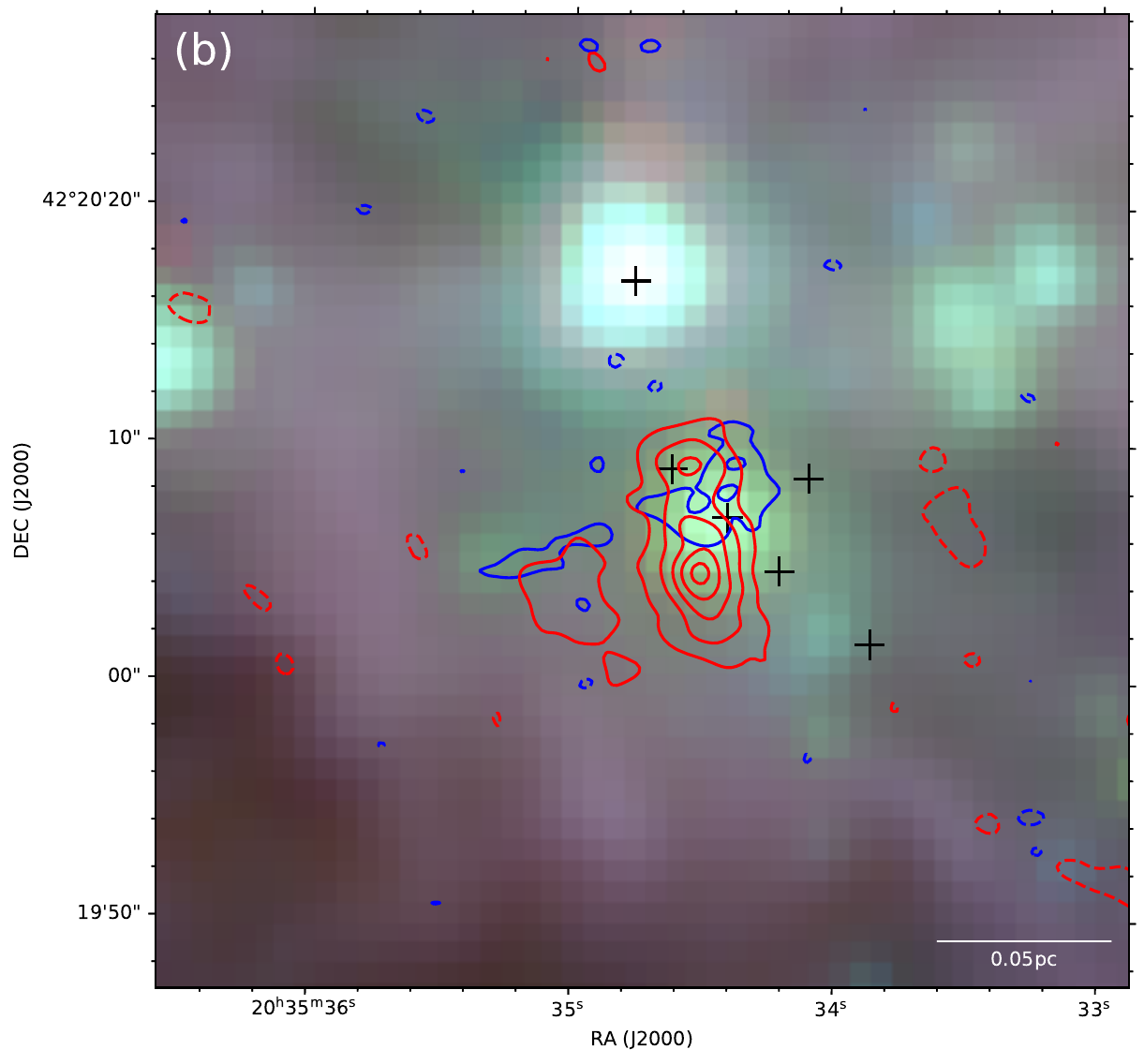}
     \caption{High-velocity ($\Delta$v $\ge$ 2.3 km s$^{-1}$) SiO (5$-$4) emission and the Spitzer IRAC three-color map for MDC 220. In each panel, a scale bar is presented at the bottom-right corner. (a) The integrated velocity ranges of the contours are presented at the top-right corner. The blue contour levels are ($-$3, 3, 6) $\times$ $\sigma$, with $\sigma$ = 0.16 Jy beam$^{-1}$ km s$^{-1}$, and the red contour levels are ($-$3, 3, 6, 9, 12, 15) $\times$ $\sigma$, with $\sigma$ = 0.30 Jy beam$^{-1}$ km s$^{-1}$. The grayscale maps represent 1.37 mm continuum emissions. The CO blueshifted and redshifted outflows are indicated by solid blue and red arrows following \citet{Pan2023}. The black and white crosses represent dust condensations identified by \citet{2021ApJ...918L...4C}. The green cross indicates the position of the protostar candidates identified in \citet{2014AJ....148...11K}. The synthesized beam is shown in the bottom-left corner. The green solid ellipses represent the full width at half maximum (FWHM) of the MDC 220 obtained from \citet{2021ApJ...918L...4C}. (b) Spitzer three-color composite image with the 3.6, 4.5, and 8.0 $\mu$m emissions coded in blue, green, and red, respectively.}
  \label{N03}
\end{figure*}

\begin{figure*}
\centering
   \includegraphics[width=240pt]{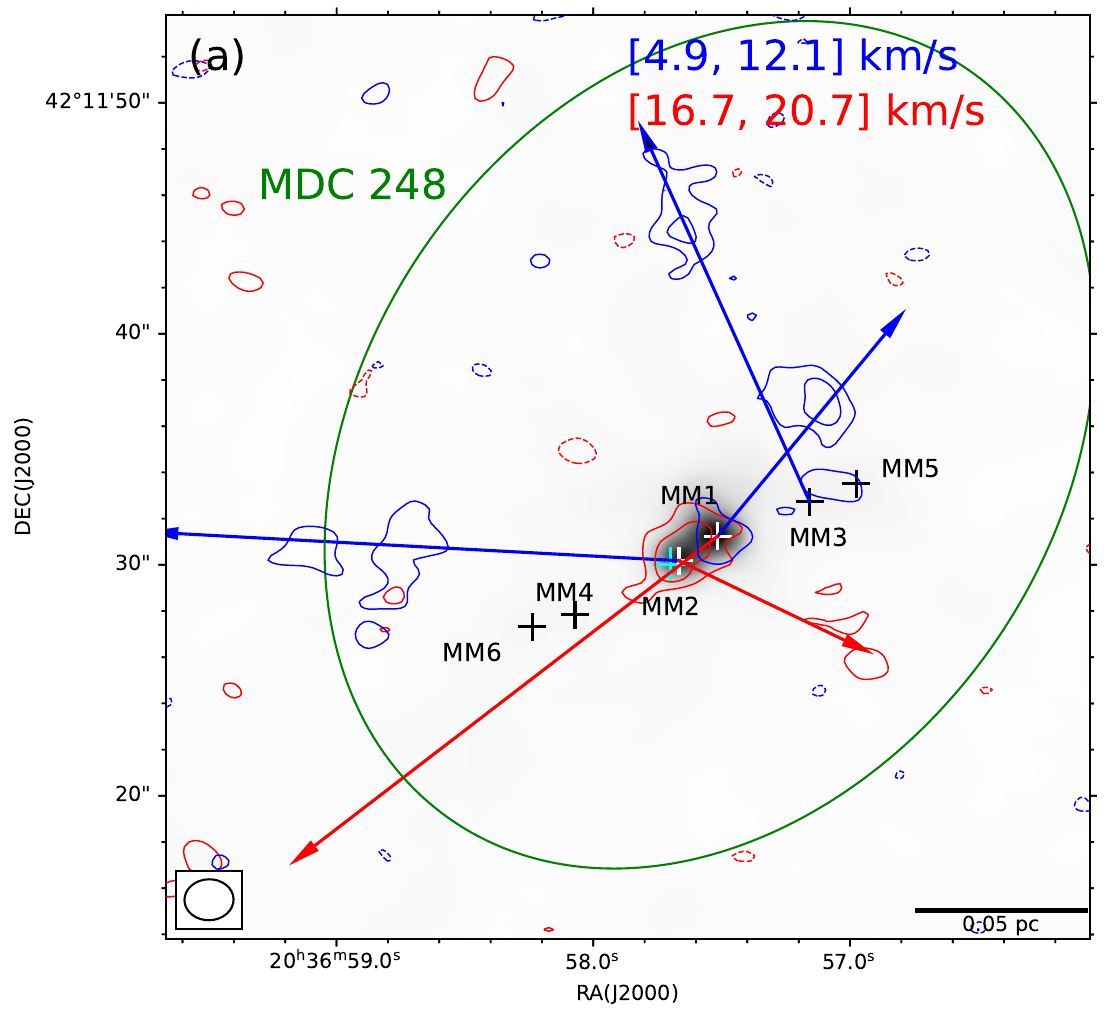} 
   \includegraphics[width=240pt]{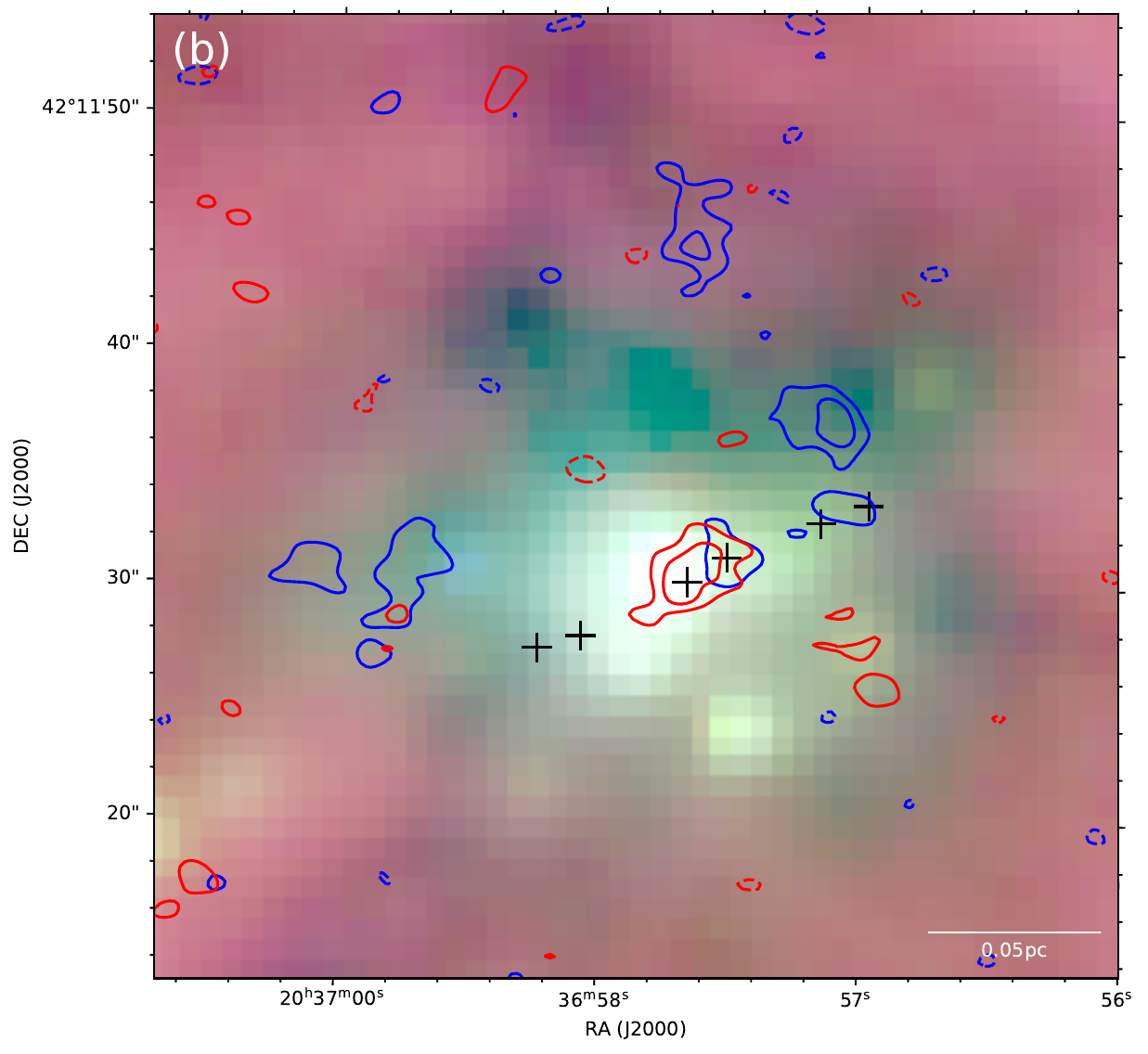}
     \caption{Same convention as Figure \ref{N03} but for MDC 248. The blue contour levels are ($-$3, 3, 6) $\times$ $\sigma$, with $\sigma$ = 0.25 Jy beam$^{-1}$ km s$^{-1}$, and the red contour levels are ($-$3, 3, 6) $\times$ $\sigma$, with $\sigma$ = 0.2 Jy beam$^{-1}$ km s$^{-1}$. (a) The cyan crosses represent water maser spots obtained from the JVLA program 17A-107 (PI: Keping Qiu). } 
  \label{N12}
\end{figure*}

\begin{figure*}
\centering
   \includegraphics[width=240pt]{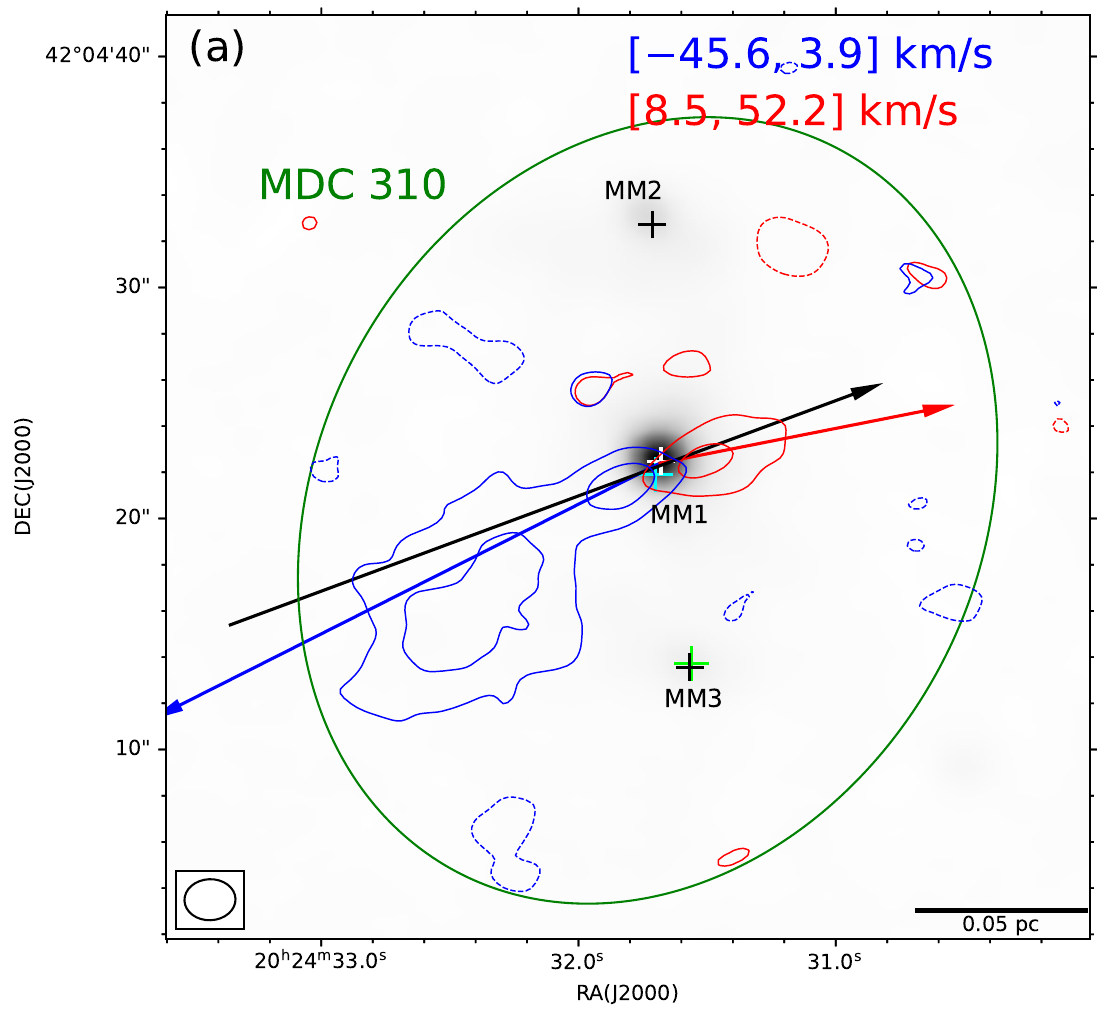} 
   \includegraphics[width=240pt]{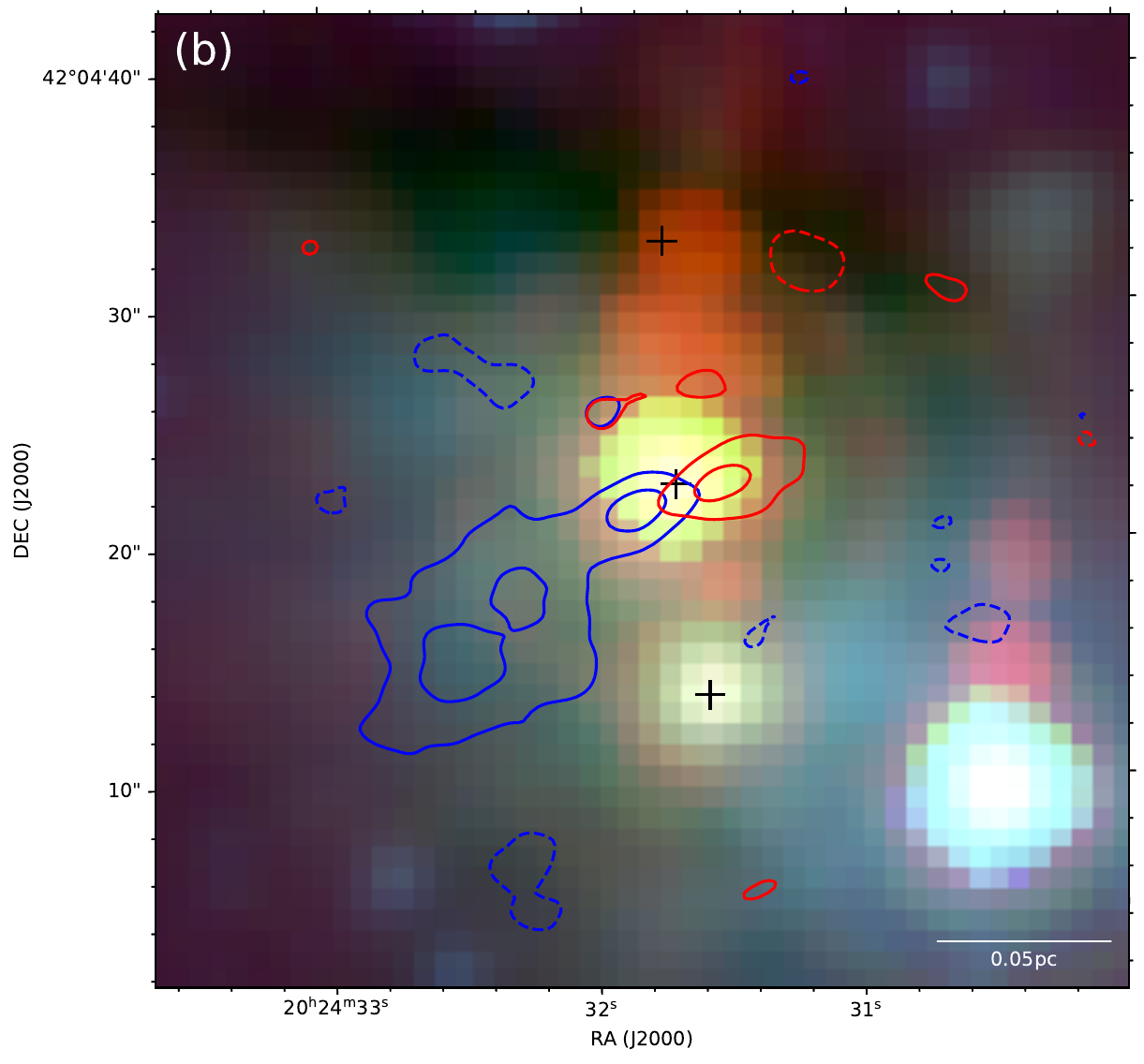}
     \caption{Same convention as Figure \ref{N03} but for MDC 310. The blue contour levels are ($-$3, 3, 6, 9) $\times$ $\sigma$, with $\sigma$ = 1.30 Jy beam$^{-1}$ km s$^{-1}$, and the red contour levels are ($-$3, 3, 6, 9) $\times$ $\sigma$, with $\sigma$ = 1.40 Jy beam$^{-1}$ km s$^{-1}$. (a) The cyan crosses represent water maser spots obtained from the JVLA program 17A-107 (PI: Keping Qiu) and the green cross points out the position of the protostar candidates identified in \citet{2014AJ....148...11K}. The black arrow line presents the PV cut path.}
  \label{NW14}
\end{figure*}

\begin{figure*}
\centering
   \includegraphics[width=240pt]{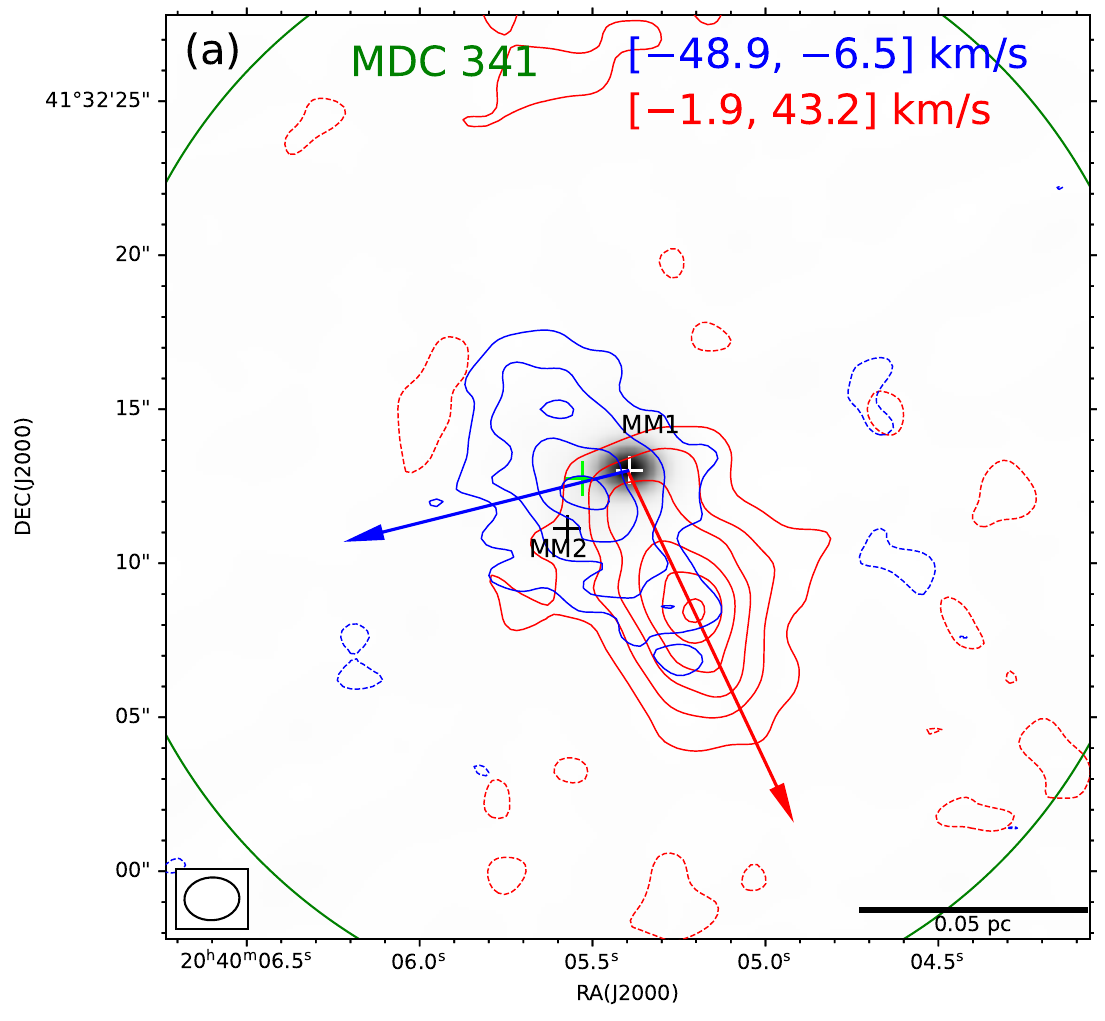} 
   \includegraphics[width=240pt]{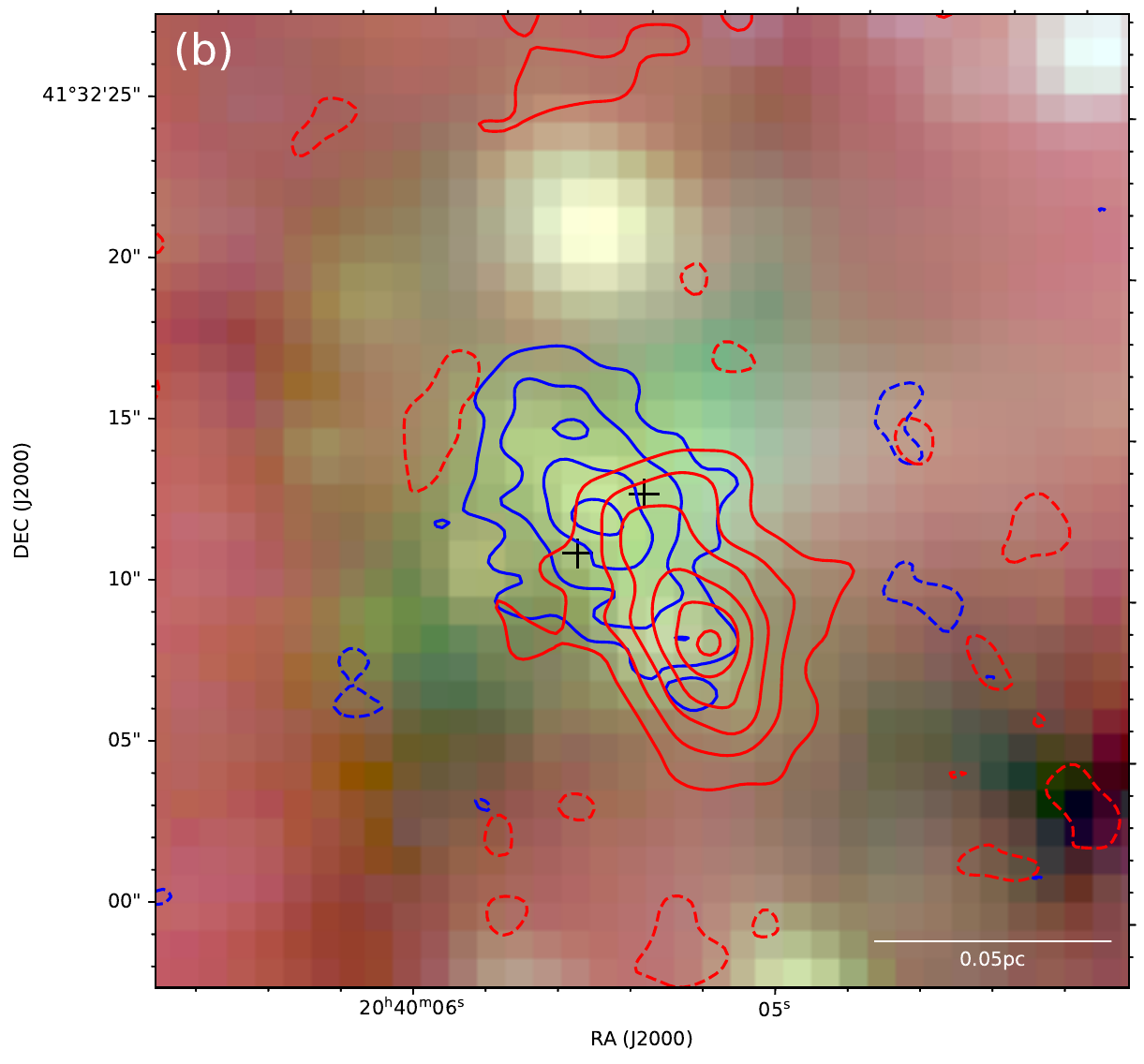}
     \caption{Same convention as Figure \ref{N03} but for MDC 341. The blue contour levels are ($-$3, 3, 6, 9, 12) $\times$ $\sigma$, with $\sigma$ = 1.00 Jy beam$^{-1}$ km s$^{-1}$, and the red contour levels are ($-$3, 3, 6, 9, 12, 15, 18) $\times$ $\sigma$, with $\sigma$ = 1.00 Jy beam$^{-1}$ km s$^{-1}$. (a) The green cross points out the position of the protostar candidates identified in \citet{2014AJ....148...11K}.}
  \label{N63}
\end{figure*}

\begin{figure*}
\centering
   \includegraphics[width=240pt]{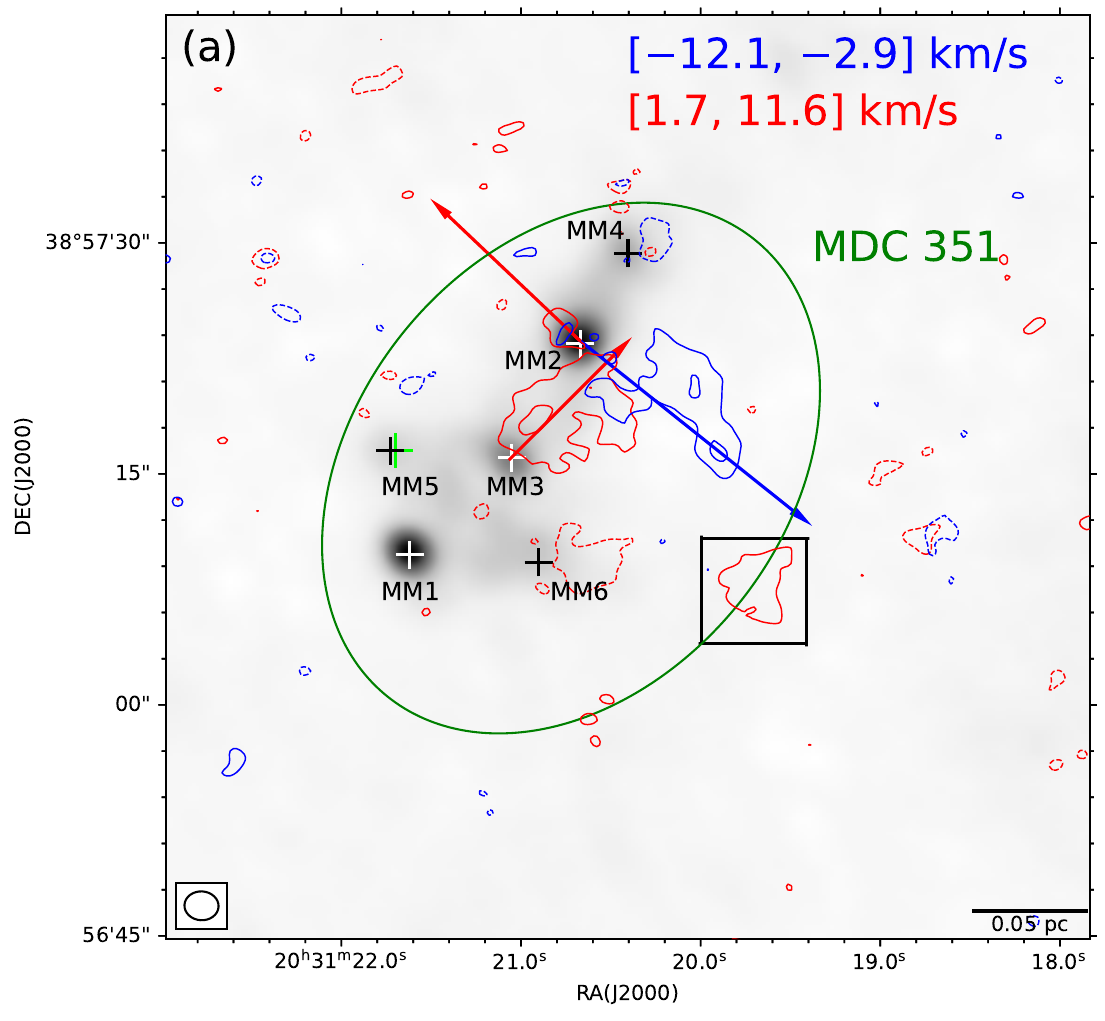} 
   \includegraphics[width=240pt]{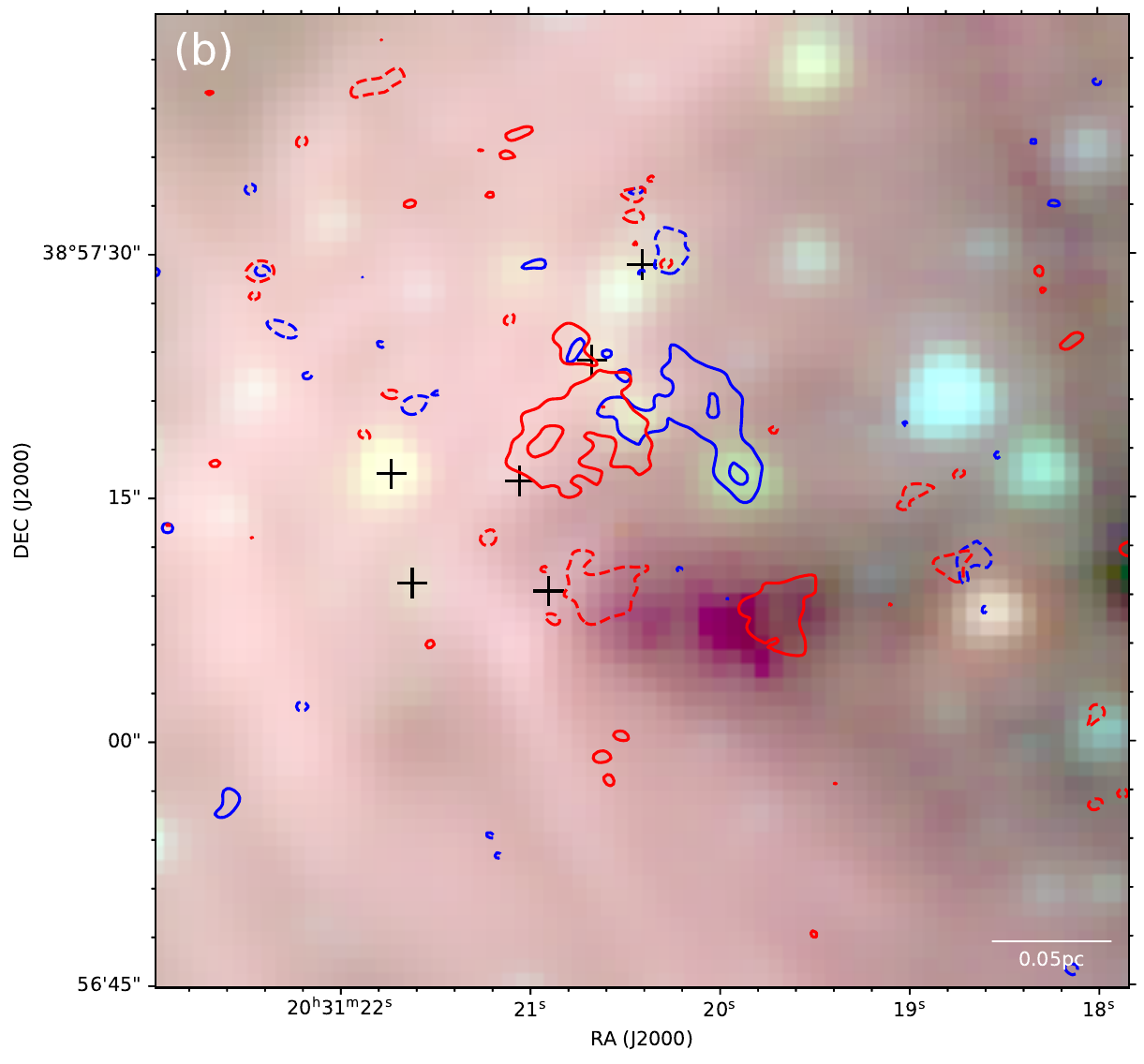}
     \caption{Same convention as Figure \ref{N03} but for MDC 351. The blue contour levels are ($-$3, 3, 6) $\times$ $\sigma$, with $\sigma$ = 0.54 Jy beam$^{-1}$ km s$^{-1}$, and the red contour levels are ($-$3, 3, 6) $\times$ $\sigma$, with $\sigma$ = 0.59 Jy beam$^{-1}$ km s$^{-1}$. (a) The green cross indicates the position of the protostar candidate identified in \citet{2014AJ....148...11K}.}
  \label{S32}
\end{figure*}

\begin{figure*}
\centering
   \includegraphics[width=220pt]{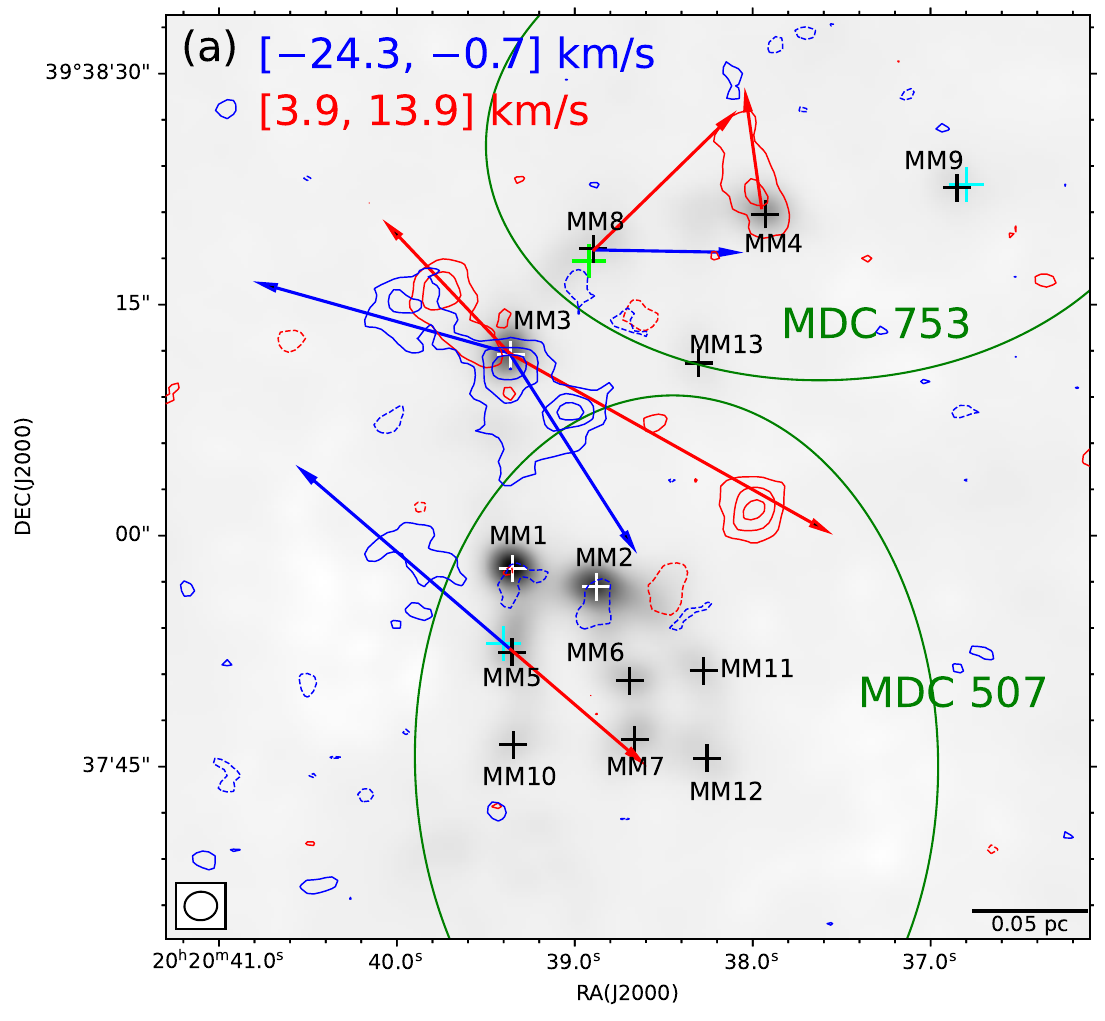} 
   \includegraphics[width=220pt]{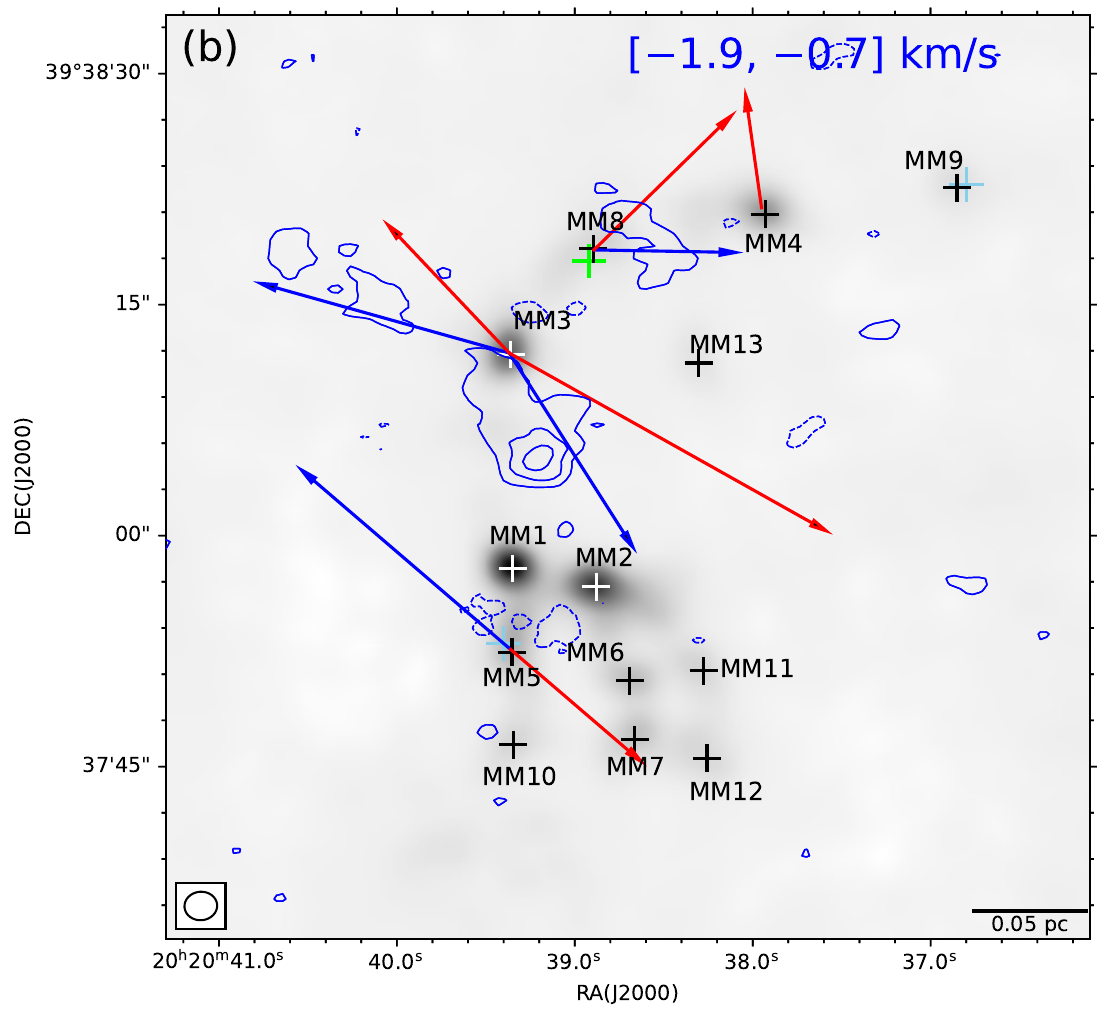} 
   
   \includegraphics[width=220pt]{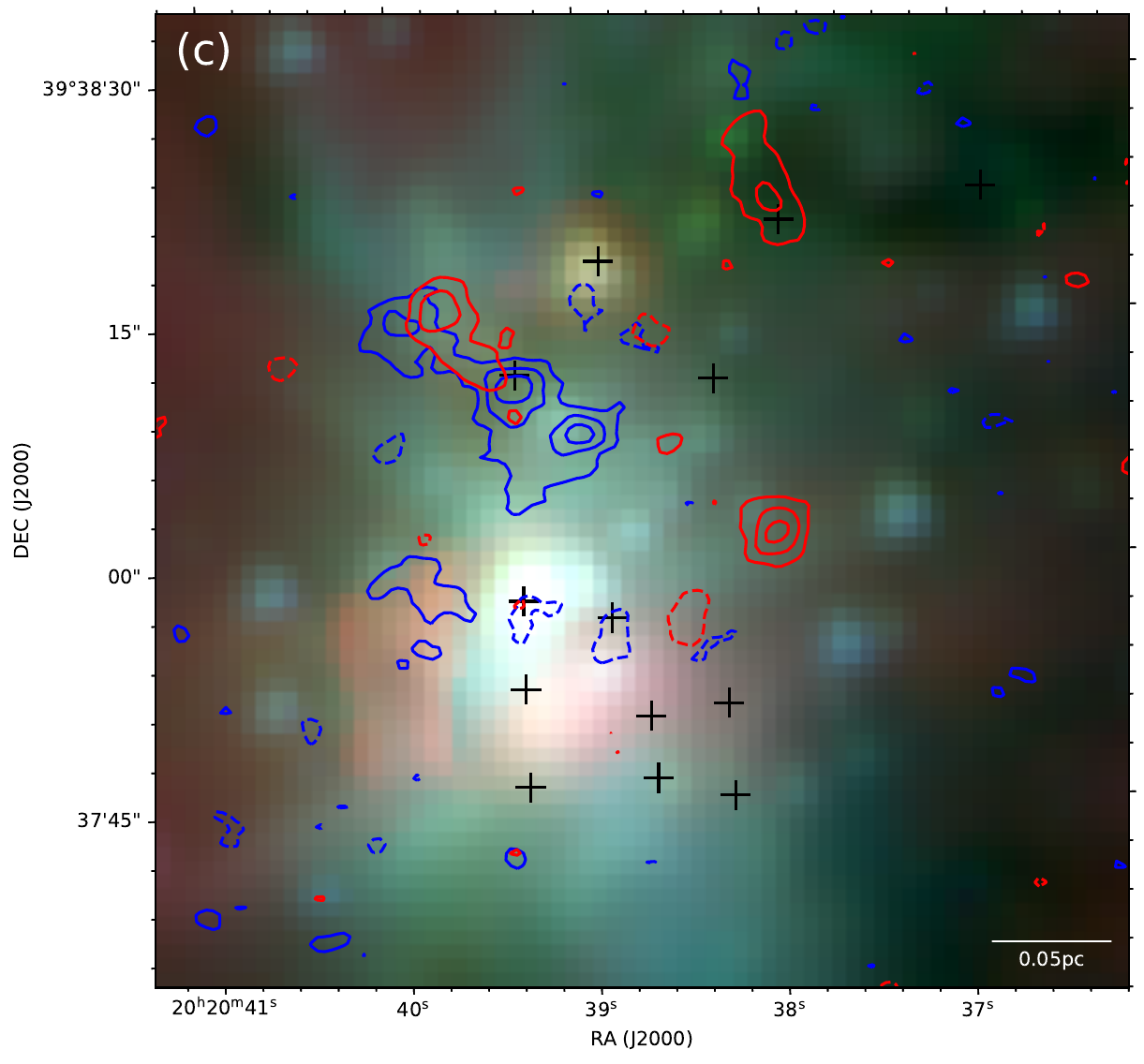}
     \caption{Same convention as Figure \ref{N03} but for MDC 507/753. (a, c) The blue contour levels are ($-$3, 3, 6, 9) $\times$ $\sigma$, with $\sigma$ = 0.48 Jy beam$^{-1}$ km s$^{-1}$, and the red contour levels are ($-$3, 3, 6, 9) $\times$ $\sigma$, with $\sigma$ = 0.29 Jy beam$^{-1}$ km s$^{-1}$. (b) The blue contour levels are ($-$3, 3, 6, 9) $\times$ $\sigma$, with $\sigma$ = 0.10 Jy beam$^{-1}$ km s$^{-1}$. (a, b) The cyan crosses represent water maser spots obtained from the JVLA program 17A-107 (PI: Keping Qiu) and the green cross shows the position of the protostar candidate identified in \citet{2014AJ....148...11K}.}
  \label{S07}
\end{figure*}

\begin{figure*}
\centering
   \includegraphics[width=240pt]{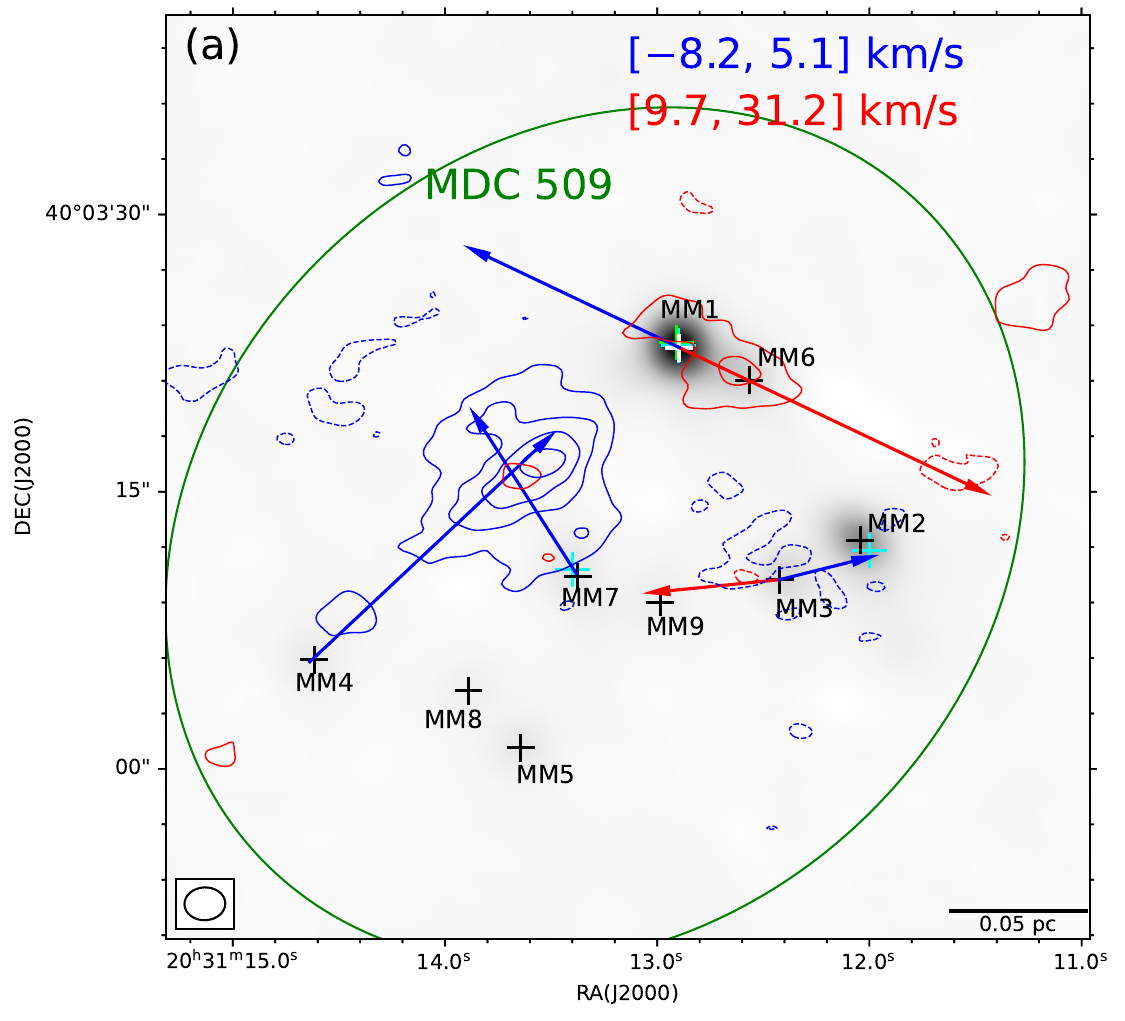} 
   \includegraphics[width=240pt]{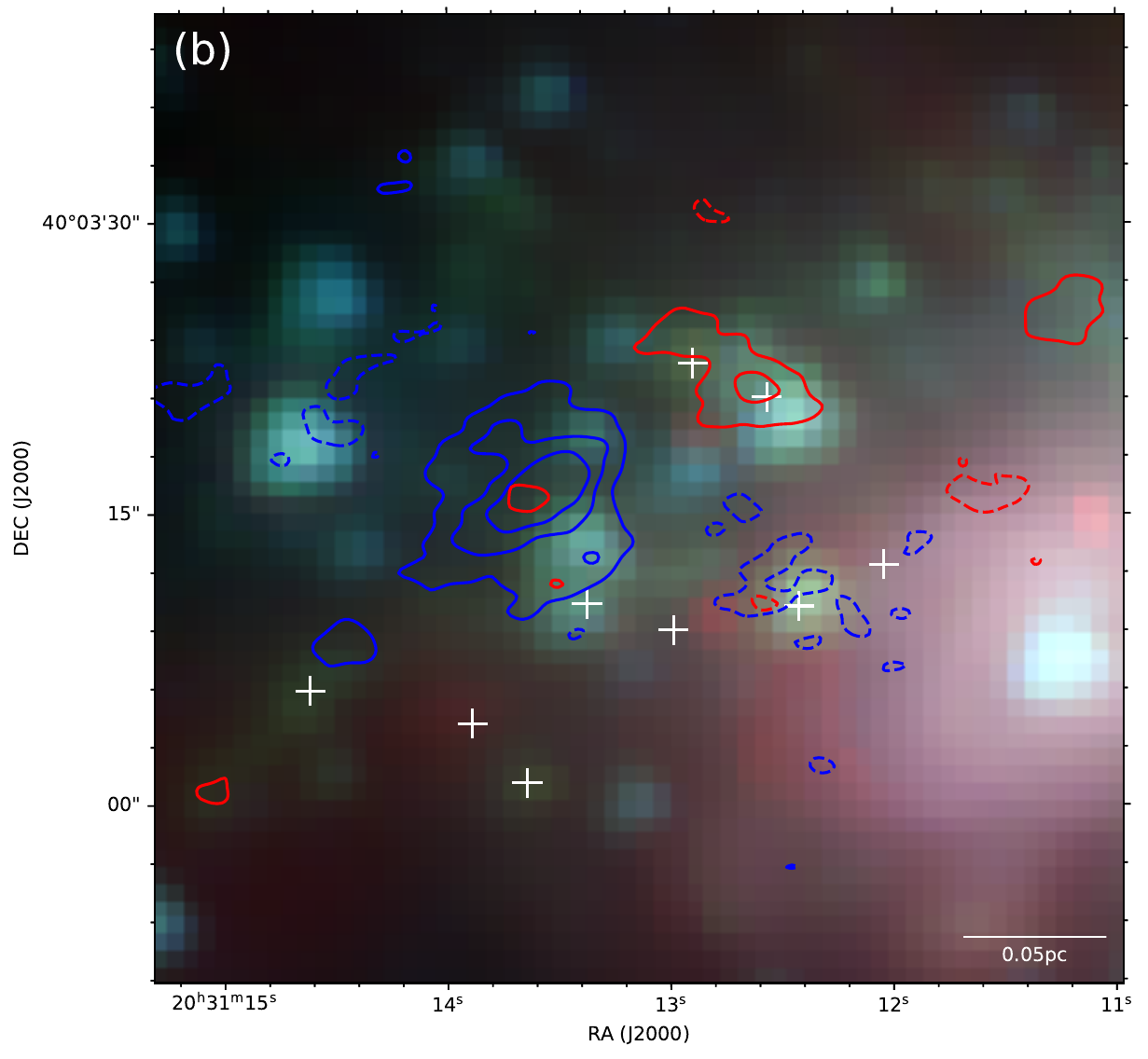}
     \caption{Same convention as Figure \ref{N03} but for MDC 509. The blue contour levels are ($-$3, 3, 6, 9, 12) $\times$ $\sigma$, with $\sigma$ = 0.92 Jy beam$^{-1}$ km s$^{-1}$, and the red contour levels are ($-$3, 3, 6) $\times$ $\sigma$, with $\sigma$ = 0.97 Jy beam$^{-1}$ km s$^{-1}$. (a) The cyan crosses represent water maser spots obtained from the JVLA program 17A-107 (PI: Keping Qiu) and the green cross shows the position of the protostar candidate identified in \citet{2014AJ....148...11K}.}
  \label{S30}
\end{figure*}

\begin{figure*}
\centering
   \includegraphics[width=240pt]{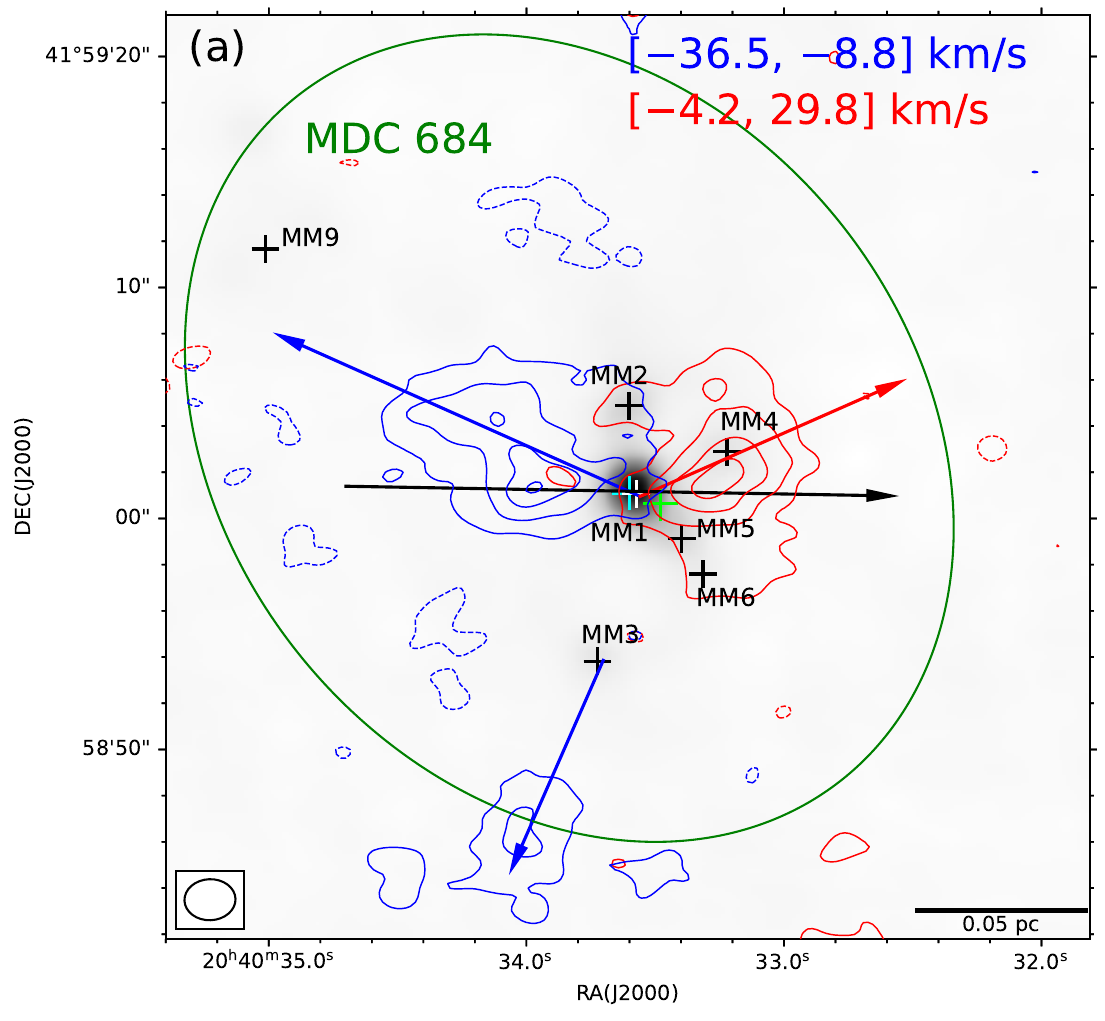} 
   \includegraphics[width=240pt]{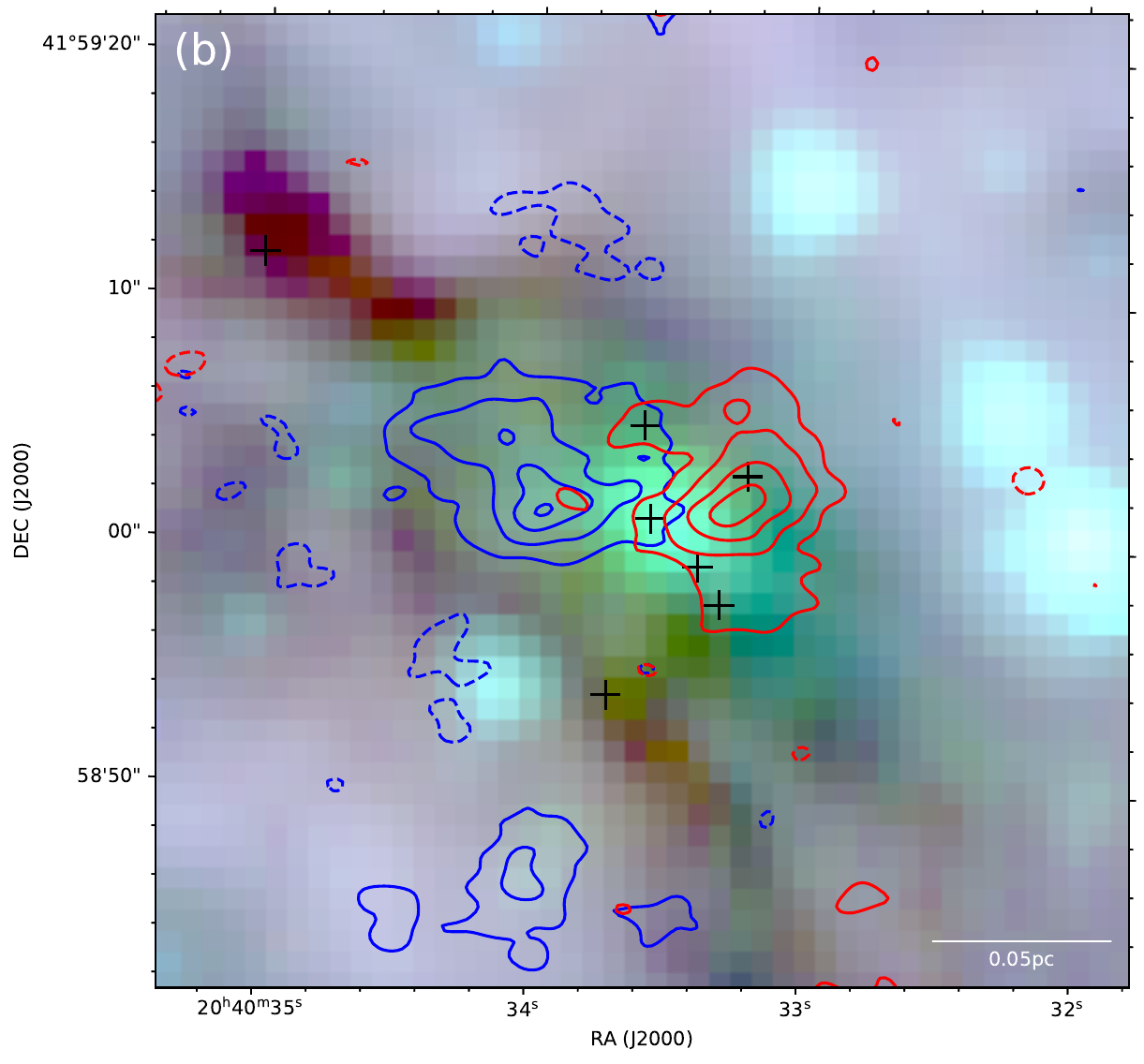}
     \caption{Same convention as Figure \ref{N03} but for MDC 684. The blue contour levels are ($-$3, 3, 6, 9, 12) $\times$ $\sigma$, with $\sigma$ = 0.88 Jy beam$^{-1}$ km s$^{-1}$, and the red contour levels are ($-$3, 3, 6, 9, 12) $\times$ $\sigma$, with $\sigma$ = 0.78 Jy beam$^{-1}$ km s$^{-1}$. (a) The cyan crosses represent water maser spots obtained from the JVLA program 17A-107 (PI: Keping Qiu) and the green cross shows the position of the protostar candidate identified in \citet{2014AJ....148...11K}. The black arrow line indicates the PV cut path.}
  \label{N68}
\end{figure*}

\begin{figure*}
\centering
   \includegraphics[width=240pt]{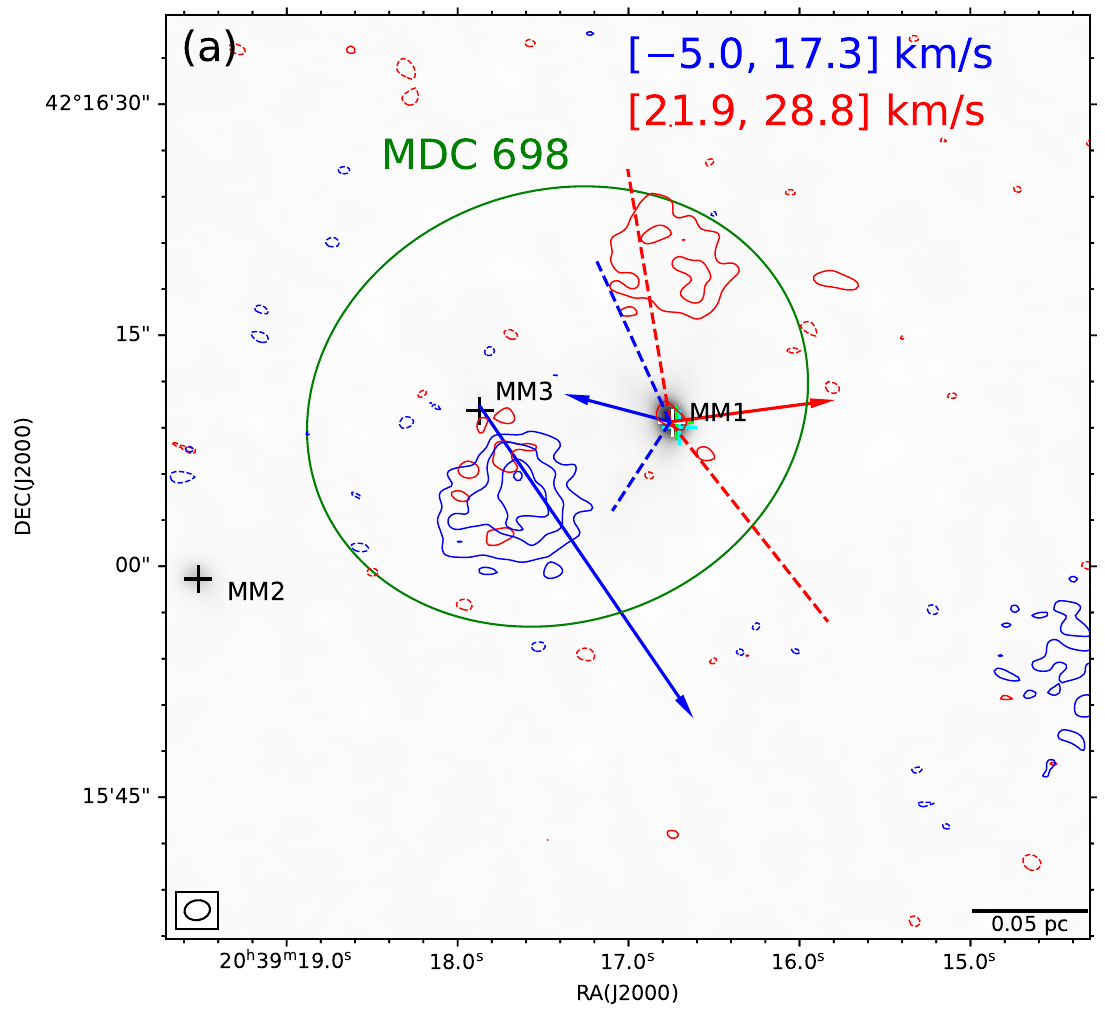} 
   \includegraphics[width=240pt]{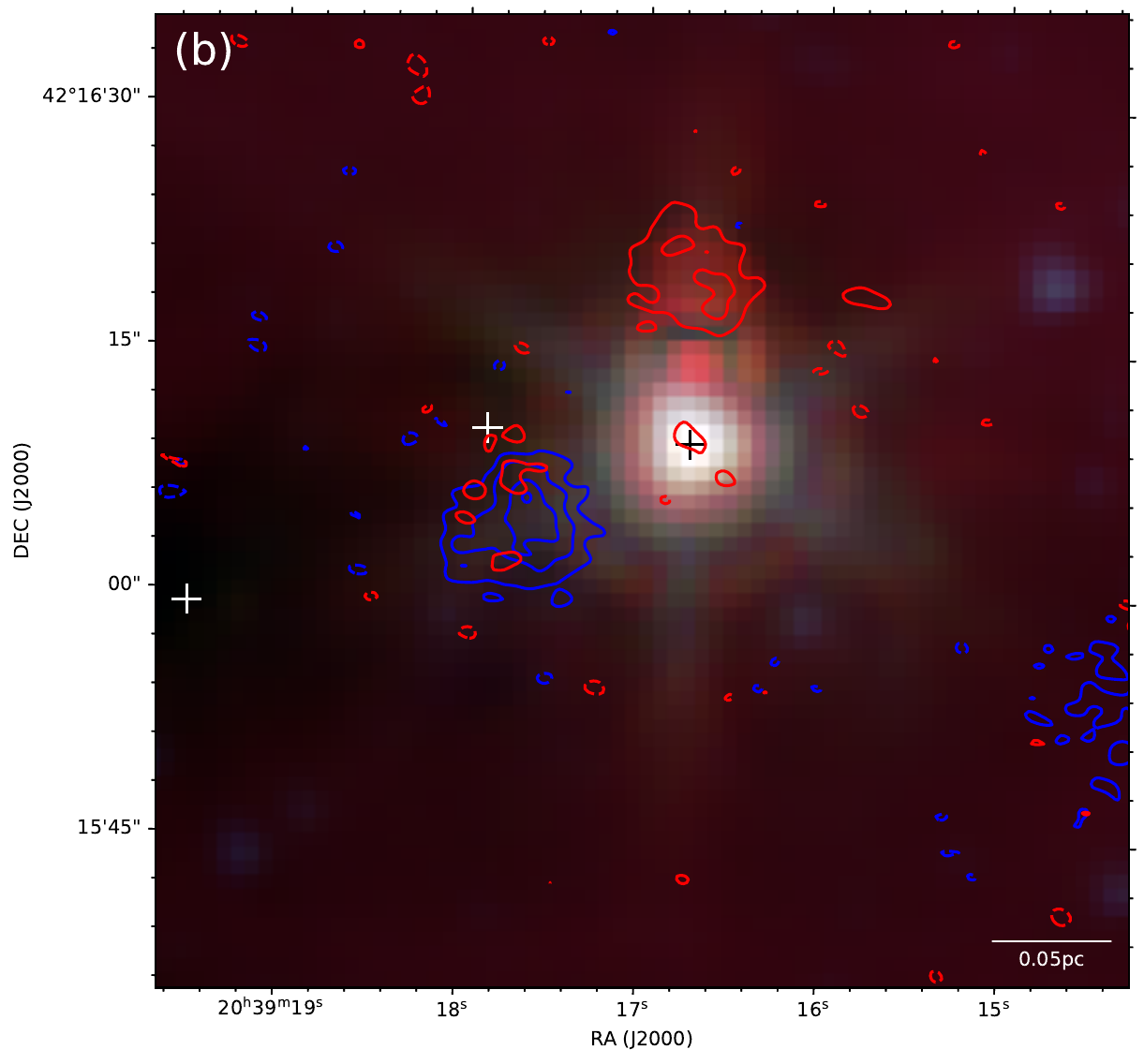}
     \caption{Same convention as Figure \ref{N03} but for MDC 698. The blue contour levels are ($-$3, 3, 6, 9, 12) $\times$ $\sigma$, with $\sigma$ = 0.69 Jy beam$^{-1}$ km s$^{-1}$, and the red contour levels are ($-$3, 3, 6) $\times$ $\sigma$, with $\sigma$ = 0.29 Jy beam$^{-1}$ km s$^{-1}$. (a) The cyan crosses represent water maser spots obtained from the JVLA program 17A-107 (PI: Keping Qiu) and the green cross points out the position of the protostar candidate identified in \citet{2014AJ....148...11K}.}
  \label{N56}
\end{figure*}

\begin{figure*}
\centering
   \includegraphics[width=240pt]{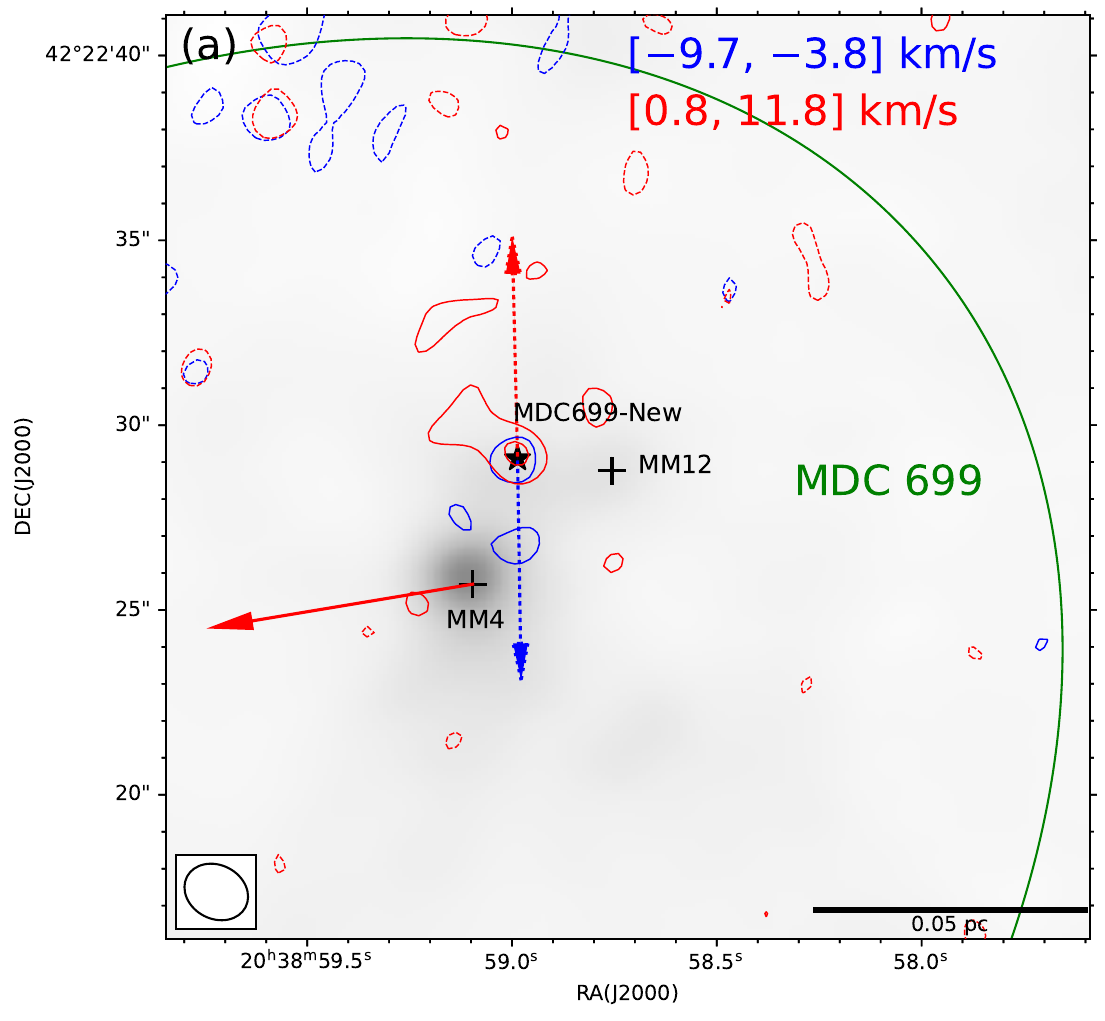} 
   \includegraphics[width=240pt]{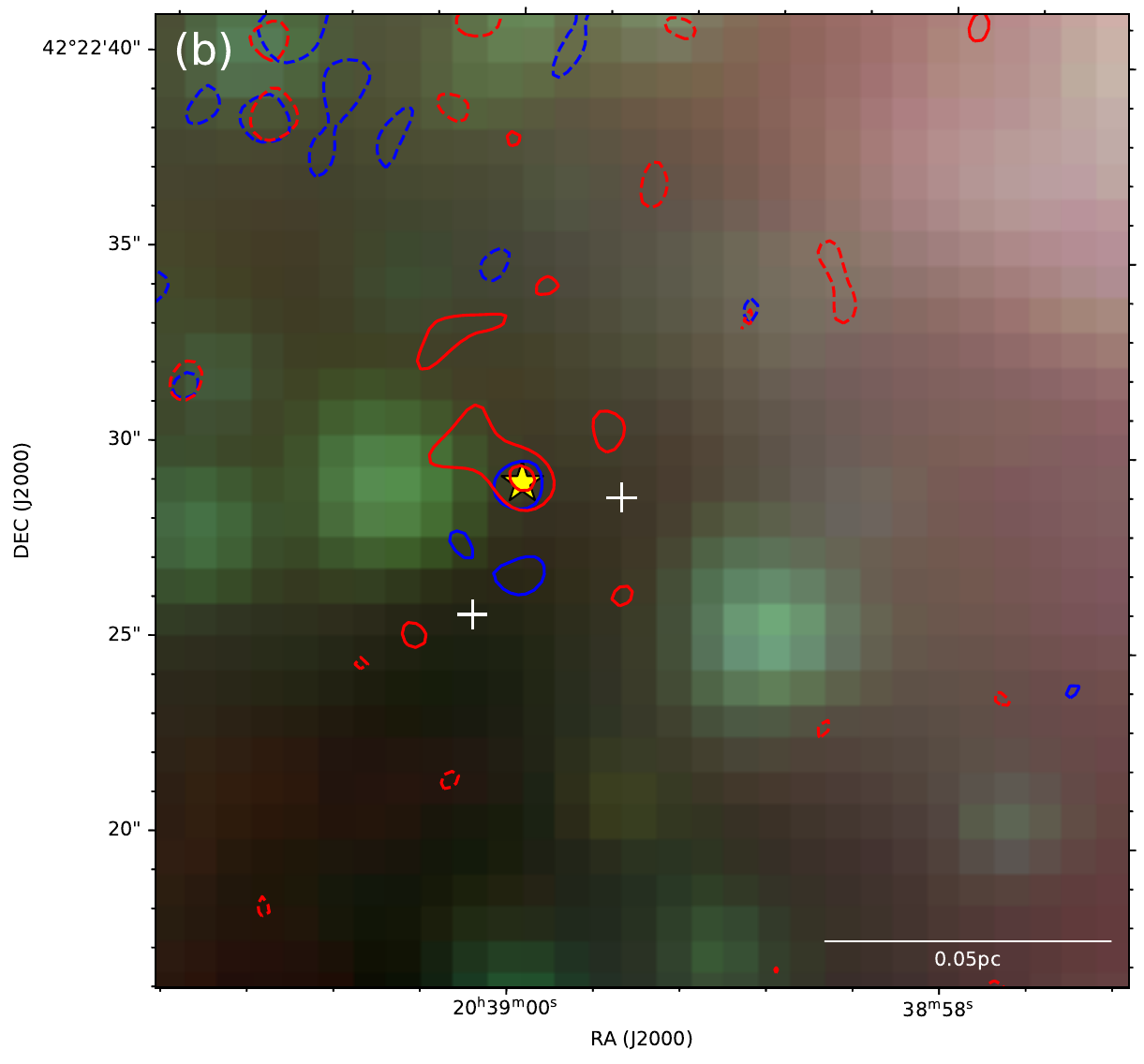}
     \caption{Same convention as Figure \ref{N03} but for MDC 699. The blue contour levels are ($-$3, 3) $\times$ $\sigma$, with $\sigma$ = 0.72 Jy beam$^{-1}$ km s$^{-1}$, and the red contour levels are ($-$3, 3, 6) $\times$ $\sigma$, with $\sigma$ = 0.90 Jy beam$^{-1}$ km s$^{-1}$.}
  \label{N38}
\end{figure*}

\begin{figure*}
\centering
   \includegraphics[width=240pt]{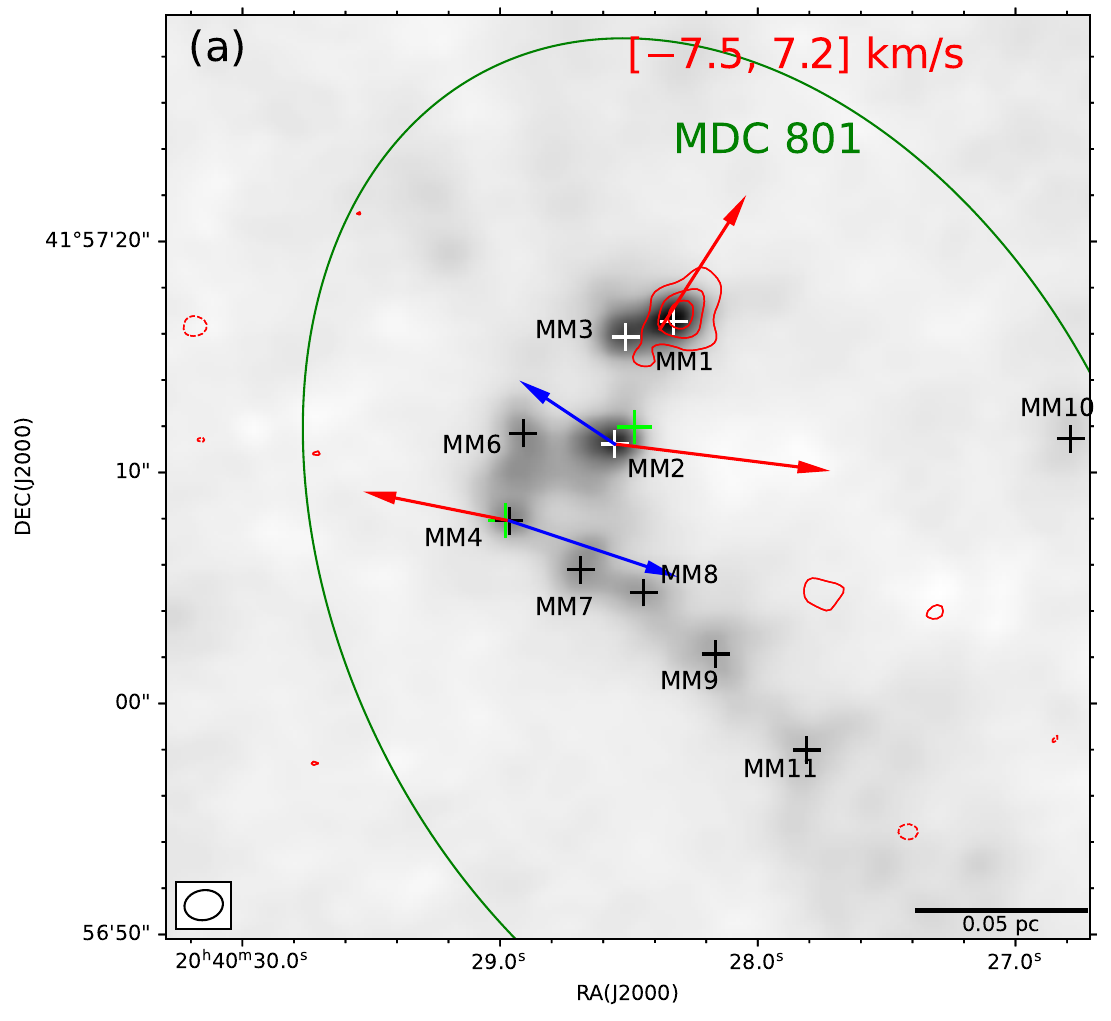} 
   \includegraphics[width=240pt]{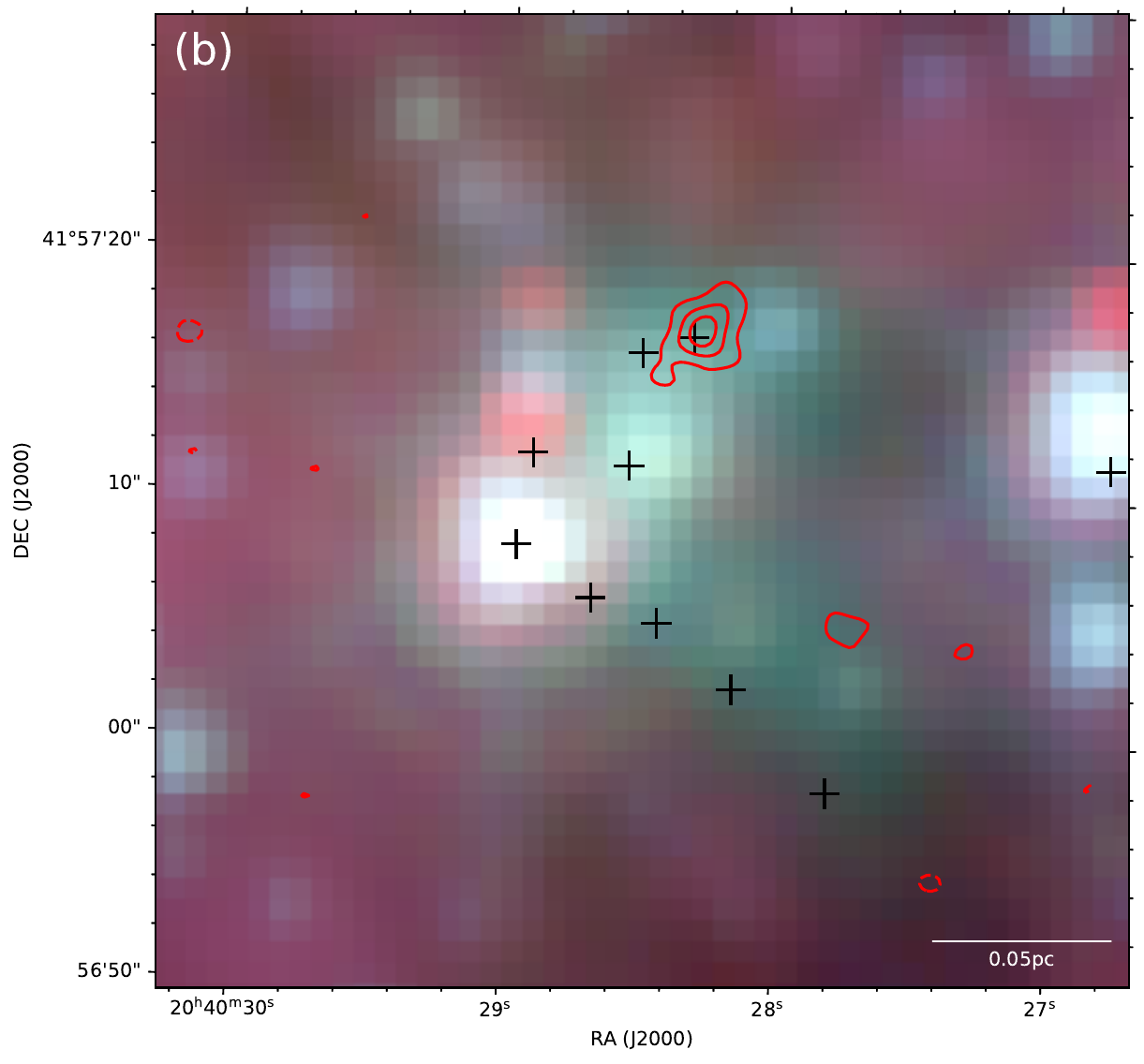}
     \caption{Same convention as Figure \ref{N03} but for MDC 801. The red contour levels are ($-$3, 3, 6, 9) $\times$ $\sigma$, with $\sigma$ = 0.45 Jy beam$^{-1}$ km s$^{-1}$. (a) The green cross points out the position of the protostar candidate identified in \citet{2014AJ....148...11K}.}
  \label{N65}
\end{figure*}

\begin{figure*}
\centering
   \includegraphics[width=240pt]{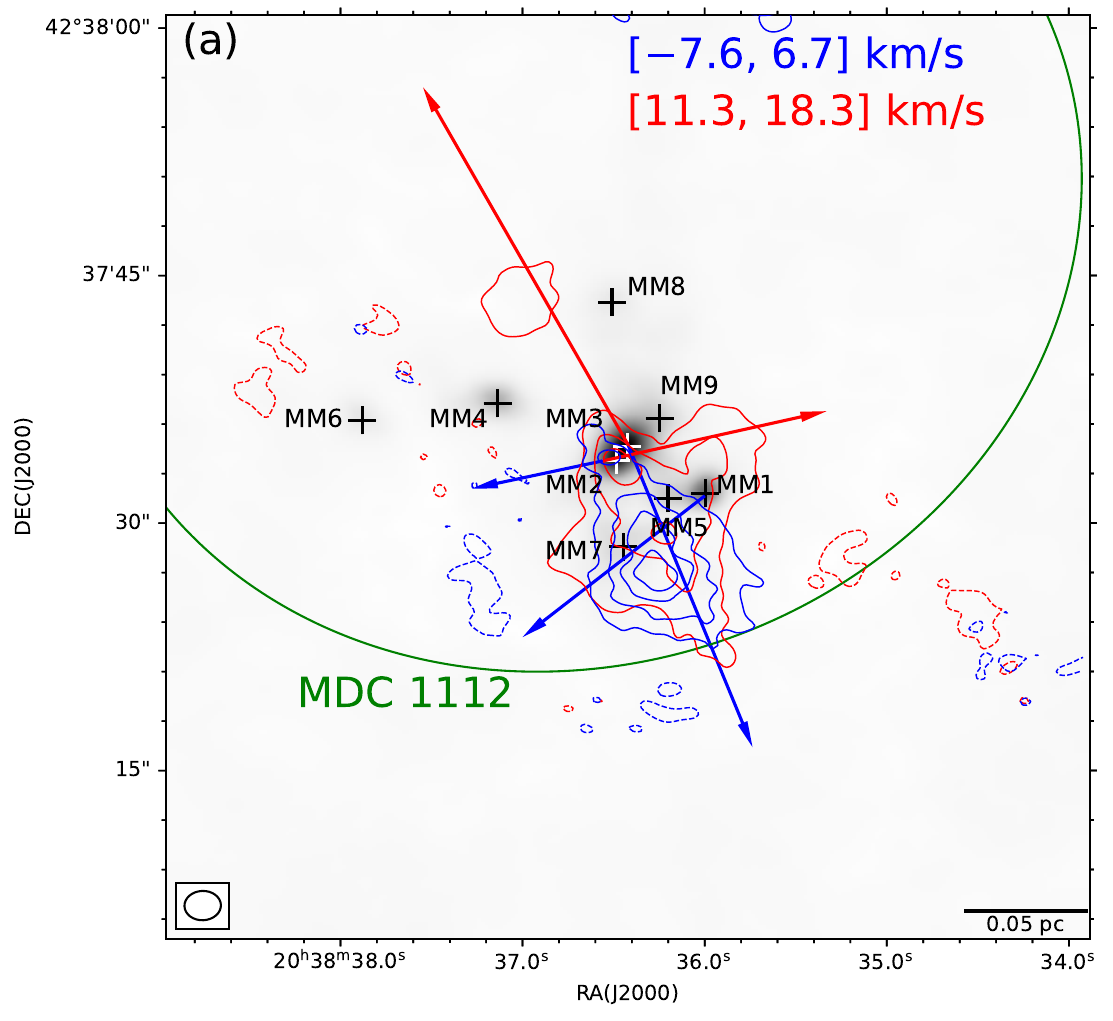} 
   \includegraphics[width=240pt]{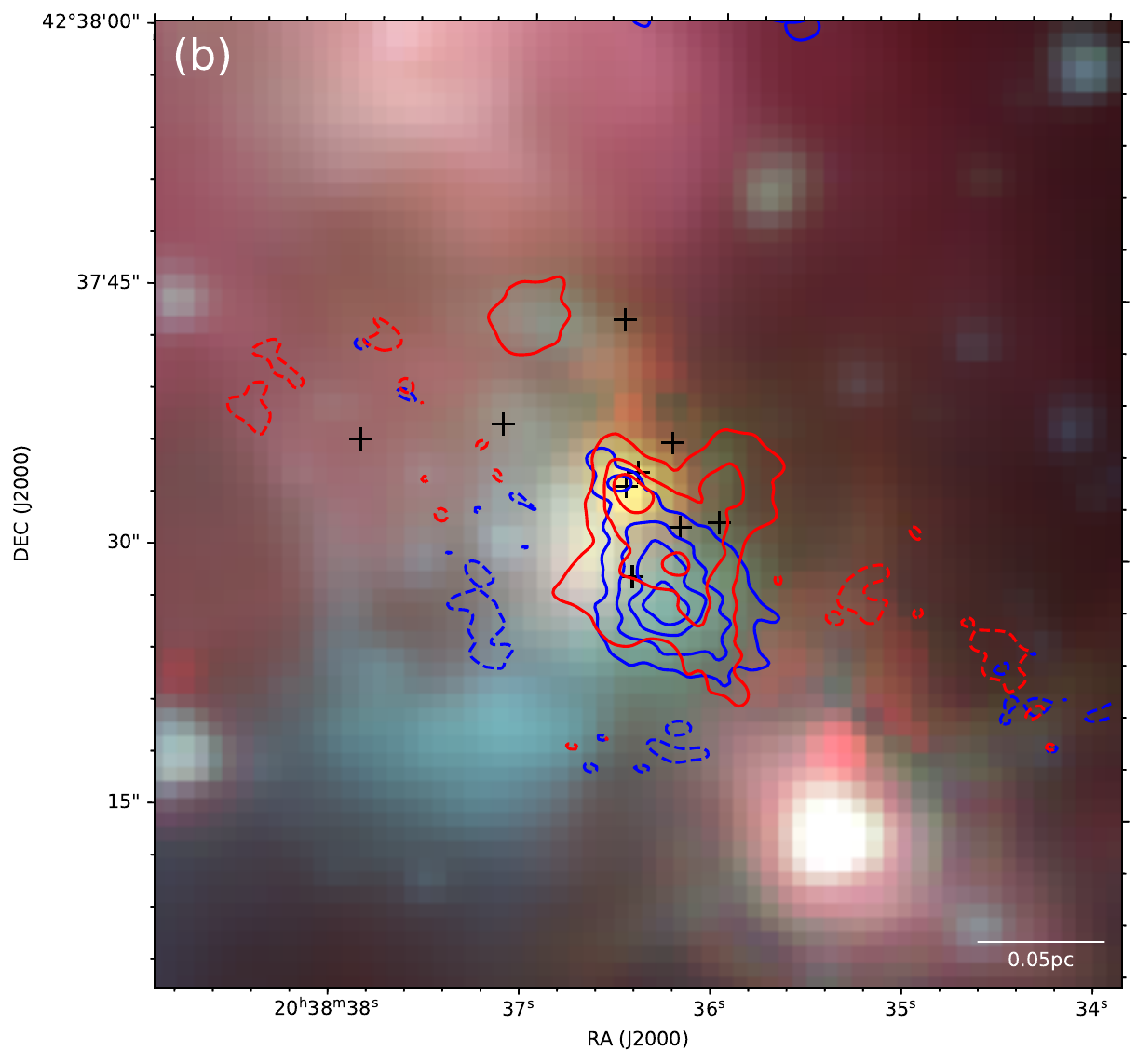}
     \caption{Same convention as Figure \ref{N03} but for MDC 1112. The blue contour levels are ($-$3, 3, 6, 9, 12) $\times$ $\sigma$, with $\sigma$ = 0.79 Jy beam$^{-1}$ km s$^{-1}$, and the red contour levels are ($-$3, 3, 6, 9) $\times$ $\sigma$, with $\sigma$ = 0.61 Jy beam$^{-1}$ km s$^{-1}$.}
  \label{N30}
\end{figure*}

\begin{figure*}
\centering
   \includegraphics[width=240pt]{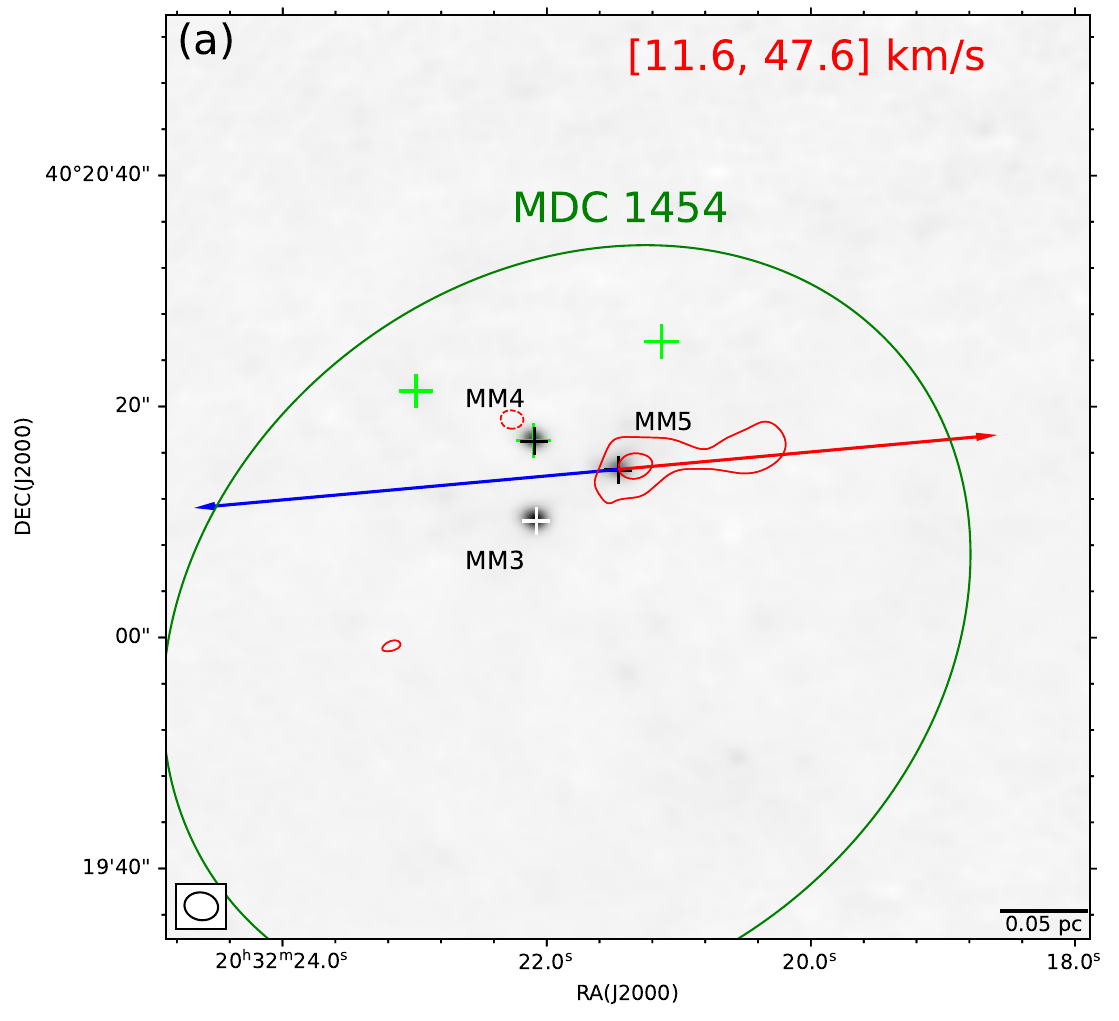} 
   \includegraphics[width=240pt]{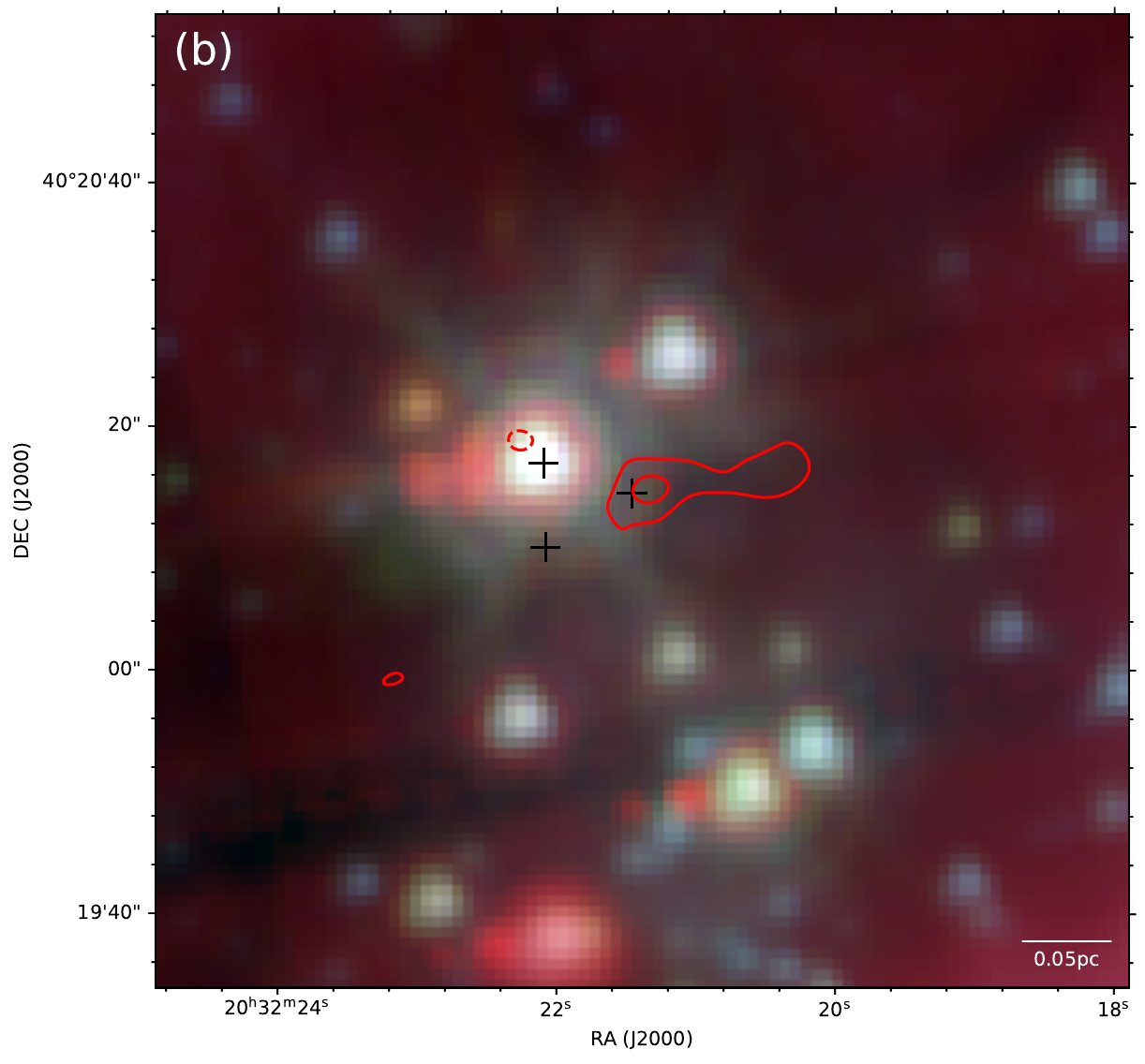}
     \caption{Same convention as Figure \ref{N03} but for MDC 1454. The red contour levels are ($-$3, 3, 6) $\times$ $\sigma$, with $\sigma$ = 1.20 Jy beam$^{-1}$ km s$^{-1}$. (a) The green cross shows the position of the protostar candidate identified in \citet{2014AJ....148...11K}.}
  \label{DR15}
\end{figure*}

\begin{figure*}
\centering
   \includegraphics[width=240pt]{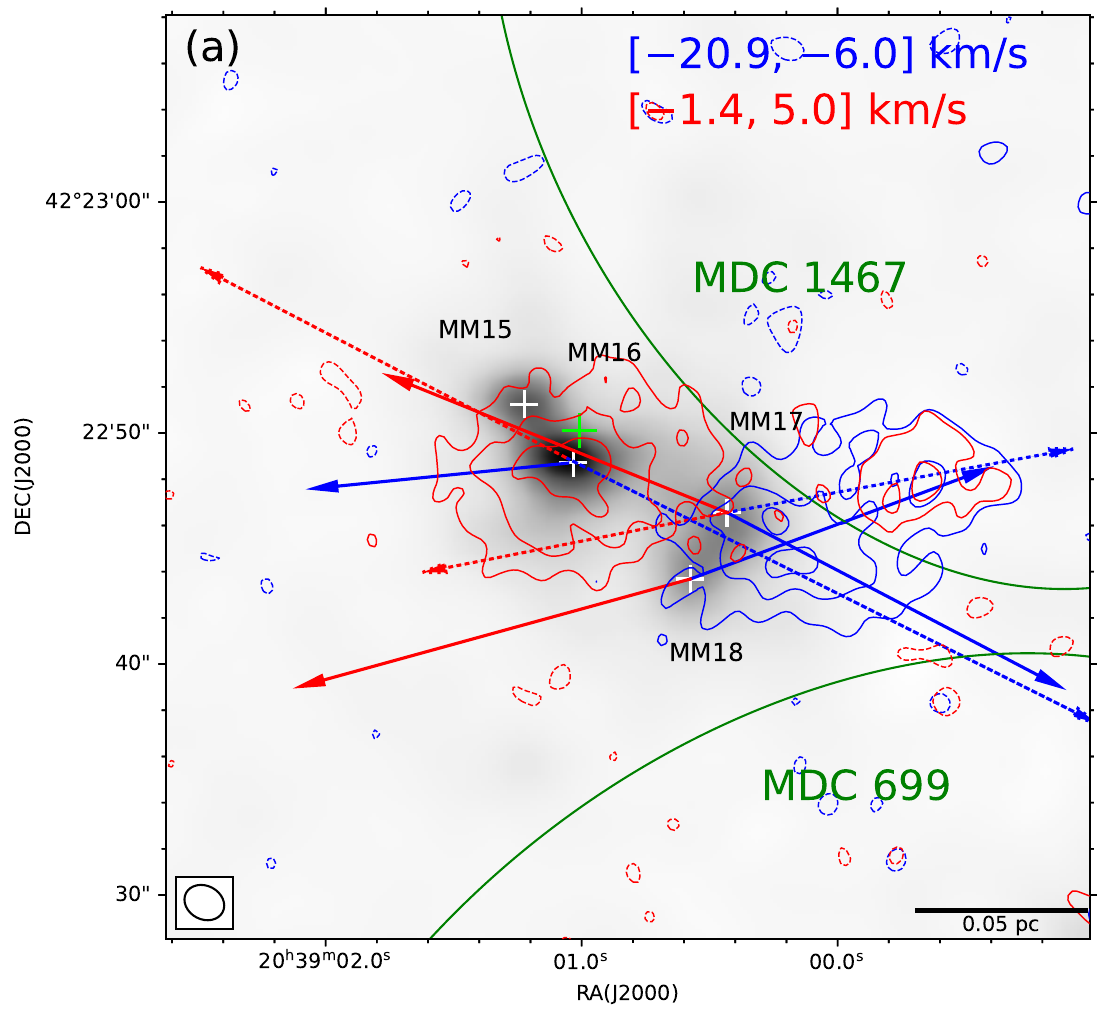} 
   \includegraphics[width=240pt]{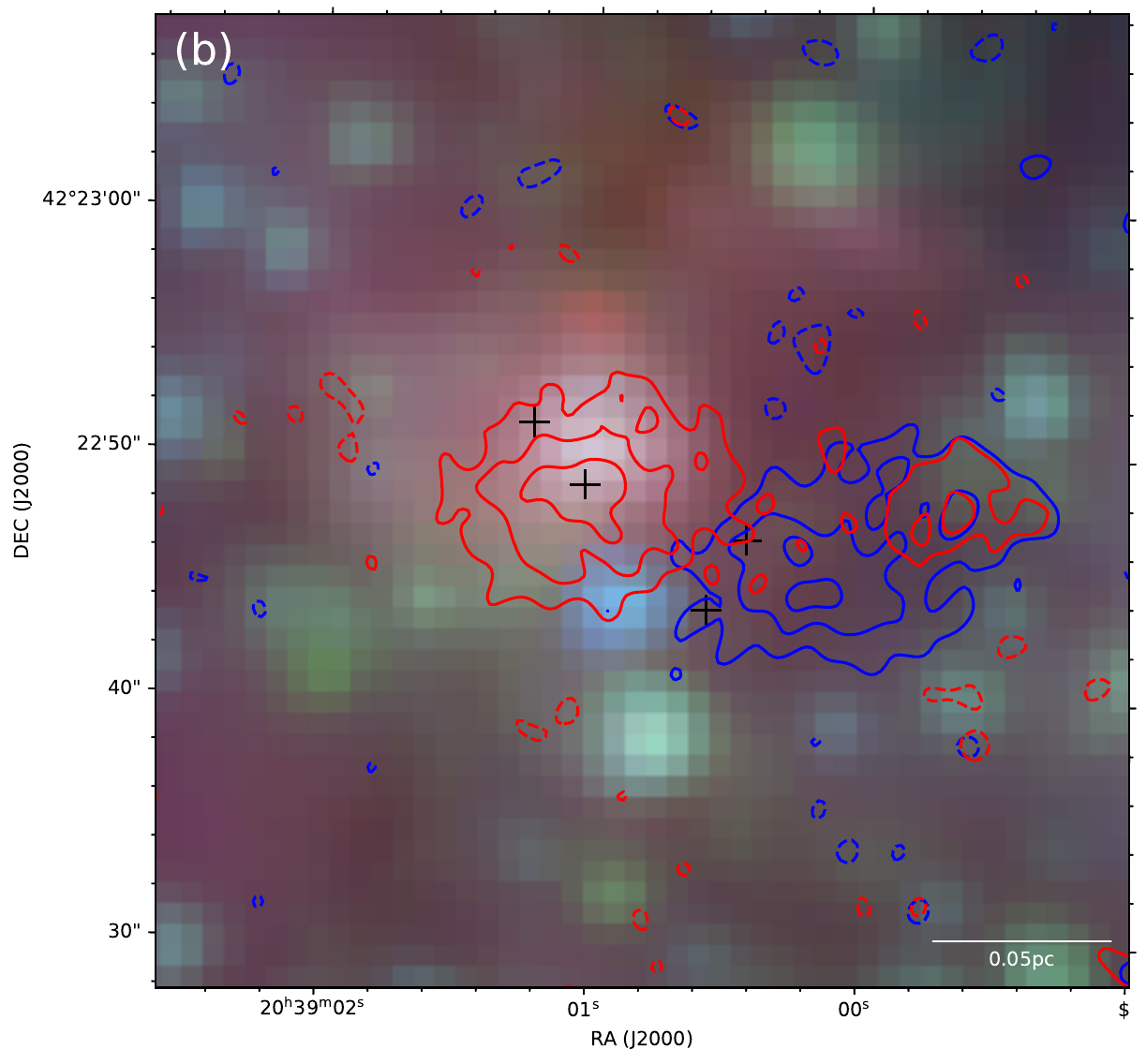}
     \caption{Same convention as Figure \ref{N03} but for MDC 1467. The blue contour levels are ($-$3, 3, 6, 9) $\times$ $\sigma$, with $\sigma$ = 1.25 Jy beam$^{-1}$ km s$^{-1}$, and the red contour levels are ($-$3, 3, 6) $\times$ $\sigma$, with $\sigma$ = 0.84 Jy beam$^{-1}$ km s$^{-1}$. (a) The green cross indicates the position of the protostar candidate identified in \citet{2014AJ....148...11K}.}
  \label{N44}
\end{figure*}

\begin{figure*}
\centering
   \includegraphics[width=240pt]{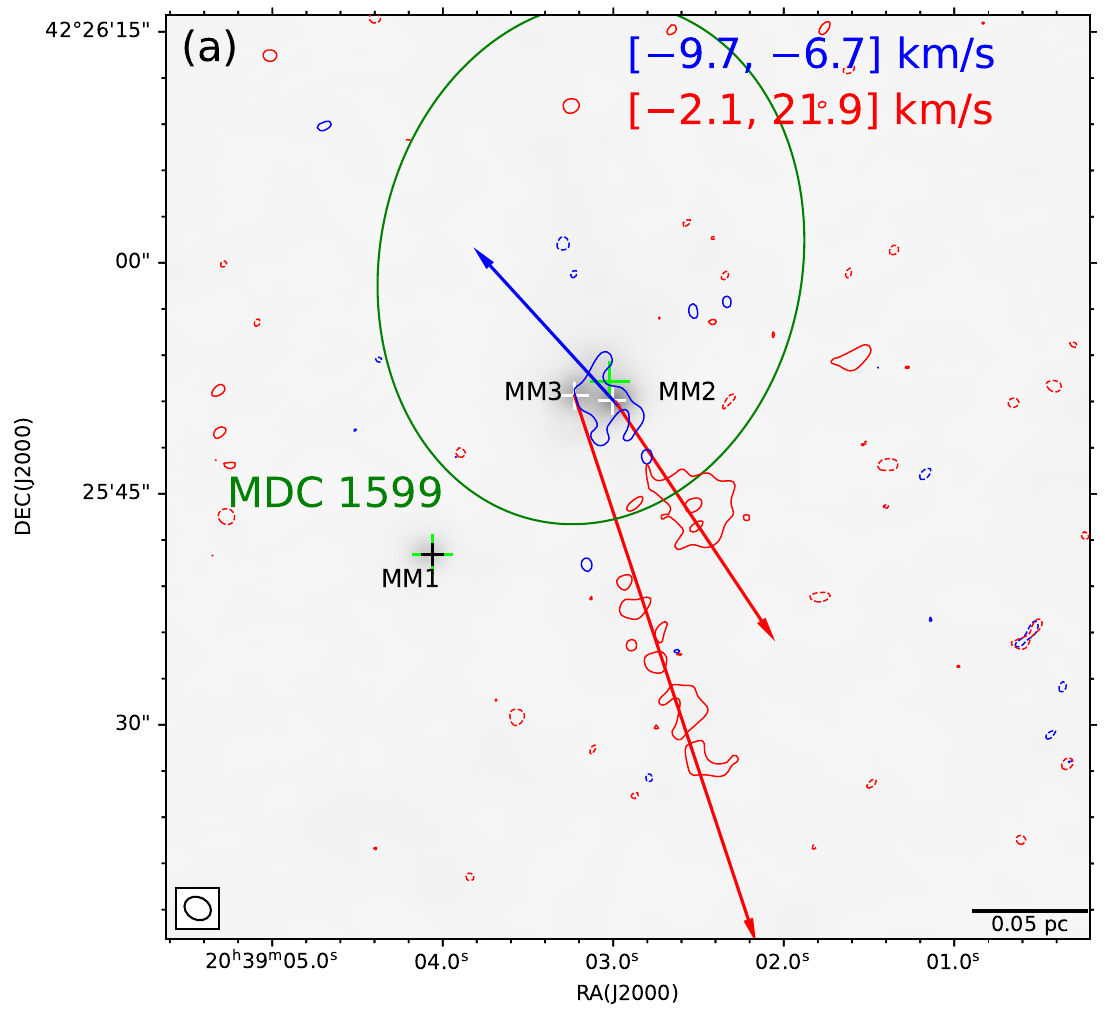} 
   \includegraphics[width=240pt]{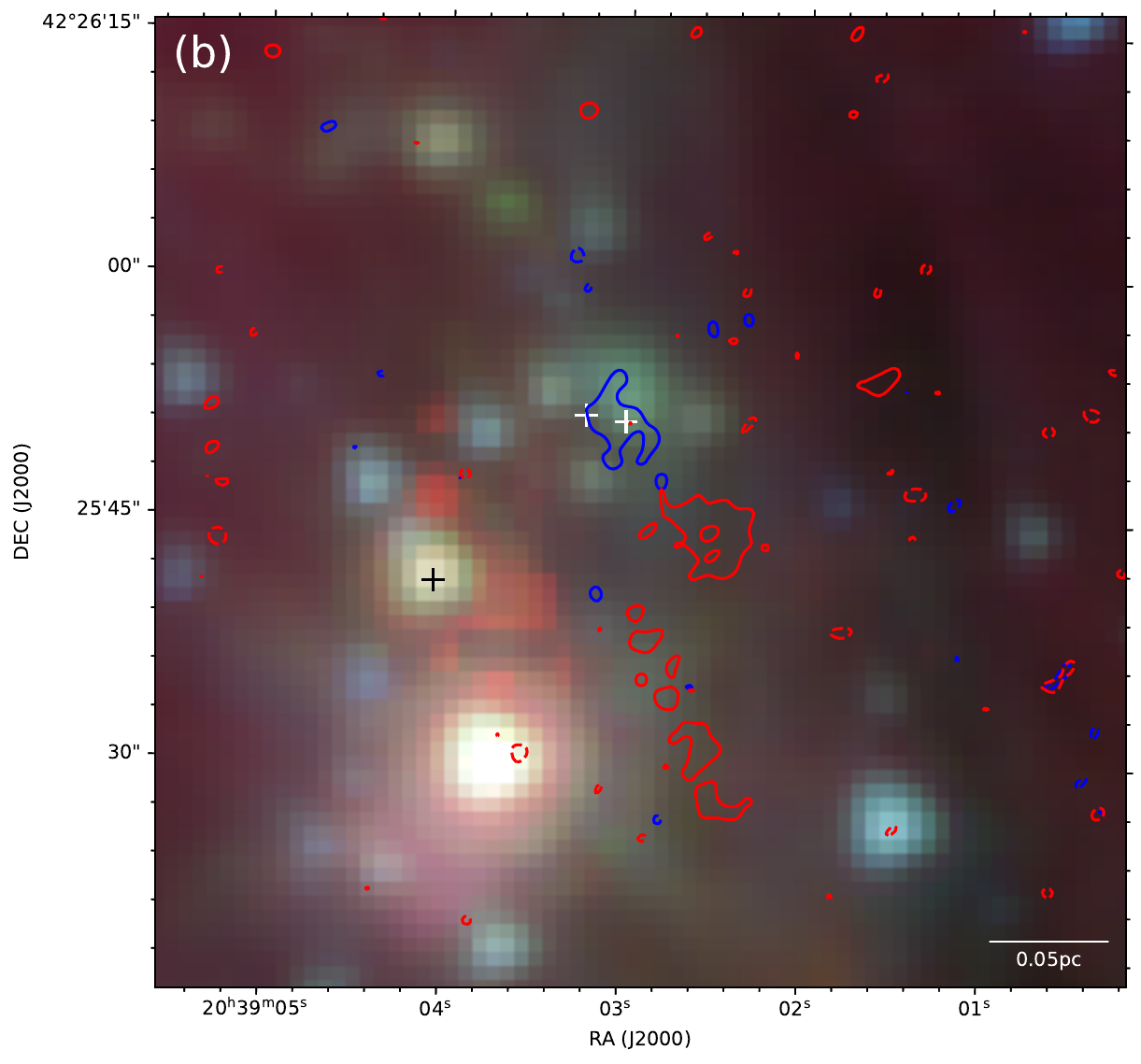}
     \caption{Same convention as Figure \ref{N03} but for MDC 1599. The blue contour levels are ($-$3, 3) $\times$ $\sigma$, with $\sigma$ = 0.42 Jy beam$^{-1}$ km s$^{-1}$, and the red contour levels are ($-$3, 3, 6) $\times$ $\sigma$, with $\sigma$ = 1.80 Jy beam$^{-1}$ km s$^{-1}$. (a) The green cross points out the position of the protostar candidate identified in \citet{2014AJ....148...11K}.}
  \label{N53}
\end{figure*}

The high-velocity SiO (5$-$4) integrated maps are presented in Figure \ref{N03}$-$\ref{N53}. We also show the Spitzer 3.6, 4.5, and 8.0 $\mu$m three-color composite maps. We identify 14 bipolar and 18 unipolar SiO outflows in 16 MDCs. The SiO (5$-$4) channel maps are shown in Appendix \ref{channel_map}.
Below we describe the details of the identification, morphology, and kinematics of the SiO emission structures in each MDC.

%-----------------------------------------------------------------------

%-----------------------------------------------------------------------

%-----------------------------------------------------------------------
\begin{figure*}
\centering
   \includegraphics[width=500pt]{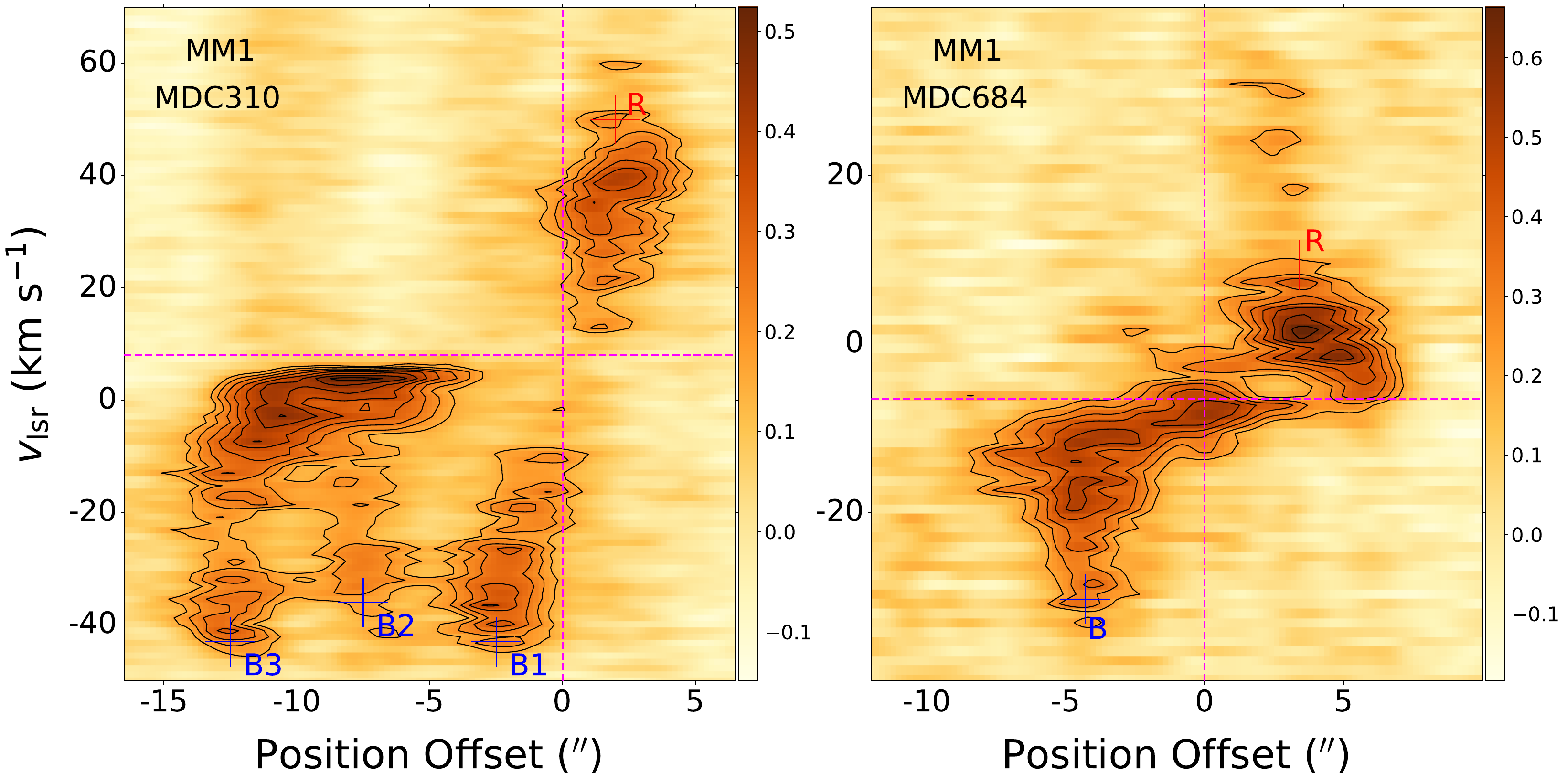}
     \caption{PV diagram of the SiO bipolar outflows. The cut of the PV diagrams is along the SiO outflows that are indicated by the arrows in Figure \ref{NW14}(a) and \ref{N68}(a). The names of the associated continuum sources and MDCs are displayed in each panel. The red and blue plus symbols mark the locations of the detected redshifted and blueshifted knots. The vertical and horizontal dashed fuchsia lines show the positions and systemic velocities of the continuum sources, respectively. The contour levels are (3 to 30 by 3 steps) $\times$ $\sigma$, with $\sigma$ = 0.05 and 0.07 Jy beam$^{-1}$, for MM1 in MDC 310 and 684, respectively.}
  \label{fig:PV}
\end{figure*}
%-----------------------------------------------------------------------

\textbf{MDC 220}: Figure \ref{N03}(a) shows the SiO (5$-$4) emissions integrated from 9.1 to 12.2 km s$^{-1}$ for the blueshifted lobe and from 16.8 to 26.1 km s$^{-1}$ for the redshifted lobe. It is difficult to determine whether the SiO emissions arise from a single outflow or multiple outflows. It is even more difficult to identify the central sources of the outflow(s). By cross-matching the CO (2$-$1) maps \citep{Pan2023}, we tentatively identify two bipolar SiO outflows originating from MM1 and MM2, respectively. For the MM1 outflow, the blueshifted lobe has an orientation consistent with that of the CO emission, and this elongated structure is also seen in the 11.4 km s$^{-1}$ velocity channel map (see Appendix \ref{channel_map}). The redshifted SiO emission from this outflow appears to be closely surrounding MM1 and is detected at 15.9 to 21.5 km s$^{-1}$. For the MM2 outflow, its redshifted lobe is the most remarkable feature in the SiO maps and is detected at 19.2 to 24.9 km s$^{-1}$ as a slightly elongated clump to the south of MM2. The blueshifted lobe of this outflow shows relatively faint SiO emission and is discernible at 10.3 and 11.4 km s$^{-1}$ immediately to the north of MM2. In Figure \ref{N03}(b), excess 4.5 $\mu$m emission is detected in the area outlined by the SiO (5$-$4) contours.

\textbf{MDC 248}: Figure \ref{N12}(a) presents the blueshifted (from 4.9 to 12.1 km s$^{-1}$) and redshifted (from 16.7 to 20.7 km s$^{-1}$) SiO (5$-$4) emission. For the redshifted emission, we find a compact structure surrounding MM1/MM2 and two tiny knots to the southwest of these continuum sources. Multiple clumpy structures are detected away from the continuum sources for the blueshifted emission. It is difficult to identify the central sources of these multiple SiO outflows. With the assistance of the CO (2$-$1) outflows, we interpret the SiO emission as two bipolar SiO outflows originating from MM1 and MM2, and one unipolar outflow arising from MM3. For the MM1 outflow, the redshifted emission extends from MM1 to the southeast; the blueshifted emission has two components, one seems to be closely surrounding MM1 and the other is detected to the northwest of MM1. For the MM2 outflow, the blueshifted and redshifted lobes are found on opposite sides of MM2 along the outflow axis. For the MM3 outflow, clumpy blueshifted SiO emission is detected to the northeast of the condensation. In Figure \ref{N12}(b), there is excess ``green'' emission detected in the area of two bipolar SiO outflows.

\textbf{MDC 310}: The integrated SiO map (Figure \ref{NW14}(a)) reveals a well-collimated bipolar outflow centered at the continuum source MM1. This bipolar outflow has a southeast--northwest (SE--NW) orientation, which is consistent with the CO (2$-$1) observation. In Figure \ref{NW14}(b), we detect significant ``green'' nebulosity surrounding MM1. The PV (position--velocity) structure (Figure \ref{fig:PV}) could be well described by Hubble Wedge features \citep{2001ApJ...554..132A}, and the multiple knots in the blueshifted lobe may result from episodic mass accretions during the formation of high-mass stars.

\textbf{MDC 341}: The SiO (5$-$4) emission (Figure \ref{N63}(a)) seems to trace a northeast--southwest (NE-SW) bipolar outflow, but neither MM1 nor MM2 coincides with the geometric center of this structure. In the channel maps (see Appendix \ref{channel_map}), the redshifted emission seems to originate from MM1 and extends to the southwest from 0.4 to 40.9 km s$^{-1}$, while the blueshifted emission appears to arise from MM1 and extends to the east from $-$34.3 to $-$20.8 km s$^{-1}$. Considering the orientations of the CO (2$-$1) outflows, it is reasonable to interpret the SiO emission as one west--east unipolar outflow and one NE-SW unipolar outflow, both centered at MM1. In Figure \ref{N63}(b), there is ``green'' emission in the central SiO-detected region. 

\begin{figure}
\centering
   \includegraphics[width=240pt]{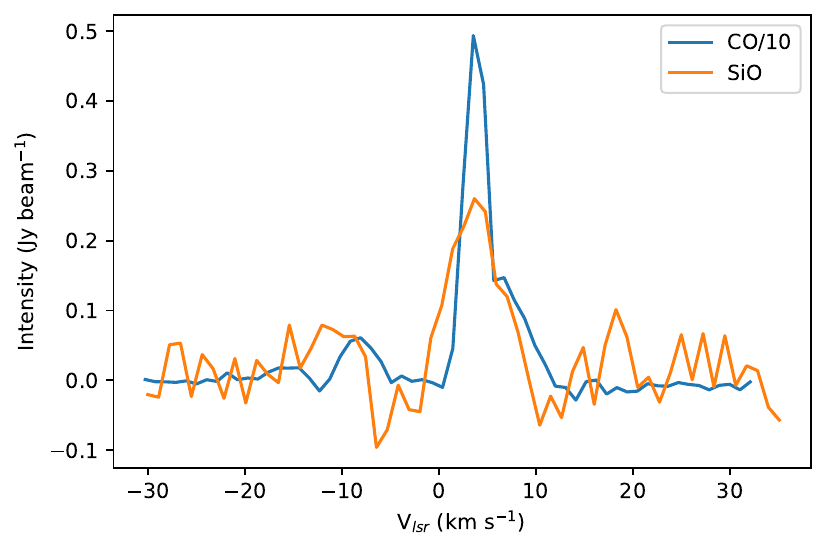}
     \caption{Averaged spectra of SiO 5$-$4 (orange) and CO 2$-$1 (blue) in the region outlined by the contours of the unidentified SiO emission in MDC 351.}
  \label{fig:s32_spec}
\end{figure}

\textbf{MDC 351}: Figure \ref{S32}(a) presents the image of blueshifted (from $-$12.1 to $-$2.9 km s$^{-1}$) and redshifted (from 1.7 to 11.6 km s$^{-1}$) SiO (5$-$4) emission. Except for the tiny redshifted emission extending from MM2 to the northeast, we cannot identify the central sources of the other SiO (5$-$4) emission. Taking the orientations of the CO (2$-$1) outflows into consideration, we believe that the other SiO (5$-$4) emission traces one bipolar outflow centered at MM2 and one unipolar outflow generated from MM3, while there is redshifted emission (pointed out with a black rectangle) not associated with any continuum sources. For the MM2 outflow, its blueshifted lobe at $-$9.8 $\sim$ $-$5.3 km s$^{-1}$ (see Appendix \ref{channel_map}) has a northeast--southwest orientation similar to that of the CO emission; the redshifted lobe of this outflow is marginally detected at 3.66 $\sim$ 9.27 km s$^{-1}$. For the unipolar outflow, the redshifted emission extends from MM3 to the northwest, which is clear at 4.8 to 9.3 km s$^{-1}$. 
The SiO (5$-$4) structure without an identified central source has a velocity range of 0.3 to 7.0 km s$^{-1}$. We also detect CO (2$-$1) emission within this velocity range with a spatial distribution comparable to that of the SiO emission. This SiO structure could have various origins: (i) one possibility is that the SiO emission arises from outflow shock generated from a low-mass protostar undetected in the SMA 1.37 mm continuum image. The sensitivity of the 1.37 mm continuum observation at a distance of 1.4 kpcs is equivalent to $\sim$1 $M_{\odot}$, and this SiO emission could trace outflow from a protostar with an envelope mass of below 1 $M_{\odot}$. We note that there is a ``dark’’ region near to this SiO structure in the Spitzer three-color map (Figure \ref{S32}(c)), and this area possibly harbors a low-mass protostar. (ii) Another possibility is that the unidentified SiO emission is related to large-scale shocks from converging flows; this scenario has been suggested for other high-mass star-forming regions by \citet{2010MNRAS.406..187J}. Figure \ref{fig:s32_spec} presents averaged spectra of SiO (5$-$4) and CO (2$-$1) in the region outlined by the SiO contours. The CO emission consists of two components; one is a blueshifted outflow from MM2, and the other could be a redshifted outflow from a potential low-mass protostar. This SiO emission could result from the collision or intersection of these two outflows. The FWHM of the SiO emission is around 5 km s$^{-1}$, which is broader than the line width of widespread SiO emission in \citet{2014A&A...570A...1D} ($\le$2.5 km s$^{-1}$) and \citet{2010MNRAS.406..187J} ($\sim$0.8 km s$^{-1}$), and this relative broad line width could be a result of the interaction between different gas components. In Figure \ref{S32}(b), we detect a knot of excess 4.5 $\mu$m emission at the head of the blueshifted SiO outflow.

%-----------------------------------------------------------------------
%-----------------------------------------------------------------------

\textbf{MDC 507/753}: In Figure \ref{S07}(a), the blueshifted (from $-$24.3 to $-$0.7 km s$^{-1}$) and redshifted (from 3.9 to 13.9 km s$^{-1}$) SiO (5$-$4) emission reveals multiple outflows. There is a redshifted outflow launching from MM4 and extending from south to north. Blueshifted emission is detected to the northeast of MM5 (with a water maser spot detected nearby) and we identify MM5 as the central source of SiO emission. The emission surrounding MM3 is detected along the northeast--southwest axis, but it is difficult to identify how many bipolar or unipolar outflows it traces. Considering the CO (2$-$1) emission, we believe this SiO (5$-$4) emission arises from two bipolar outflows. For one bipolar outflow with redshifted emission detected to the southwest of MM3, this redshifted lobe is clear at 4.8 to 9.4 km s$^{-1}$ in the channel maps (see Appendix \ref{channel_map}), and its blueshifted lobe is detected at $-$9.7 to $-$0.7 km s$^{-1}$ as an elongated clump to the northeast of MM3. For the other bipolar outflow, the redshifted and blueshifted lobes are found along the outflow axis with an orientation of NE--SW. In Figure \ref{S07}(b), we detect a blueshifted outflow originating from MM8 in a narrower velocity range (from $-$1.9 to $-$0.7 km s$^{-1}$). In Figure \ref{S07}(c), excess 4.5 $\mu$m emission is detected surrounding the bipolar outflows, and ``green'' knots are found at the fronts of three unipolar outflows.

\textbf{MDC 509}: Figure \ref{S30}(a) shows the image of blueshifted (from $-$8.2 to 5.1 km s$^{-1}$) and redshifted (9.7 to 31.2 km s$^{-1}$) SiO (5$-$4) emission. The redshifted lobe seems to originate from MM1 and to extend to the southwest. Also, a tiny blueshifted lobe is detected to the northwest of MM4 and appears to arise from this continuum source. For the remarkable blueshifted lobe close to MM7, the emission at 1.9 $\sim$ 4.2 km s$^{-1}$ in the channel maps (see Appendix \ref{channel_map}) appears to be associated with MM7, which is a continuum source with water maser detection, and the emission at $-$1.4 $\sim$ 0.8 seems to originate from MM4. Therefore, we tentatively regard this blueshifted lobe as a combination of two outflows from MM4 and MM7, and the orientations of the outflows are consistent with those of the CO (2$-$1) observation. In Figure \ref{S30}(b), there is excess 4.5 $\mu$m emission detected in the area outlined by the redshifted SiO (5$-$4) contours and two ``green'' tips are found nearby MM4 and MM7.

\textbf{MDC 684}: The high-velocity blueshifted (from $-$36.5 to $-$8.8 km s$^{-1}$)  SiO (5$-$4) emission in Figure \ref{N68}(a) reveals a bipolar outflow and the redshifted (from $-$4.2 to 29.8 km s$^{-1}$) SiO (5$-$4) emission reveals an isolated blueshifted outflow. The bipolar outflow is centered at MM1 and has an east--west orientation. Although there are a few instances of SiO emission generated from MM2 at $-$14.0 $\sim$ $-$11.8 and $-$1.7 $\sim$ 0.6 km s$^{-1}$ in the channel maps (see Appendix \ref{channel_map}), we still regard MM1 as the central source of this bipolar SiO outflow. For the southern blueshifted lobe, it is difficult to identify its central source. Considering the CO (2$-$1) observation, we interpret MM3 as the central source of this outflow. In Figure \ref{N68}(b), the 4.5 $\mu$m excess and the SiO emission show good spatial consistency. The PV diagram (Figure \ref{fig:PV}) shows a Hubble Law feature \citep{1996ApJ...459..638L}.

\textbf{MDC 698}: Figure \ref{N56}(a) presents both blueshifted (from $-$5.0 to 17.3 km s$^{-1}$) and redshifted (from 21.9 to 28.8 km s$^{-1}$) SiO (5$-$4) emission. We detect a blueshifted lobe to the south of MM3 and a redshifted lobe to the north of MM1, and it is difficult to identify their central sources. Considering the CO (2$-$1) observations, we tentatively identify one blueshifted outflow generated from MM3 and one redshifted outflow arising from MM1. In Figure \ref{N56}(b), the emission from MM1 contributes to most of the infrared emission in this region, and we cannot find 4.5 $\mu$m excess in the SiO detected area.

\textbf{MDC 699}: Figure \ref{N38}(a) shows both blueshifted (from $-$9.7 to $-$3.8 km s$^{-1}$) and redshifted (from 0.8 to 11.8 km s$^{-1}$) SiO (5$-$4) emission. Due to their marginal detections, we could not identify the central source or the orientations of the outflows traced by the SiO emission. In the CO observations, we only find a unipolar outflow to the southeast of MM4 (an arrow with a solid line) which is not associated with the SiO emission. With the assistance of the CO outflows identified by \citet{2022A&A...660A..39S} (arrows with dashed lines), the SiO emission is believed to trace a south--north bipolar outflow. Although no continuum sources are identified at the center of this bipolar outflow, a source with the continuum emission under our detection limit cannot be ruled out. In Figure \ref{N38}(b), we cannot find ``green'' excess spatially associated with the SiO emissions.

\textbf{MDC 801}: Figure \ref{N65}(a) presents the redshifted SiO (5$-$4) emission integrated from $-$7.5 to 7.2 km s$^{-1}$. The redshifted SiO lobe arises from MM1 and extends to the northwest, which is consistent with the CO (2$-$1) observation in orientation. In Figure \ref{N65}(b), the distribution of the SiO emission is coincident with that of the extended ``green'' emission.

\textbf{MDC 1112}: MDC 1112 is one of the most well-studied star-forming regions in the Cygnus X complex, and is also known as W75N(B). Figure \ref{N30}(a) shows blueshifted (from $-$7.6 to 6.7 km s$^{-1}$) and redshifted (from 11.3 to 18.3 km s$^{-1}$) SiO (5$-$4) emission. It is difficult to identify how many outflows the SiO emission traces and it is even more difficult to determine the central sources of the outflows. By cross-matching our CO (2$-$1) observations and identified outflows in \citet{2004ApJ...601..952S}, we tentatively identify two bipolar SiO outflows originating from MM2 and MM3, respectively, and one unipolar outflow arising from MM1. For the MM3 outflow, the blueshifted lobe is discernible to the southwest of this continuum source at $-$7.6 $\sim$ 1.4 km s$^{-1}$ in the channel maps (see Appendix \ref{channel_map}), and the redshifted emission is detected to the northeast at 10.4 $\sim$ 17.1 km s$^{-1}$. For the MM2 outflow, its blueshifted emission is relatively faint at 5.9 km s$^{-1}$. The redshifted lobe of this outflow is detected at 11.5 $\sim$ 13.8 km s$^{-1}$. For the MM1 unipolar outflow, the redshifted emission is oriented northwest--southeast, and this elongated structure can be seen in the 11.5 $\sim$ 12.6 km s$^{-1}$ velocity channels. In Figure \ref{N30}(b), we detect a ``green'' tip associated with the northern redshifted SiO outflow, and the area outlined by the SiO contours to the south of MM3 is embedded in the widespread ``green'' emission.

\textbf{MDC 1454}: MDC 1454 is in DR15, which is one of the most prominent star-forming regions in the southern Cygnus-X complex. Figure \ref{DR15}(a) presents the redshifted SiO (5$-$4) emission integrated from 11.6 to 47.6 km s$^{-1}$. This redshifted outflow launches from MM5 and extends to the west, which is consistent with that of the CO (2$-$1) outflow in orientation. There is no SiO detection around the systematic velocity of 1.9 km s$^{-1}$. In Figure \ref{DR15}(b), there is no 4.5 $\mu$m excess found associated with the SiO emission.

\textbf{MDC 1467}: This MDC is located in the DR21(OH) filament, and a number of studies have focused on it (e.g., \citealt{2010A&A...520A..49S}; \citealt{2013ApJ...772...69G}; \citealt{2022ApJ...927..106C}). In Figure \ref{N44}(a), the redshifted SiO (5$-$4) emission (from $-$0.7 to 0.5 km s$^{-1}$) appears to be closely surrounding MM16, and the blueshifted SiO (5$-$4) emission (from $-$20.9 to $-$6.2 km s$^{-1}$) is detected immediately to the west of the redshifted lobe. It is difficult to identify whether the SiO emission comes from a single outflow or multiple outflows, and even more difficult to find the central sources of the outflow(s). We cannot determine the outflows from our CO observations either. Considering the identified outflows in \citet{2013ApJ...772...69G} (arrows with dashed lines), the SiO emissions could come from a bipolar outflow centered at MM16 and a blueshifted outflow centered at MM17. The MM17 unipolar outflow could be seen at $-$11.9 to $-$7.4 km s$^{-1}$ in the channel maps (see Appendix \ref{channel_map}). For the MM16 outflow, the blueshifted emission is detected to the southwest of this continuum source at $-$14.1 to $-$6.2 km s$^{-1}$. The redshifted emission of this outflow is found surrounding MM16, which is unusual for high-velocity outflow emission. Another interpretation (arrows with solid lines) for this outflow may offer some clues: the SiO (5$-$4) emission arises from a bipolar outflow centered at MM17 and a blueshifted outflow centered at MM18. The emission surrounding MM16 could be well explained by a redshifted lobe of the MM17 outflow. \citet{Pan2023} detected high-density molecular line emission surrounding MM17 with a velocity gradient along the NE--SW direction, which is consistent with the direction of the MM17 outflow axis in the latter interpretation, and we therefore favor this explanation. In Figure \ref{N44}(b), ``green'' knots are detected at the fronts of the blueshifted outflows.

\textbf{MDC 1599}: Figure \ref{N53}(a) presents both blueshifted (from $-$9.7 to $-$6.7 km s$^{-1}$) and redshifted (from $-$2.1 to 21.9 km s$^{-1}$) SiO (5$-$4) emission. It is difficult to identify the original source in both cases. Considering the CO (2$-$1) observation, we tentatively identify one bipolar outflow generated from MM2 and one unipolar outflow arising from MM3. For the MM2 outflow, its blueshifted emission is detected between MM2 and MM3. The redshifted emission of this outflow is detected at $-$0.7 to 7.2 km s$^{-1}$ (see Appendix \ref{channel_map}). The MM3 outflow is marginally detected with an elongated morphology to the northwest of this continuum source. In Figure \ref{N53}(b), we find successive excess 4.5 $\mu$m emission associated with the SiO emission.

Overall, we identify 14 bipolar and 18 unipolar SiO (5$-$4) outflows associated with 29 central sources. All the central sources are identified from the 1.37 mm continuum maps. Two bipolar SiO outflows share a joint central source, namely MDC507-MM3, and two unipolar outflows share a common central source, MDC341-MM1. We have not found a continuum source associated with the bipolar outflow in MDC699.

Although many lines (e.g., polycyclic aromatic hydrocarbons) could contribute to the IRAC 4.5 $\mu$m band emission, excess 4.5 $\mu$m emission reveals shocked H$_{2}$ emission associated with outflows in many cases \citep{2005MNRAS.357.1370S, 2006ApJ...645.1264S, 2010AJ....140..196D}. Therefore, such IRAC composite images can be a useful diagnostic tool for shocked H$_{2}$ emission in outflows (e.g., \citealt{2004ApJS..154..352N}; \citealt{2004ApJS..154..346R}; \citealt{2008ApJ...685.1005Q}; \citealt{2009ApJ...702.1615C}; \citealt{2018RAA....18...19X}). We find that 11 bipolar and 13 unipolar outflows are associated with excess 4.5 $\mu$m emission.

\subsection{Low-velocity SiO components} \label{res:low}

\begin{figure*}
\centering
   \includegraphics[width=240pt]{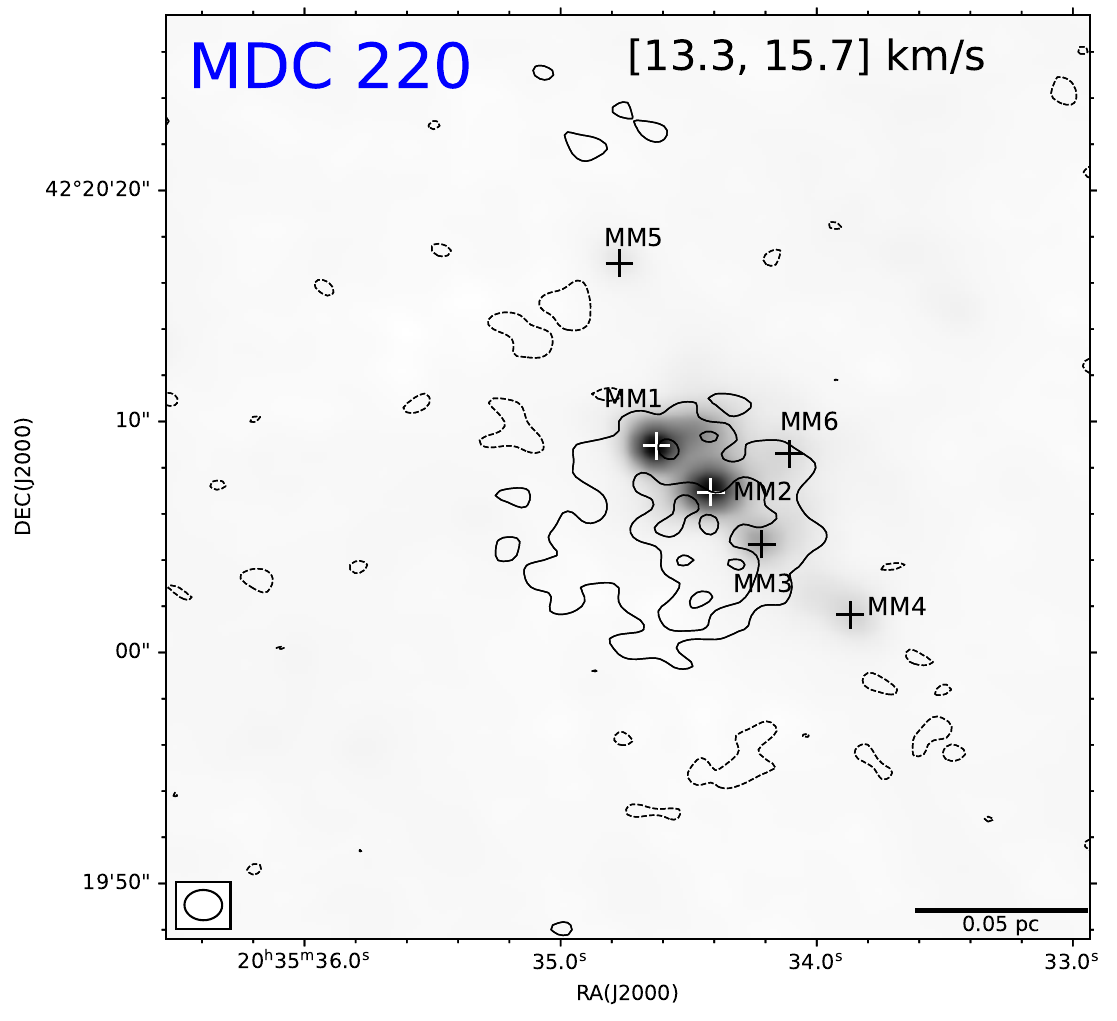} 
   \includegraphics[width=240pt]{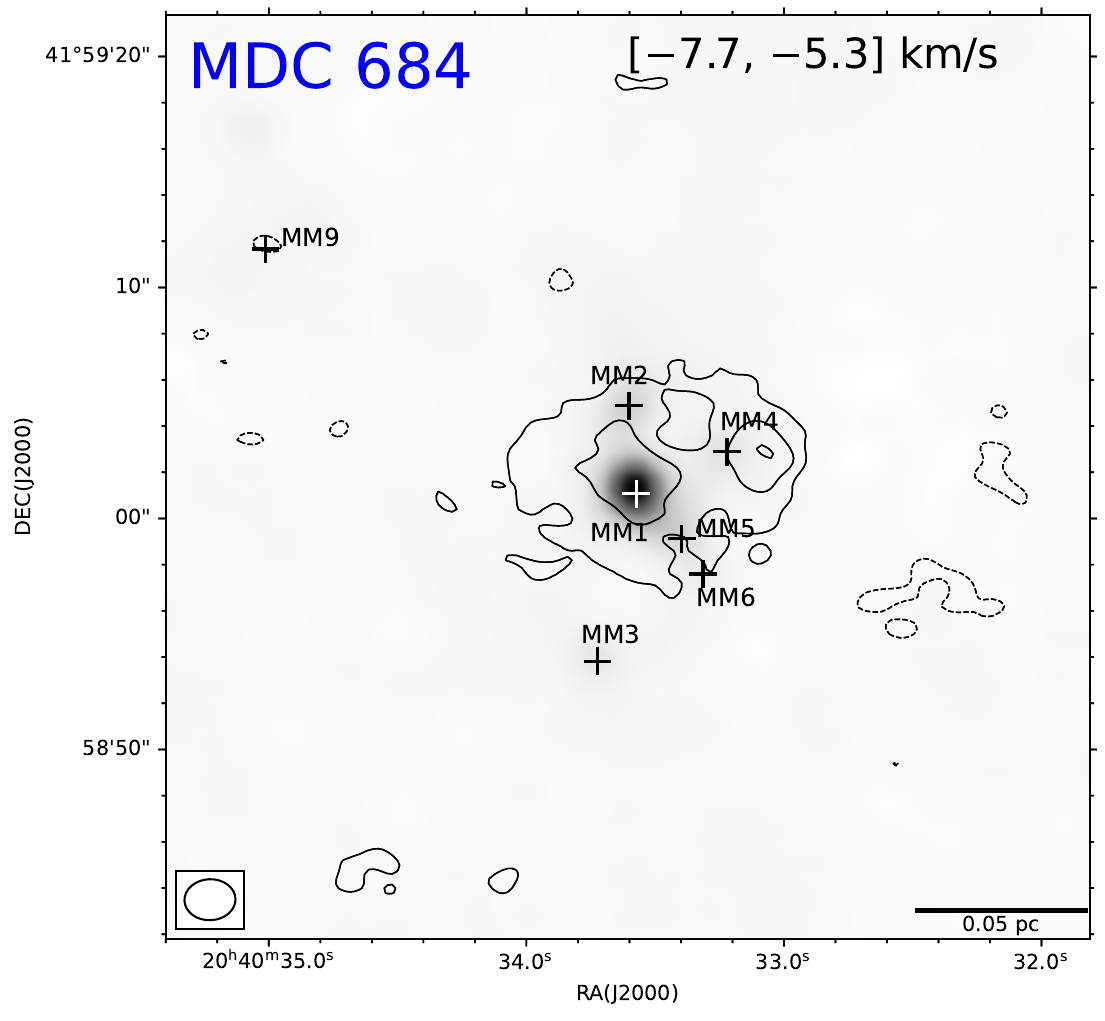}
     \caption{Low-velocity ($\Delta_{v}$ $\le$ 1.2 km s$^{-1}$) SiO (5$-$4) emission for MDC220 (left) and MDC684 (right). The integrated velocity ranges of the contours are presented at the top-right corner. The black contour levels are ($-$3, 3, 6, 9) $\times$ $\sigma$, with $\sigma$ = 0.13 Jy beam$^{-1}$ km s$^{-1}$, for both MDCs. The black and white crosses represent dust condensations identified by \citet{2021ApJ...918L...4C}. The grayscale maps represent 1.37 mm continuum emission. The scale bar is presented in the bottom-right corner. The synthesized beam is shown in the bottom-left corner.}
  \label{low_velocity}
\end{figure*}

We investigated the distribution of the SiO (5$-$4) emission close to the MDC systemic velocities ($\Delta{v}$ $\le$ one channel width of $\sim$ 1.2 km s$^{-1}$). Among the 16 SiO-detected MDCs, the low-velocity SiO emission is closely linked with the high-velocity emission in 13 MDCs, which suggests that these low-velocity components are largely dominated by outflows. For the remaining 3 MDCs, MDC 1454 has no low-velocity SiO (5$-$4) detection, and MDC 220 and MDC 684 show relatively diffuse low-velocity SiO emission closely surrounding the continuum sources (see Figure \ref{low_velocity}). Such low-velocity SiO emission could have various origins, and we explore some possibilities below.

%-----------------------------------------------------------------------
%-----------------------------------------------------------------------

One explanation is that the low-velocity SiO emission traces outflow shocks, which are largely decelerated due to the interaction with the surrounding quiescent gas. This scenario was first proposed by \citet{1999A&A...343..585C}, who estimated a deceleration timescale of $\sim$10$^{4}$ yr, which is the same as the timescale for the depletion of gas-phase Si. The narrow line profile and relatively low intensity of SiO (5$-$4) are consistent with this hypothesis. To figure out whether this could be the case for our detection, we estimated CO outflow dynamical ages, finding 0.98 and 0.76 $\times$ 10$^{4}$ yr for MDC220-MM2 and MDC684-MM1, respectively. Considering the average gas density, which is of the order of 10$^{5}$ cm$^{-3}$ \citep{2019ApJS..241....1C}, both MDCs fit this scenario.

Another possible explanation for these low-velocity SiO components was proposed by \citet{2010MNRAS.406..187J}, who suggest that the emission is generated from large-scale shocks during global cloud collapse. \citet{2013ApJ...775...88N} mentioned that SiO could be efficiently formed by low-velocity shocks when there are some Si atoms in the gas phase or grain mantles. This explanation is accepted by \citet{2014A&A...570A...1D}, who made PdBI SiO (2$-$1) observations toward six MDCs in Cygnus-X. These latter authors proposed a picture of evolutionary stages: At early stages, extended and low-velocity SiO emission would trace shocks from the global collapse of material onto the MDC. At later stages, the protostars are formed, and star formation becomes more active. The SiO emission is then dominated by outflows, and the weaker low-velocity SiO components may arise from the interaction between the last collapse of the MDC and the dense envelope of the protostar. 

%-----------------------------------------------------------------------

\section{Discussion} \label{sec:dis}

\subsection{Detection of the SiO (5$-$4) emission and its general nature}

In our SMA observations, we detect SiO (5$-$4) toward 16 out of 48 MDCs. Of the 194 identified dust condensations in these 48 MDCs \citep{2021ApJ...918L...4C}, 29 are regarded as SiO central sources. For the SiO (5$-$4) high-velocity components, we find 14 bipolar and 18 unipolar outflows. For the SiO (5$-$4) low-velocity components, most of them are detected arising from outflows, while the emission in MDC 220 and MDC 684 is unique. The low-velocity emission associated with MDC220-MM2 and MDC684-MM1 is found closely surrounding their central sources, both of which have a different morphology from the cone-shaped or jet-like morphology of the high-velocity outflows. Considering its low velocity, this emission could have a different origin from the outflows (discussed in Section \ref{res:low}). In addition, because large-scale and dispersed emission is filtered out in our interferometric observations, there could be more low-velocity emission in our target regions.

Based on our observations of SiO (5$-$4) and CO emission (our CO (2$-$1)  and CO (3$-$2) observations toward MDC 699 in \citep{2022A&A...660A..39S}), we identify 14 bipolar and 18 unipolar SiO outflows. However, the central sources of five unipolar SiO outflows, namely MDC507-MM5, MDC509-MM1, MDC698-MM1, MDC1454-MM5, and MDC1467-MM17/MM18, are found to generate CO bipolar outflows. Furthermore, we find no SiO emission associated with the  central
sources of seven CO outflows. The morphology of the SiO outflows in MDC310-MM1, MDC341-MM1, and MDC684-MM1 is highly consistent   with that of the CO outflows. However, in MDC 248, MDC 698, and MDC 1599, only patches of SiO emission are found along the CO outflows. In particular, compact SiO emission is detected toward MDC220-MM1, MDC351-MM2, MDC801-MM1, and MDC1599-MM2 with the emission peaks immediately around the central sources. The differences in morphology between the SiO and CO outflows could be attributable to the following factors: (i) SiO (5$-$4) has a higher critical density ($\sim$2.5 $\times$ 10$^{6}$ cm$^{-3}$) and a higher excitation temperature (E$_{u}$ = 31.26 K) than CO (2$-$1), which suggests that SiO (5$-$4) can only be detected in the shocked regions with relatively high density and high temperature. Also, SiO (5$-$4) is more likely to arise from the dense material surrounding the central sources, which could provide an explanation for the compact SiO emission found immediately around the continuum sources. (ii) The SiO models of \citet{2008A&A...482..809G,2008A&A...490..695G} show that a lower initial percentage of silicon in grain mantles or shocks with lower velocities could lead to fewer Si atoms released into the gas phase, which would make the SiO (5$-$4) intensities too low to be detected by the SMA observation.

\subsection{SiO outflow parameters} \label{mass}

\begin{figure*}
\centering
   \includegraphics[width=160pt]{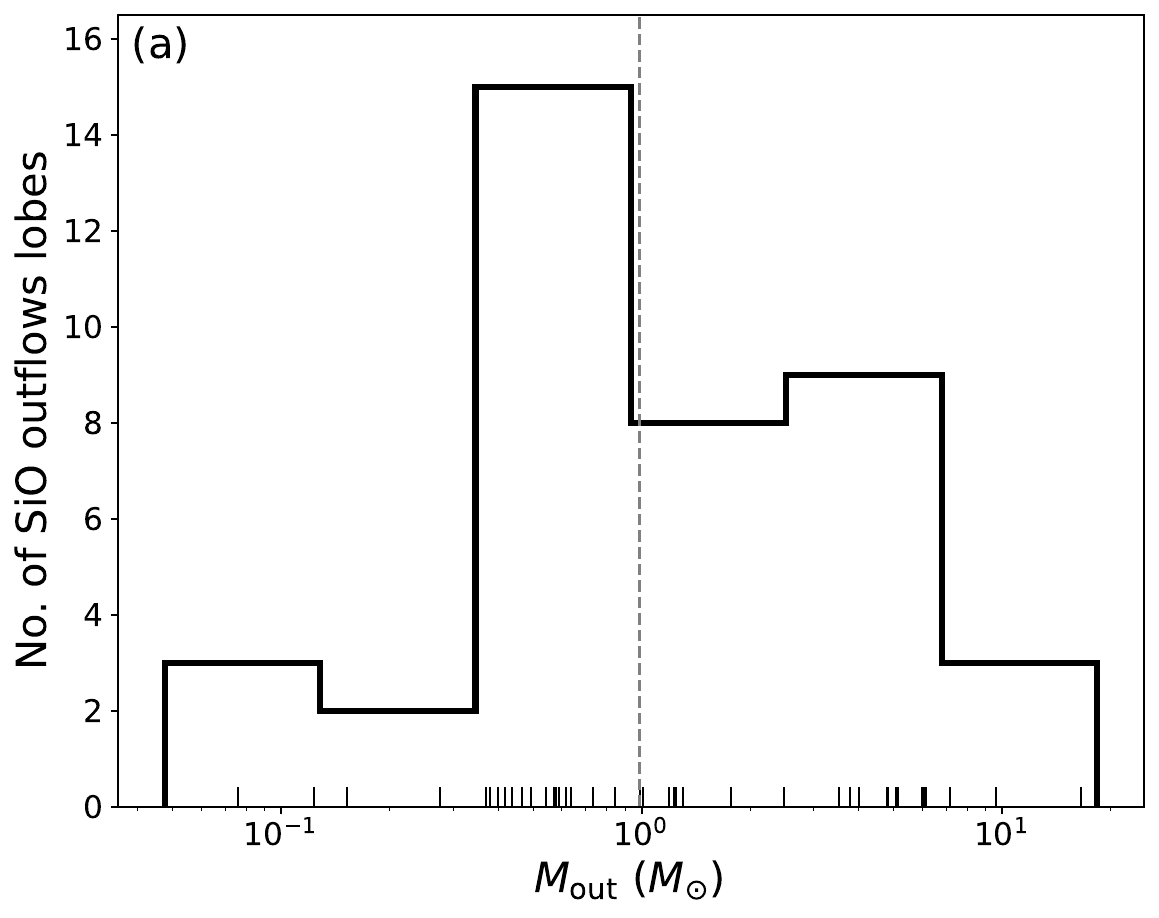}
   \includegraphics[width=160pt]{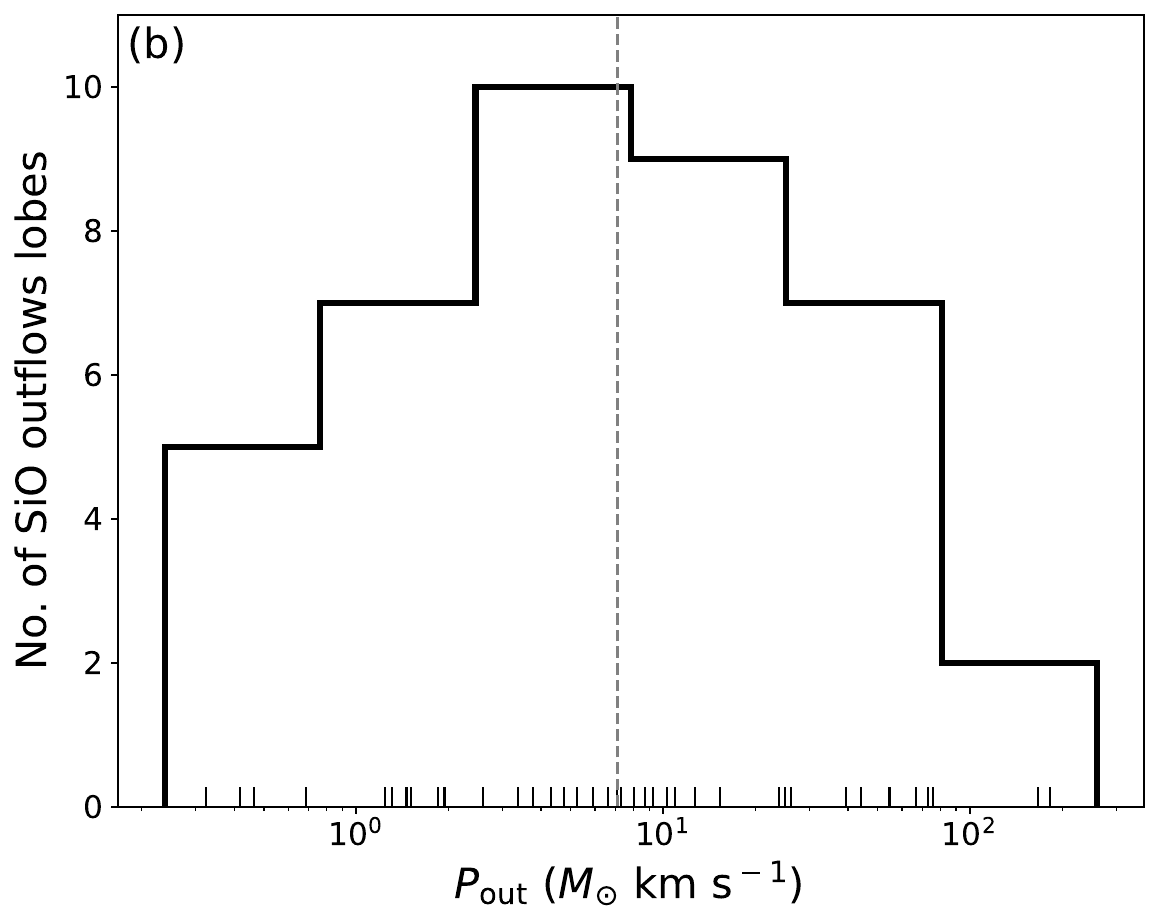} 
   \includegraphics[width=160pt]{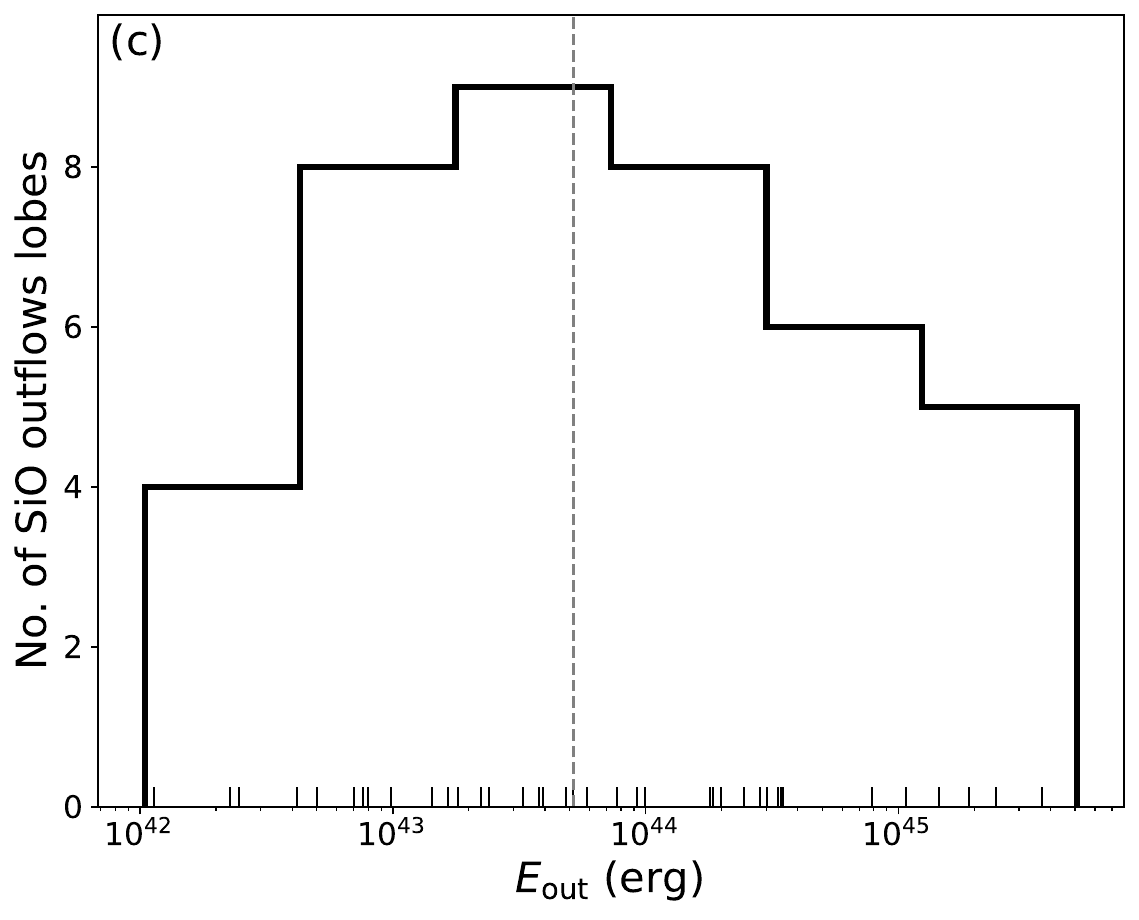} 
   
   \includegraphics[width=160pt]{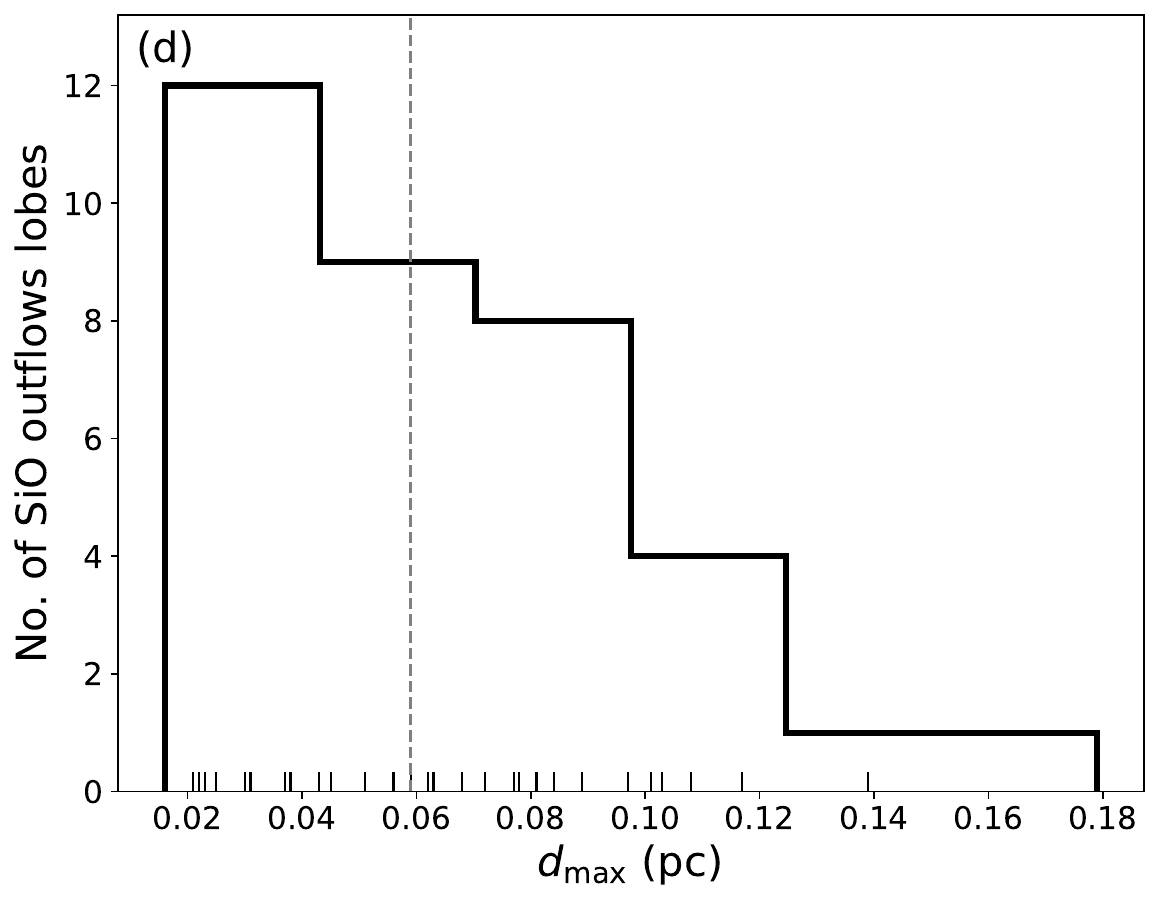}
   \includegraphics[width=160pt]{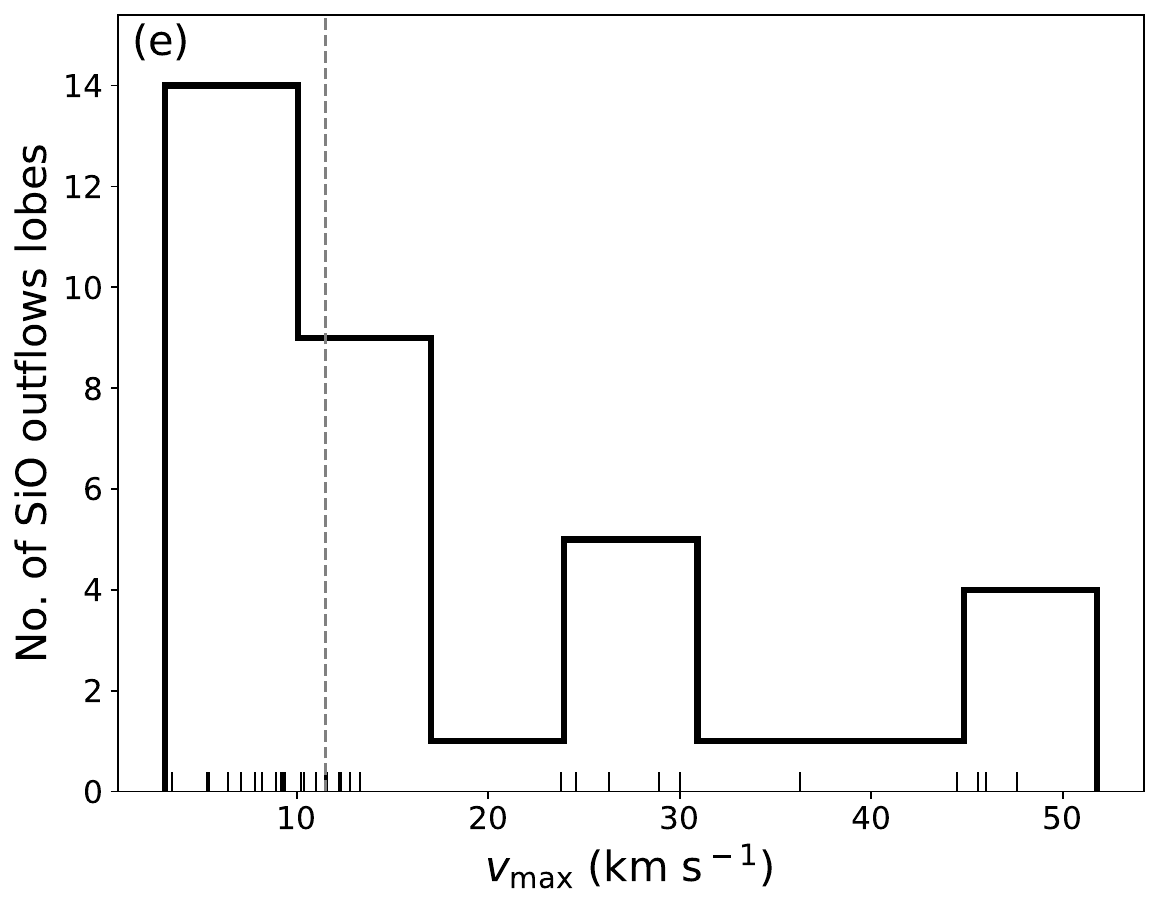} 
   \includegraphics[width=160pt]{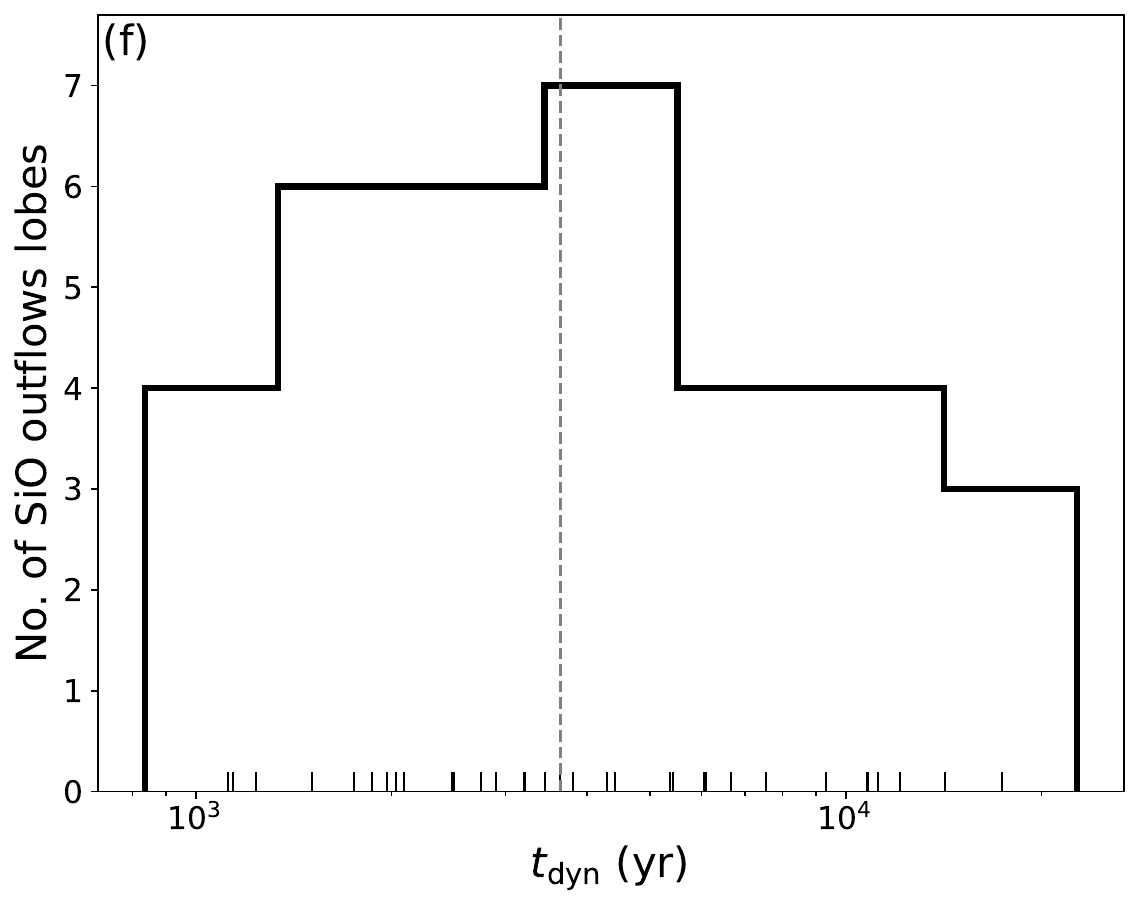} 
   
   \includegraphics[width=160pt]{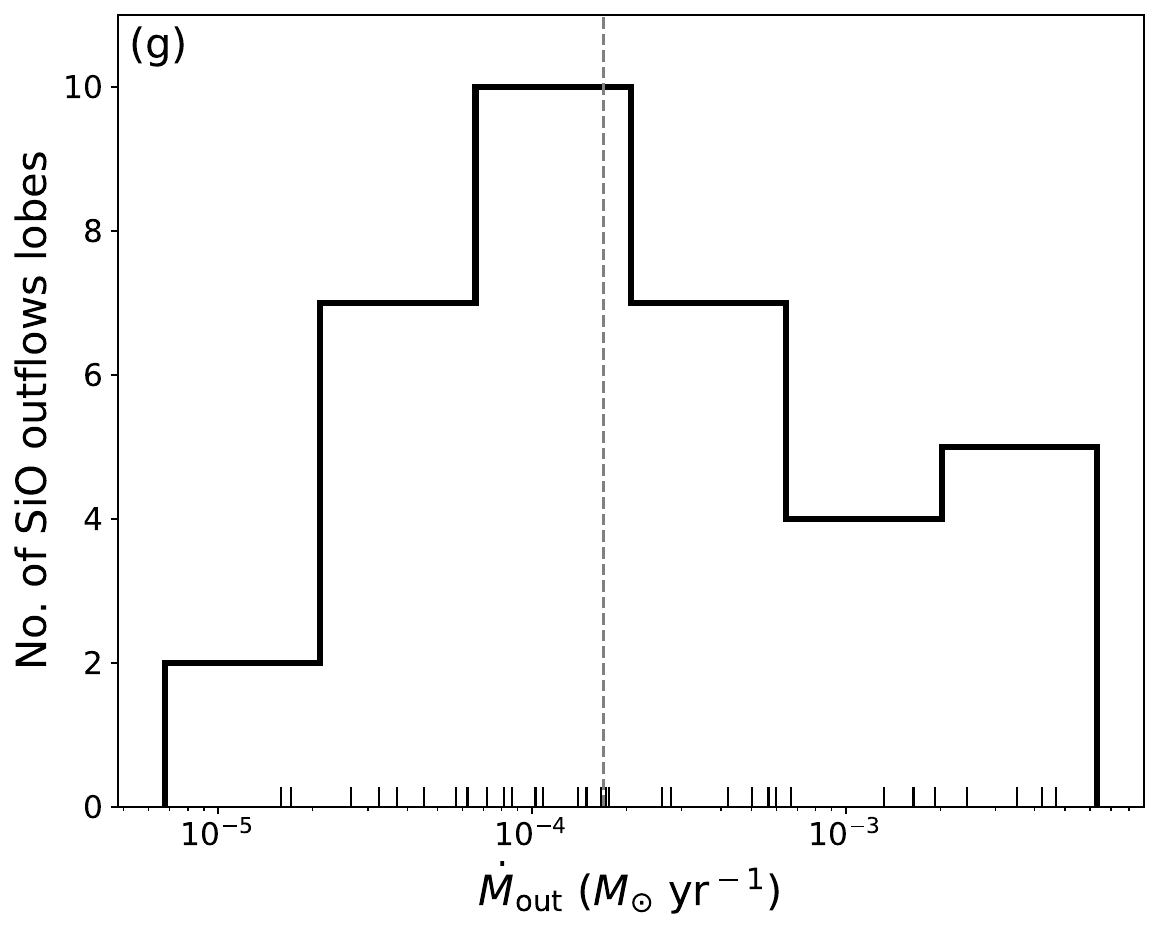}
   \includegraphics[width=160pt]{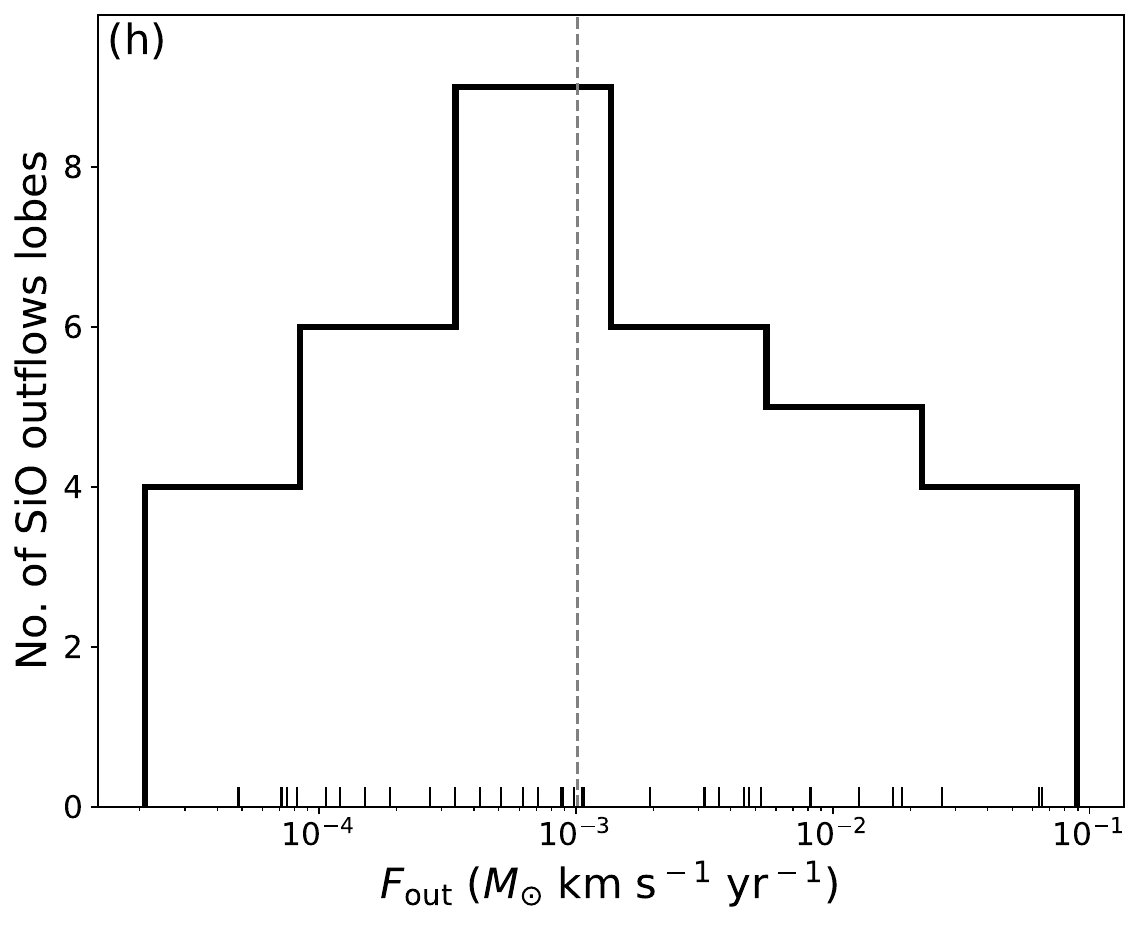} 
   \includegraphics[width=160pt]{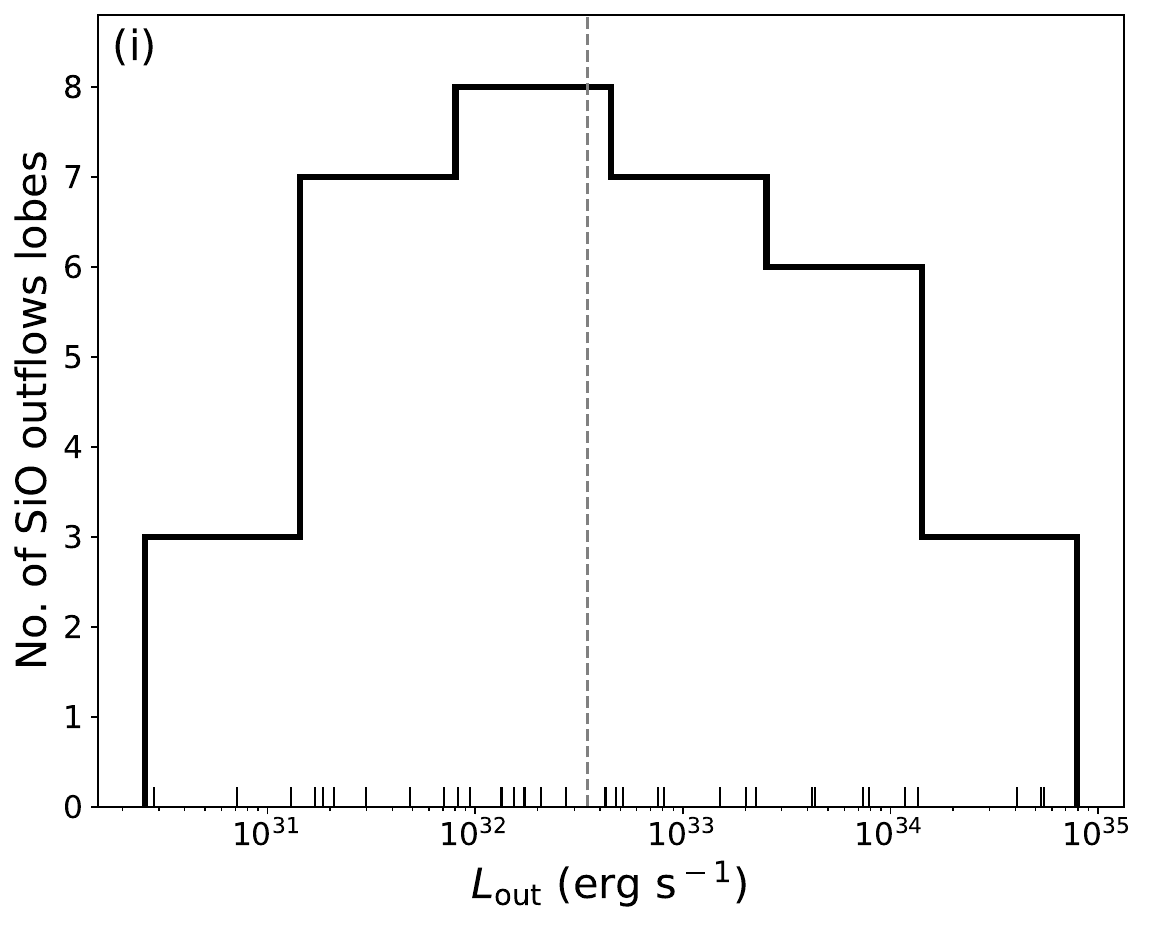} 
     \caption{Distributions of SiO outflow parameters. Panels (a) to (i) show the distributions of SiO outflow mass ($M_{\rm out}$), outflow momentum ($P_{\rm out}$), outflow energy ($E_{\rm out}$), maximum projected distance ($d_{\rm max}$), maximum velocity ($v_{\rm max}$), outflow dynamical timescale ($t_{\rm dyn}$), mass-outflow rate ($\dot{M}_{\rm out}$), mechanical force ($F_{\rm out}$), and outflow luminosity ($L_{\rm out}$), respectively. The black dashed lines indicate the median value of each parameter.}
  \label{fig:relations}
\end{figure*}

With the assumption of optically thin thermal SiO (5$-$4) line emission in local thermodynamic equilibrium (LTE) and the SiO (5$-$4) line excitation temperature of 30 K, we can estimate the SiO (5$-$4) column density ($N_{\rm tot}$) and outflow mass ($M_{\rm out}$) \citep{1999ApJ...517..209G}:
\begin{equation}
N_{\rm u} = \frac{4 \pi}{hcA_{\rm ul}} \int(S_{\rm \nu}/\Omega)\,dv , 
\end{equation}
\begin{equation}
N_{\rm tot} = N_{\rm u} \frac{Q(T_{\rm ex})}{g_{u}} e^{\frac{E_{\rm u}}{kT_{\rm ex}}} ,
\end{equation}
\begin{equation}
M_{\rm out} = N_{\rm tot} \left[\frac{\rm H_{2}}{\rm SiO} \right] \mu_{g} m_{\rm H_{2}} d^{\rm 2} ,
\end{equation}
where $N_{\rm u}$ is the upper state column density of SiO (5$-$4), $A_{\rm ul}$ is the Einstein transition probability coefficient, ${S_{\rm \nu}/\Omega}$ is the SiO intensity at frequency ${\rm \nu}$, $dv$ is the velocity interval in km s$^{-1}$, $T_{\rm ex}$ is the line excitation temperature, $Q(T_{\rm ex})$ is the partition function, $g_{u}$ is the degeneracy of the upper state, $E_{\rm u}$ is the upper level energy, $\mu_{g}$ = 1.36 is the mean atomic weight, $m_{\rm H_{2}}$ is the mass of hydrogen molecule, and $d$ is the source distance of 1.4 kpc. The abundance of SiO varies by several orders of magnitude within a range of 10$^{-12}$ to 10$^{-7}$ \citep{1989ApJ...343..201Z, 1998ApJ...509..768G, 2005MNRAS.361..244C, 2007A&A...462..163N, 2012ApJ...756...60S, 2021ApJ...909..177L} and it may decrease with time because of decaying jet activity \citep{2010A&A...517A..66L, 2013A&A...557A..94S}. Of the 16 SiO-detected MDCs, 12 are identified as infrared-quiet MDCs based on the 24 $\mu$m flux in \citet{2022ApJ...927..185W}, and we adopt a [SiO]/[H$_2$] ratio of 10$^{-8}$, which is a typical value for mid-infrared-dark sources in \citet{2013A&A...557A..94S}. We then compute the outflow momentum ($P_{\rm out}$) and energy ($E_{\rm out}$):
\begin{equation}
P_{\rm out} = \sum M_{\rm out} (\Delta v) \Delta v ,
\end{equation}
\begin{equation}
E_{\rm out} = \frac{1}{2} \sum M_{\rm out} (\Delta v) \Delta v^{2} ,
\end{equation}
where $\Delta v$ is the outflow velocity relative to $v_{\rm lsr}$. Using the projected distance ($d_{\rm max}$) from the  SiO outflow to the central source, we derive the outflow dynamical timescale ($t_{\rm dyn}$), mass-outflow rate ($\dot{M}_{\rm out}$), outflow luminosity ($L_{\rm out}$), and mechanical force ($F_{\rm out}$):
\begin{equation}
t_{\rm dyn} = d_{\rm max} / v_{\rm max} ,
\end{equation}
\begin{equation}
\dot{M}_{\rm out} = M_{\rm out} / t_{\rm dyn} ,
\end{equation}
\begin{equation}
L_{\rm out} = E_{\rm out} / t_{\rm dyn} ,
\end{equation}
\begin{equation}
F_{\rm out} = P_{\rm out} / t_{\rm dyn} .
\end{equation}
Here, $v_{\rm max}$ is the maximum SiO velocity relative to $v_{\rm lsr}$. These results are shown in Table \ref{tab:parameters} and Figure \ref{fig:relations}. For the overlapping areas of SiO (5$-$4) emission generated from different central sources, we only estimate their SiO column densities, outflow masses, momentums, and energies here.

The derived outflow masses range from 0.05 to 18.3 M$_{\odot}$. The outflow dynamical timescales, mass-outflow rates, mechanical force, and outflow luminosity have medians of 3\,720 years, 1.69 $\times$ 10$^{-4}$ M$_{\odot}$ yr$^{-1}$, 1.02 $\times$ 10$^{-3}$ M$_{\odot}$ km s$^{-1}$ yr$^{-1}$, and 3.49 $\times$ 10$^{32}$ erg s$^{-1}$, respectively.

%-----------------------------------------------------------------------
\begin{table*}
\begin{center}
\scriptsize
\begin{tabular}{cc r c c ccc cc ccc}
\hline\hline
\noalign{\smallskip}
 MDC ID & Continuum &  & $\Delta v$ & $N_{\rm SiO}$ & $M_{\rm out}$ & $P_{\rm out}$ & $E_{\rm out}$ & $d_{\rm max}$ & $t_{\rm dyn}$ & $\dot{M}_{\rm out}$ & $F_{\rm out}$ & $L_{\rm out}$ \\ 
  & Source ID &  &  & $\times$10$^{14}$ &  &  & $\times$10$^{43}$ &  & $\times$10$^{3}$ & $\times$10$^{-4}$ & $\times$10$^{-3}$ & $\times$10$^{32}$ \\
  &  &  & (km s$^{-1}$) & (cm$^{-2}$) & (M$_{\odot}$) & (M$_{\odot}$ km s$^{-1}$) & (erg) & (pc) & (yr) & (M$_{\odot}$ yr$^{-1}$) & (M$_{\odot}$ km s$^{-1}$ yr$^{-1}$) & (erg s$^{-1}$) \\
% (1) & (2) & (3) & (4) & (5) & (6) & (7) & (8) & (9) & (10) & (11) & (12) & (13) & (14) \\
 \hline
220 & MM1 & blue & [9.1, 12.2]  & 0.89  & 0.08  & 0.24  & 0.10   & 0.062 & 11.2 & 0.067  & 0.021  & 0.029 \\
    &             & red  & [16.8, 24.9] & 22.6  & 1.23  & 7.30  & 5.18  & 0.022 & 2.03 & 6.00    & 3.59    & 8.12 \\
\cline{2-13}
    & MM2 & blue & [9.1, 12.2]  & 3.98  & 0.58  & 1.45  & 0.50  & 0.030 & 5.37 & 1.08   & 0.270  & 0.299 \\
    &          & red  & [16.8, 26.1] & 6.24  & 1.01  & 7.08  & 5.88  & 0.043 & 3.64 & 2.77   & 1.95    & 5.16 \\
%N03
\hline
248 & MM1 & blue & [4.9, 12.1]  & 1.61  & 0.49  & 2.58  & 1.81  & 0.059 & 6.09 & 0.814  & 0.424  & 0.944 \\
      &           & red  & [16.7, 20.7] & 4.09  & 0.57  & 1.94  & 0.99  & 0.025 & 3.80 & 1.49    & 0.508  & 0.826 \\
\cline{2-13}
    & MM2  & blue & [4.9, 12.1]  & 1.54  & 0.40  & 1.46  & 0.80  & 0.117 & 12.1 & 0.325   & 0.121  & 0.210 \\
    &           & red  & [16.7, 18.5] & 11.3  & 0.85  & 1.84  & 0.50  & 0.072 & 22.7 & 0.371   & 0.082  & 0.071 \\
\cline{2-13}
    & MM3 & blue & [7.2, 12.1]  & 2.98   & 0.38  & 1.51  & 0.76  & 0.103 & 14.2 & 0.265   & 0.106  & 0.169 \\
%N12
\hline
 310 & MM1 & blue & [-45.6, 3.9] & 11.8  & 9.60  & 182   & 510   & 0.108 & 1.75 & 46.8  & 89.3   & 794 \\
      &            & red  & [8.5, 52.2]  & 2.82  & 0.47   & 10.3  & 30.2  & 0.038 & 1.12 & 5.67  & 12.6   & 118 \\
%NW14
\hline
 341 & MM1 & blue & [$-$48.9, $-$6.5] & 11.2  & 3.52  & 54.6   & 144  & 0.038 & 0.835 & 42.2  & 65.5  & 551 \\
        &           & red  &  [$-$1.9, 43.2]   &   19.2  & 4.00  & 72.7   & 191  & 0.056 & 1.14  & 35.1   & 63.7  & 532 \\
%N63
\hline
351 & MM2 & blue & [$-$12.1, $-$2.9]  & 13.8  & 3.79  & 23.8  & 18.0  & 0.089 & 7.53 & 5.04  & 3.16  & 7.60 \\
           &           & red  & [1.7, 10.4]        & 15.2  & 0.99  & 9.25  & 9.28  & 0.022 & 1.97 & 5.03  & 4.71  & 15.1 \\
\cline{2-13}
            & MM3 & red  & [1.7, 11.6]        & 23.8  & 7.14  & 44.0  & 34.4  & 0.068 & 5.42 & 13.2  & 8.11  & 20.2 \\
%S32
\hline
 507/753  & MM3 & blue & [$-$24.3, $-$0.7] & 27.0  & 16.5  & 166  & 243   &       &      &       &       &  \\
          &           & red  & [3.9, 13.9]       & 4.24  & 0.73  & 4.74  & 3.78  & 0.056 & 4.41 & 1.66    & 1.07     & 2.73 \\
          &           & red  & [3.9, 9.4]        & 2.39   & 0.28  & 1.24  & 0.70  & 0.139 & 17.4 & 0.159  & 0.071   & 0.129 \\
\cline{2-13}
          & MM4 & red  & [3.9, 10.5]      & 3.38    & 0.59   & 3.36  & 2.24   & 0.031 & 3.45 & 1.72    & 0.979   & 2.07 \\
\cline{2-13}
          & MM5 & blue & [$-$8.6, $-$6.3] & 8.01  & 0.44  & 3.76  & 3.29  & 0.063 & 6.06 & 0.720  & 0.622  & 1.73 \\
\cline{2-13}
          & MM8 & blue & [$-$1.9, $-$0.7]    & 3.76    & 0.64  & 0.68  & 0.11  & 0.051 & 14.2 & 0.451  & 0.048  & 0.026 \\
%S07
\hline
 509 & MM1 & red  & [9.7, 31.2] & 5.82  & 1.24  & 15.3   & 24.6  & 0.045 & 1.87 & 6.69  & 8.22  & 42.0 \\
\cline{2-13}
        &    & blue & [$-$8.2, 5.1]   & 5.22   & 4.81  & 25.0  & 18.5  &  &  &  &  & \\
%S30
\hline
 684 & MM1 & blue & [$-$36.5, $-$8.8] & 13.3  & 6.08  & 66.5  & 107   & 0.077 & 2.50 & 24.3  & 26.6  & 136 \\
        &           & red  & [$-$4.2, 29.8]     & 10.5   & 2.48  & 26.0  & 35.1  & 0.056 & 1.51 & 16.4  & 17.2  & 74.0 \\
\cline{2-13}
        & MM3 & blue & [$-$35.4, $-$8.8] & 7.66  & 1.19  & 12.7  & 20.0  & 0.084 & 3.21 & 4.20  & 4.49  & 22.5 \\
%N68
\hline
 698 & MM1 & red  &  [21.9, 28.8]  & 1.82  & 0.62  & 4.32 & 2.40   & 0.101 & 10.8 & 0.573  & 0.336  & 0.712 \\
\cline{2-13}
     & MM3 & blue & [$-$5.0, 17.3] & 17.7   & 6.16  & 75.6 & 78.9   & 0.081 & 3.20 & 19.2   & 18.6    & 78.7 \\
%N56  
\hline
699 & New & blue & [$-$9.7, $-$3.8] & 1.88 & 0.15 & 0.47 & 0.23 & 0.021 & 2.48 & 0.620 & 0.188 & 0.299 \\
    &      & red  & [0.8, 11.8]      & 0.22 & 0.05 & 0.42 & 0.42 & 0.037 & 2.75 & 0.170 & 0.151 & 0.487 \\
%N38
\hline
801 & MM1 & red & [$-$7.5, 7.2] & 3.11  & 0.12  & 1.31  & 1.65  & 0.016 & 1.24 & 1.03   & 1.05  & 4.25 \\
%N65
\hline
 1112 & MM3 & red & [11.3, 18.3] & 12.4  & 1.30  & 6.63  & 3.93  & 0.089 & 9.33 & 1.40  & 0.711  & 1.34 \\
\cline{2-13}
      &    & blue & [$-$7.6, 6.7]     & 9.02    & 5.99  & 39.5  & 33.5  &  &  &  &  & \\
      &    & red  & [11.3, 18.3]        & 5.80   & 5.13  & 5.23  & 7.70  &  &  &  &  & \\
%N30
\hline
 1454 & MM5 & red & [10.0, 47.5] & 0.96  & 0.54  & 10.9  & 28.4  & 0.097 & 2.09 & 2.59  & 5.23  & 43.3  \\
%DR15
\hline
1467 & MM17 & red  & [$-$1.4, 5.0]  & 4.27  & 1.77  & 8.72  & 4.83  & 0.089 & 10.0 & 1.76  & 0.870  & 1.54 \\
\cline{2-13}
         &    & blue & [$-$20.9, $-$6.0] & 4.31   & 5.07  & 8.05  & 1.43  &  &  &  &  & \\
%N44
\hline
1599 & MM2 & blue & [$-$9.7, $-$6.7] & 9.41 & 0.37  & 0.32  & 0.25  & 0.023 & 4.29 & 0.864  & 0.075  & 0.184 \\
         &           & red  & [$-$2.1, 21.9]   & 14.1  & 18.3  & 259   & 369   & 0.078 & 2.90 & 63.3    & 89.5    & 406 \\
\cline{2-13}
         & MM3 & red  & [$-$2.1, 21.9]   & 2.23   & 0.42  & 5.91  & 9.98  & 0.179 & 6.67 & 0.627  & 0.886  & 4.76 \\
%DR21/N53
\hline
  Mean  &  &  &    & 8.08 & 2.83 & 28.7 & 47.7 & 0.066 & 5.87 & 8.74 & 12.0 & 79.9 \\
\hline
Median  &  &  &    & 5.80 & 0.99 & 7.08 & 5.18 & 0.061 & 3.72 & 1.69 & 1.02 & 3.49 \\
\hline
Minimum &  &  &    & 0.22 & 0.05 & 0.24 & 0.10 & 0.016 & 0.817 & 0.067 & 0.021 & 0.026 \\
\hline
Maximum &  &  &    & 27.0 & 18.3 & 259  & 510  & 0.179 & 22.7  & 63.3  & 89.5  & 794 \\
\hline\hline
\end{tabular}
\end{center}
\caption{Estimated physical parameters of identified SiO outflows.} \label{tab:parameters}
\end{table*}
%-----------------------------------------------------------------------

%-----------------------------------------------------------------------
%-----------------------------------------------------------------------

\subsection{SiO mass--velocity diagrams of the outflows} \label{M-V}
The mass--velocity diagram of molecular outflows often shows a power-law relation, $M(v) \propto v^{-\gamma}$, which can serve as a diagnostic tool for the interaction between outflows and their ambient gas, although its physical origin is not well established \citep{1996ApJ...471..308C,1996ApJ...459..638L,2000prpl.conf..867R,2001A&A...378..495R,2004ApJ...604..258S,2007prpl.conf..245A,2007ApJ...654..361Q,2009ApJ...702L..66Q,2011ApJ...743L..25Q}. The slope often changes at a velocity of between 6 and 12 km s$^{-1}$ and becomes steeper at higher velocities \citep{2007prpl.conf..245A}. The break velocity falls in a broader range of 4 to $\sim$20 km s$^{-1}$ in \citet{2018MNRAS.473.4220L}, and a weak, young outflow or a large inclination angle could contribute to a low break velocity. 

\begin{figure*}
\centering
   \includegraphics[width=250pt]{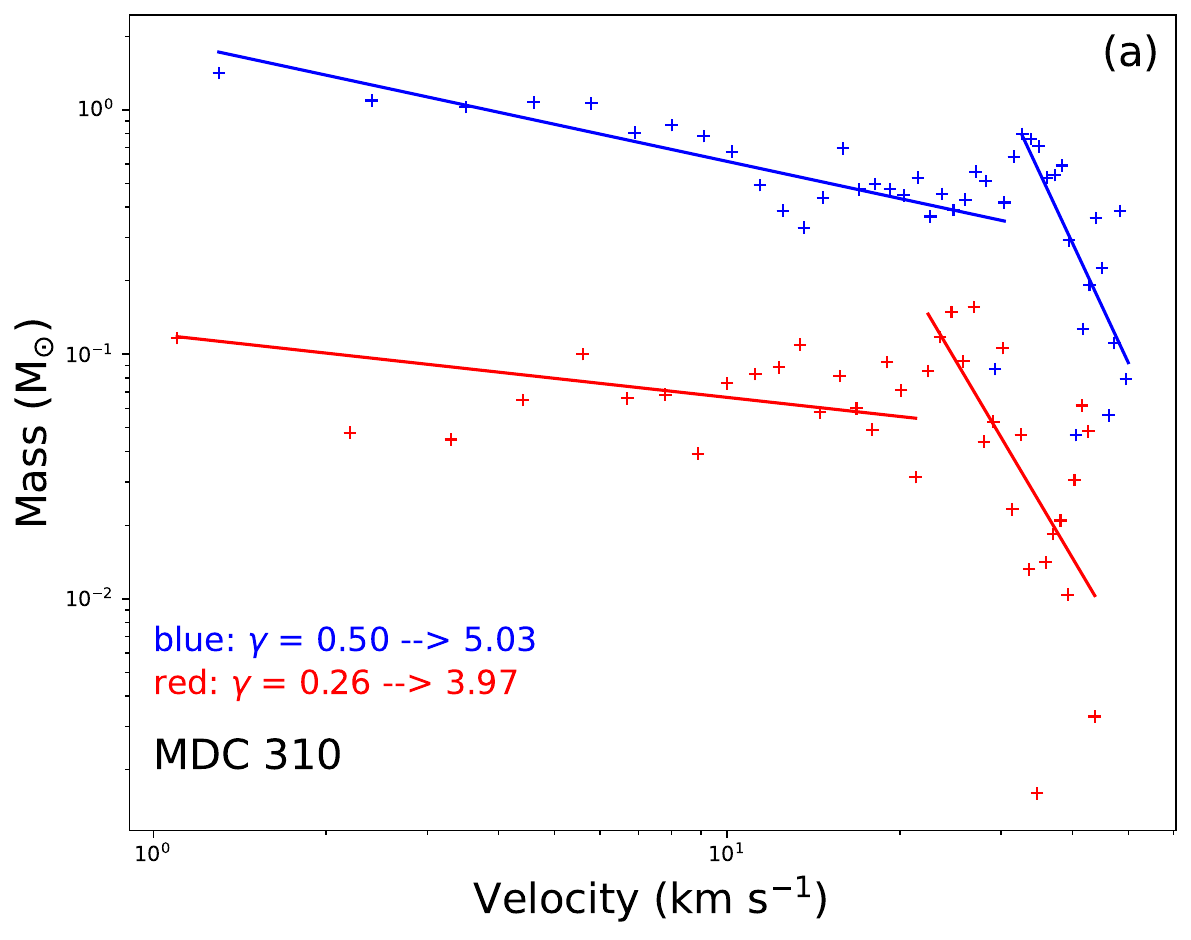}
   \includegraphics[width=250pt]{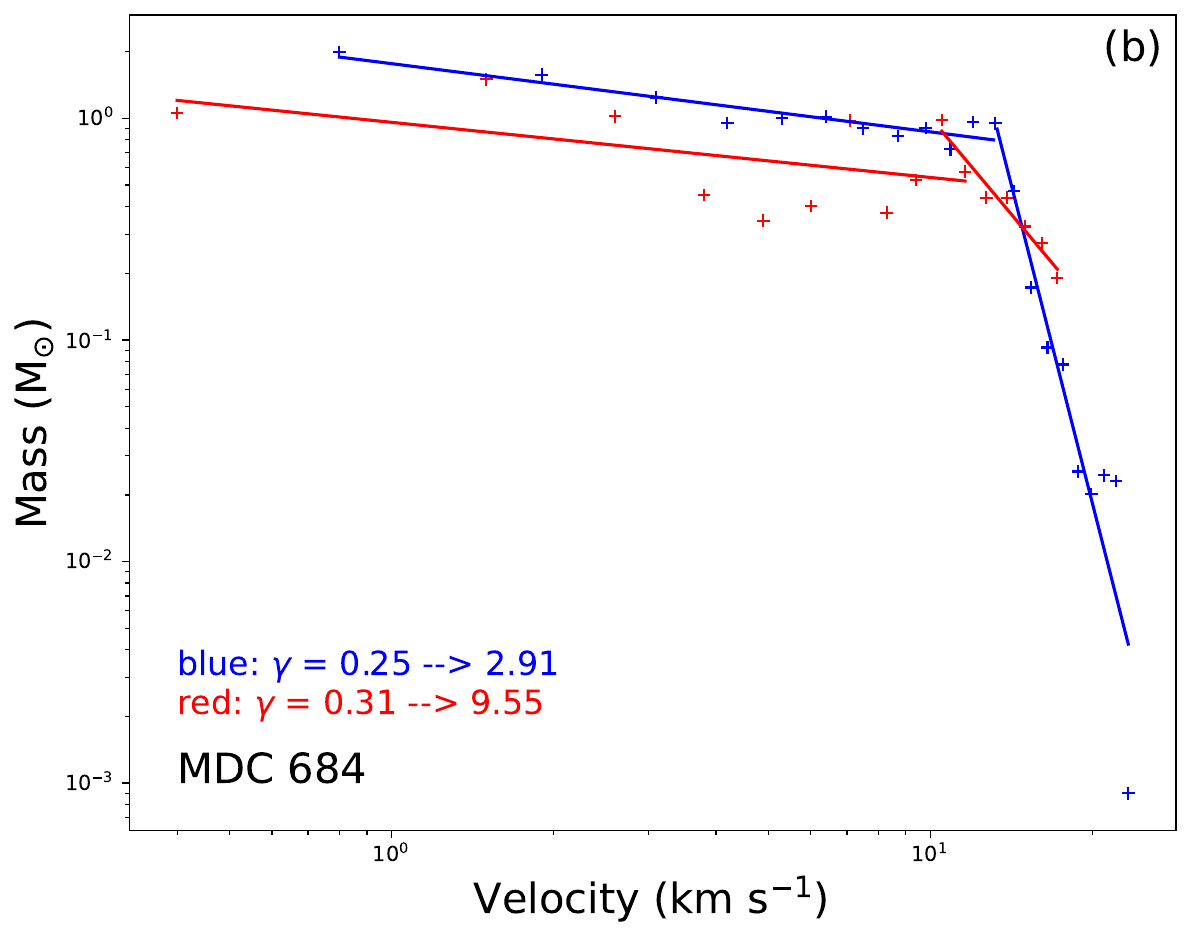}
     \caption{Mass--velocity relation of the SiO (5$-$4) outflow associated with MM1 in MDC 310 (a) and MM1 in MDC 684 (b), calculated from the SMA data for the gas mass in each channel, with each channel being 1.12 km s$^{-1}$ in width. The blue plus symbol denotes the measurement for the blueshifted lobe, and the red plus symbol indicates that for the redshifted lobe. Solid blue and red lines indicate broken power-law fittings to the blueshifted and redshifted lobes, respectively.}
  \label{fig:MV}
\end{figure*}

Figure \ref{fig:MV} presents the mass--velocity diagrams of the only two isolated SiO (5$-$4) bipolar outflows, and the outflow masses are calculated in each velocity channel. These two SiO bipolar outflows are associated with MM1 in MDC 310 and MM1 in MDC 684. For the mass--velocity diagram of the bipolar outflow centered at MM1 in MDC 310 (Figure \ref{fig:MV}(a)), both lobes can be fitted with a broken power law, with $\gamma$ steepening from 0.50 to 5.03 at 31.5 km s$^{-1}$ for the blueshifted lobe and from 0.26 to 3.97 at 21.9 km s$^{-1}$ for the redshifted lobe. For the mass--velocity diagram of the bipolar outflow related to MM1 in MDC 684 (Figure \ref{fig:MV}(b)), a broken power law can also be fitted for both lobes and $\gamma$ steepens from 0.25 to 2.91 at 13.2 km s$^{-1}$ for the blueshifted lobe and from 0.31 to 9.55 at 11.0 km s$^{-1}$ for the redshifted lobe. 

The fitted $\gamma$s at low velocities are consistent with the fitting values of the SiO outflows in \citet{2017ApJ...849...25L}, but are flatter than $\gamma$ values fitted from CO outflows \citep{1996ApJ...471..308C, 1996ApJ...459..638L, 2001A&A...378..495R, 2007ApJ...654..361Q, 2009ApJ...702L..66Q, 2018RAA....18...19X, 2019ApJ...873...73Z}. These latter values often range from 1 to 3. Simulations of molecular outflows in \citet{2003A&A...403..135D} and \citet{2005A&A...437..517K} suggest that the mass-spectrum slope is flatter at a younger age. Considering the estimated SiO outflow dynamical ages of 1.75 and 1.12 $\times$ 10$^{3}$ yr for blueshifted and redshifted lobes of the MDC310-MM1 outflow, and 2.50 and 1.51 $\times$ 10$^{3}$ yr for blueshifted and redshifted lobes of the MDC684-MM1 outflow, the flatter $\gamma$ values at low velocities appear to be mainly attributable to the young ages of the SiO outflows. In addition, the SiO emission can be moderately optically thick for the velocity range of $\Delta{v}$ $\le$ 10 km s$^{-1}$, and our assumption of optically thin SiO emission in LTE will underestimate the SiO outflow mass at low velocities, which will make the power-law index smaller.

%-----------------------------------------------------------------------
%-----------------------------------------------------------------------

%-----------------------------------------------------------------------
\begin{figure*}
   \includegraphics[width=165pt]{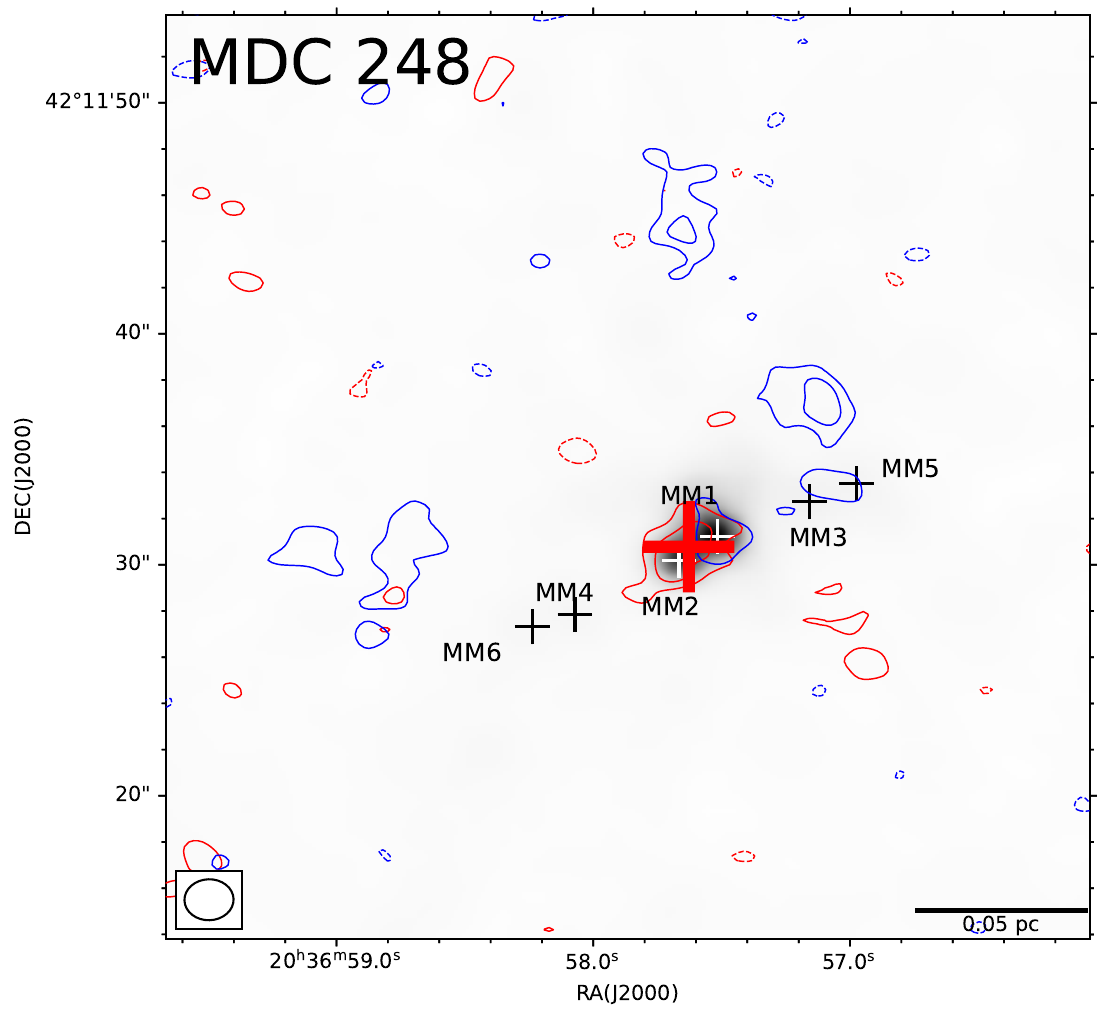}
   \includegraphics[width=165pt]{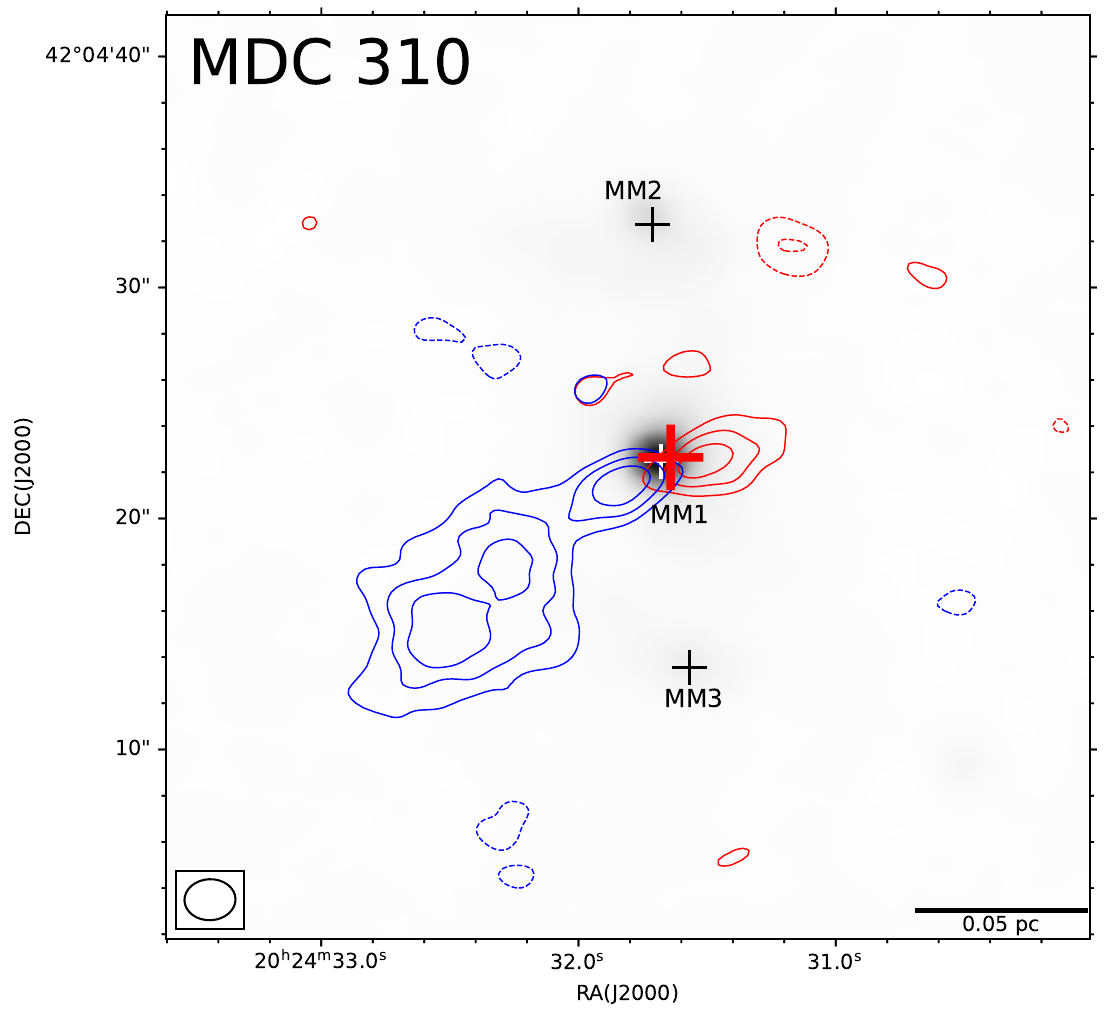} 
   \includegraphics[width=165pt]{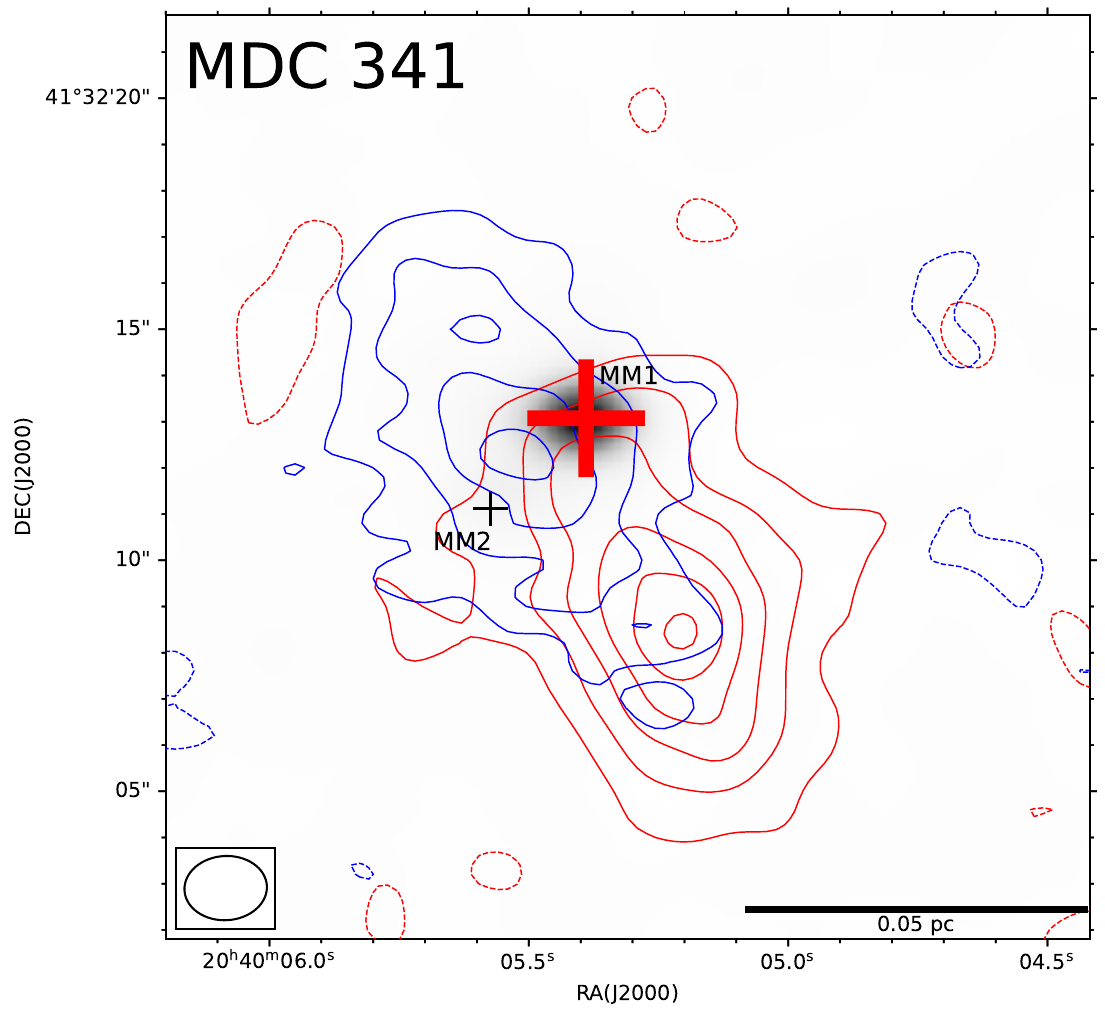} 
   
   \includegraphics[width=165pt]{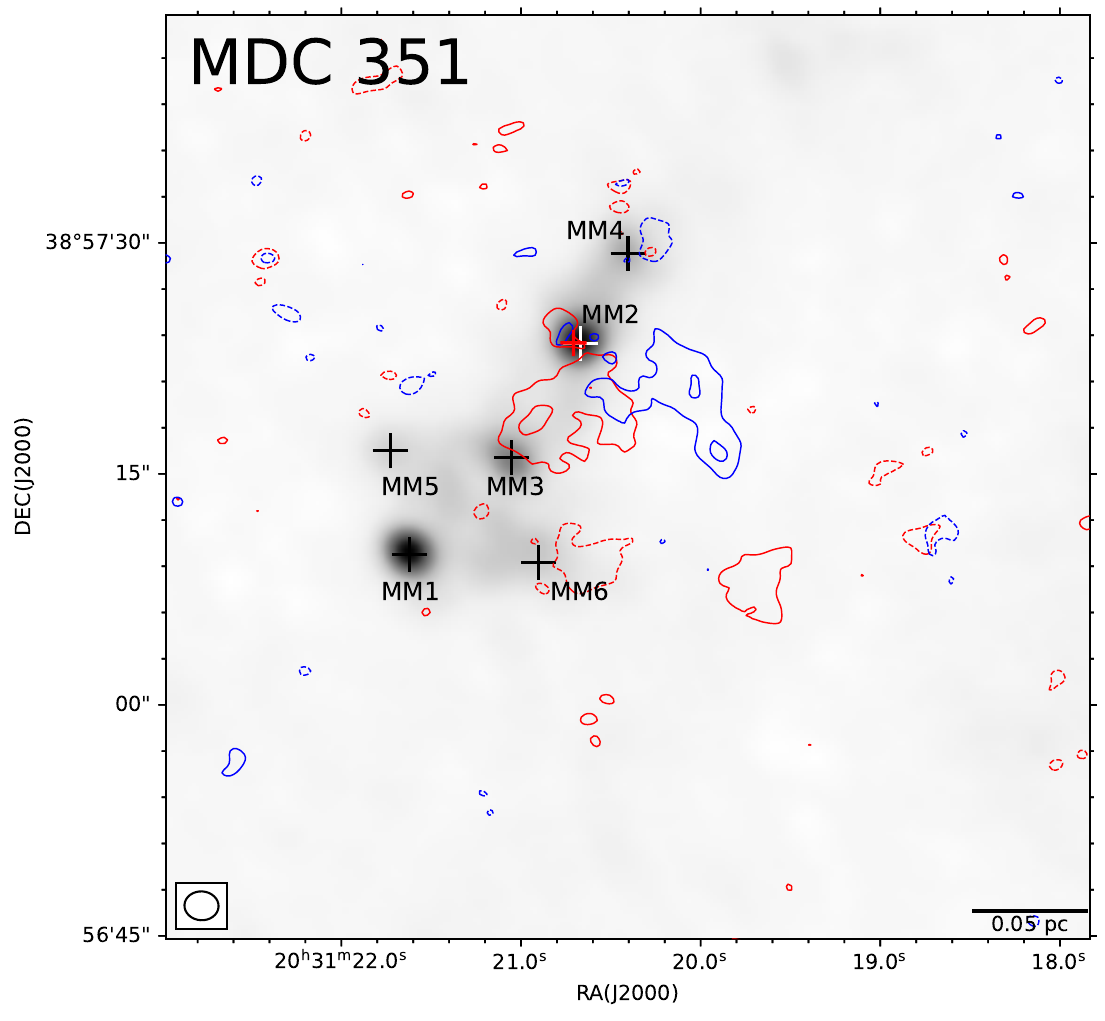} 
   \includegraphics[width=165pt]{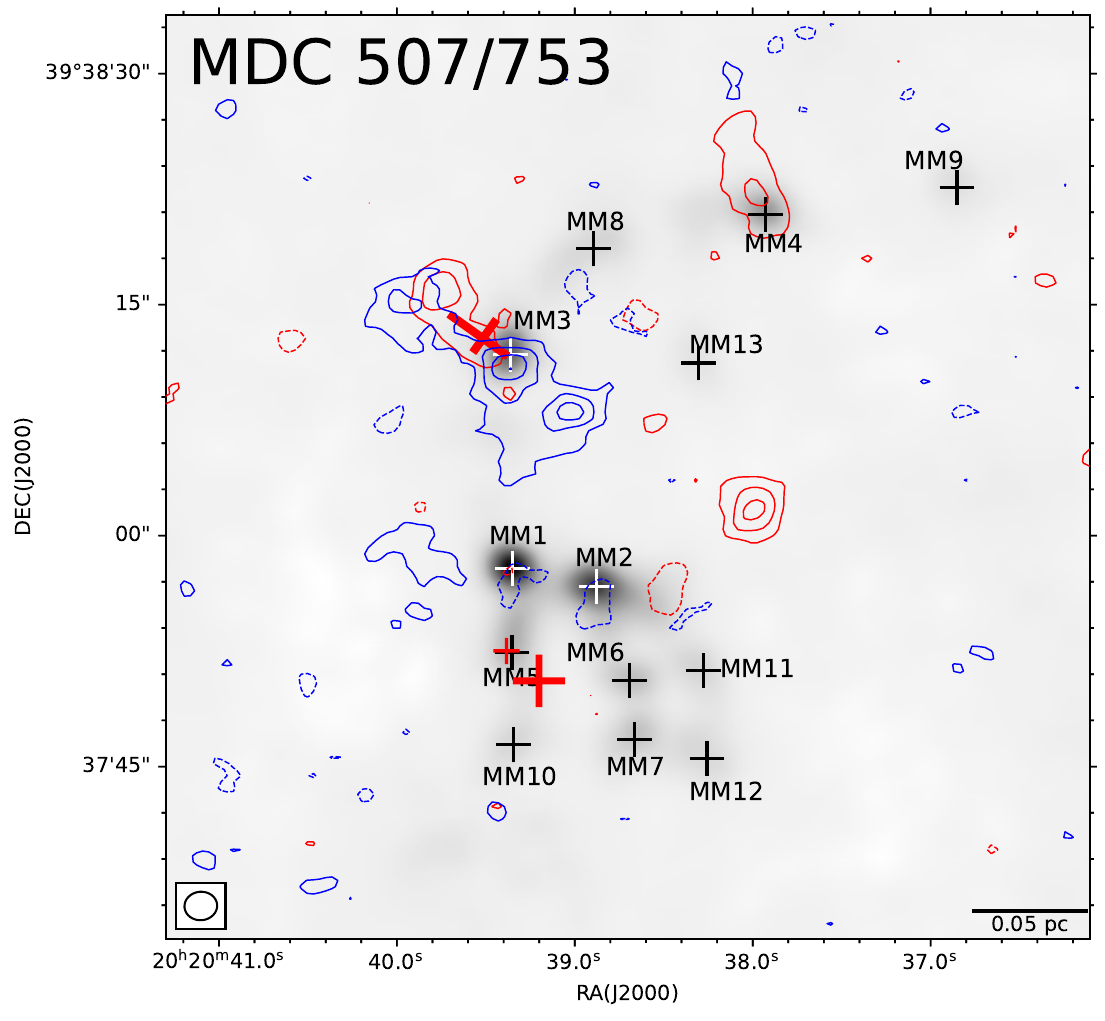} 
   \includegraphics[width=165pt]{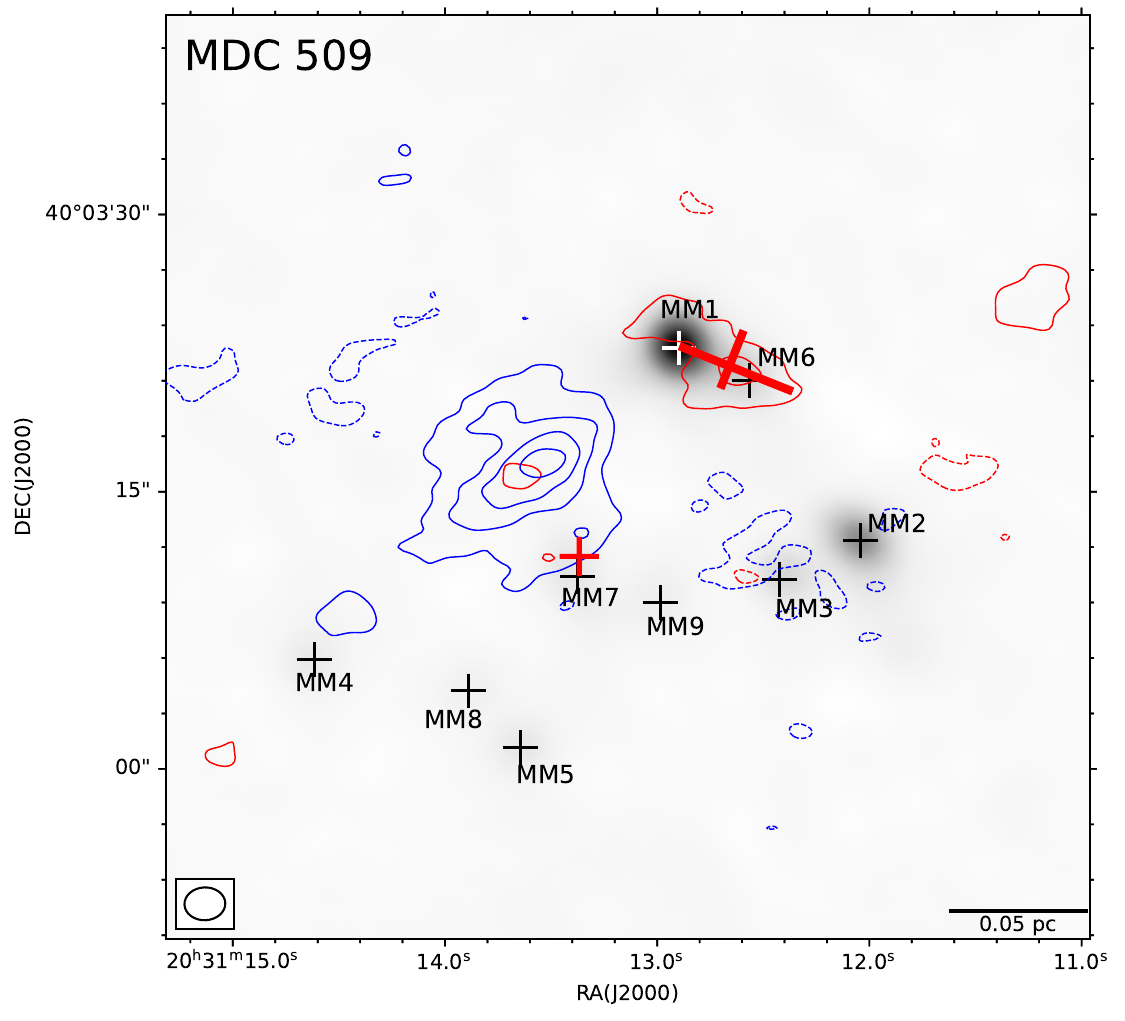} 
   
   \includegraphics[width=165pt]{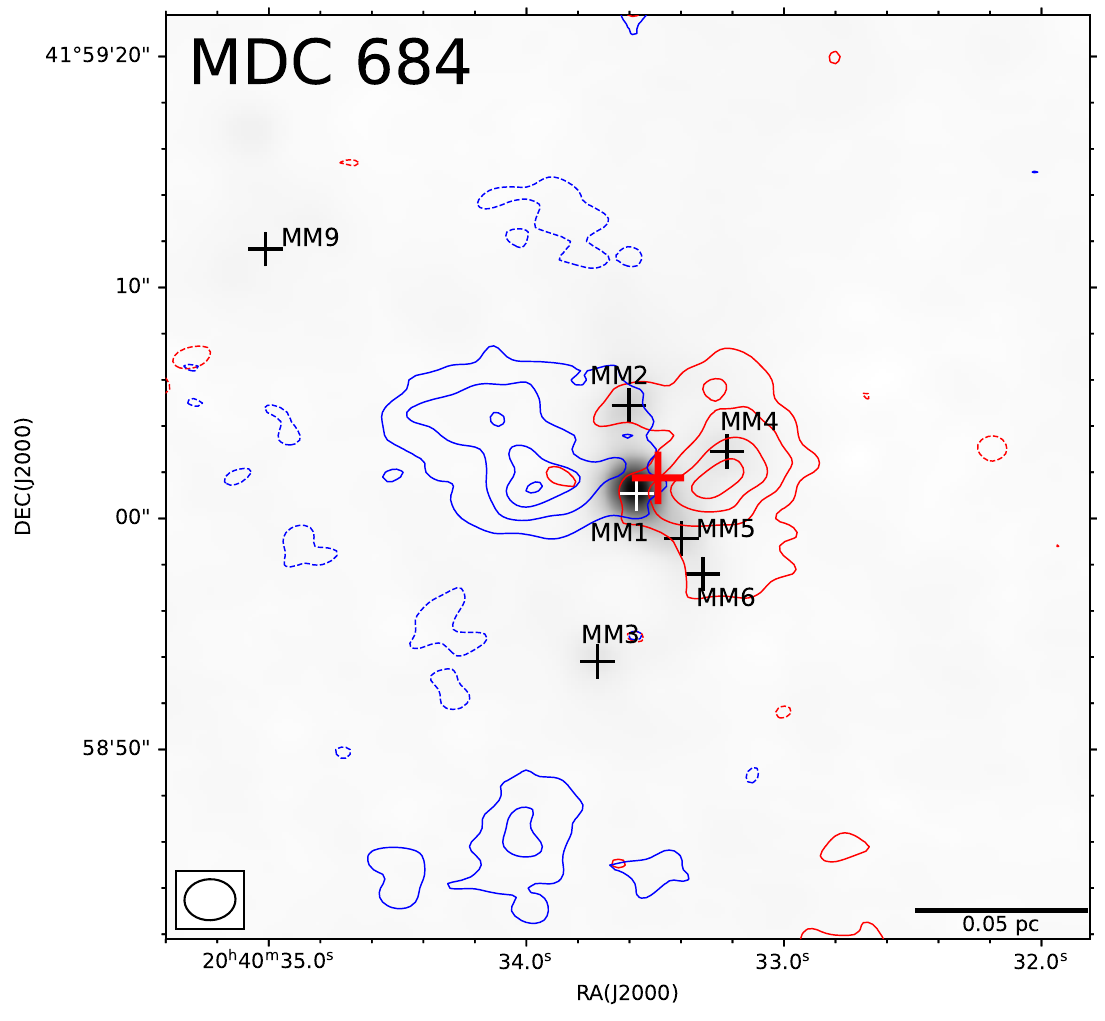} 
   \includegraphics[width=165pt]{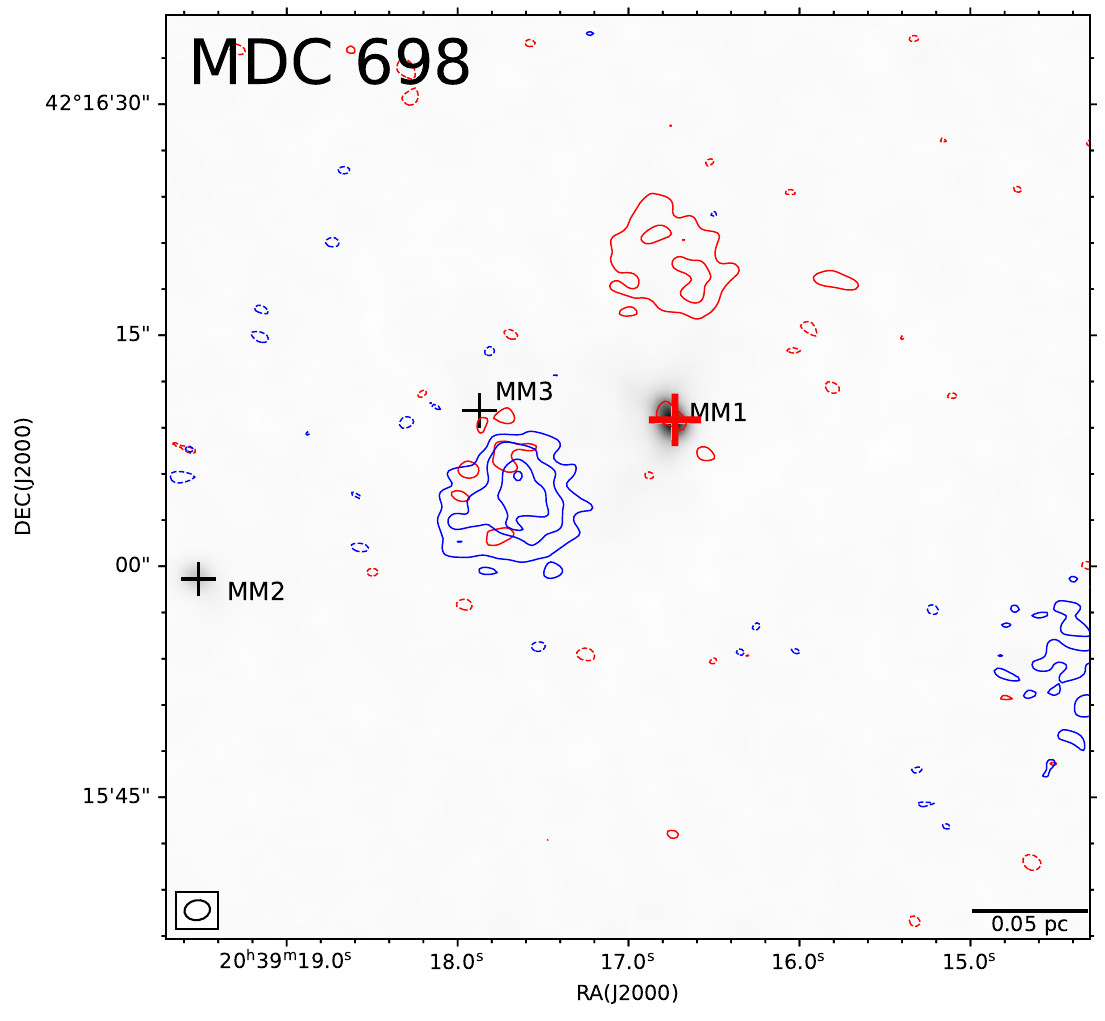} 
   \includegraphics[width=165pt]{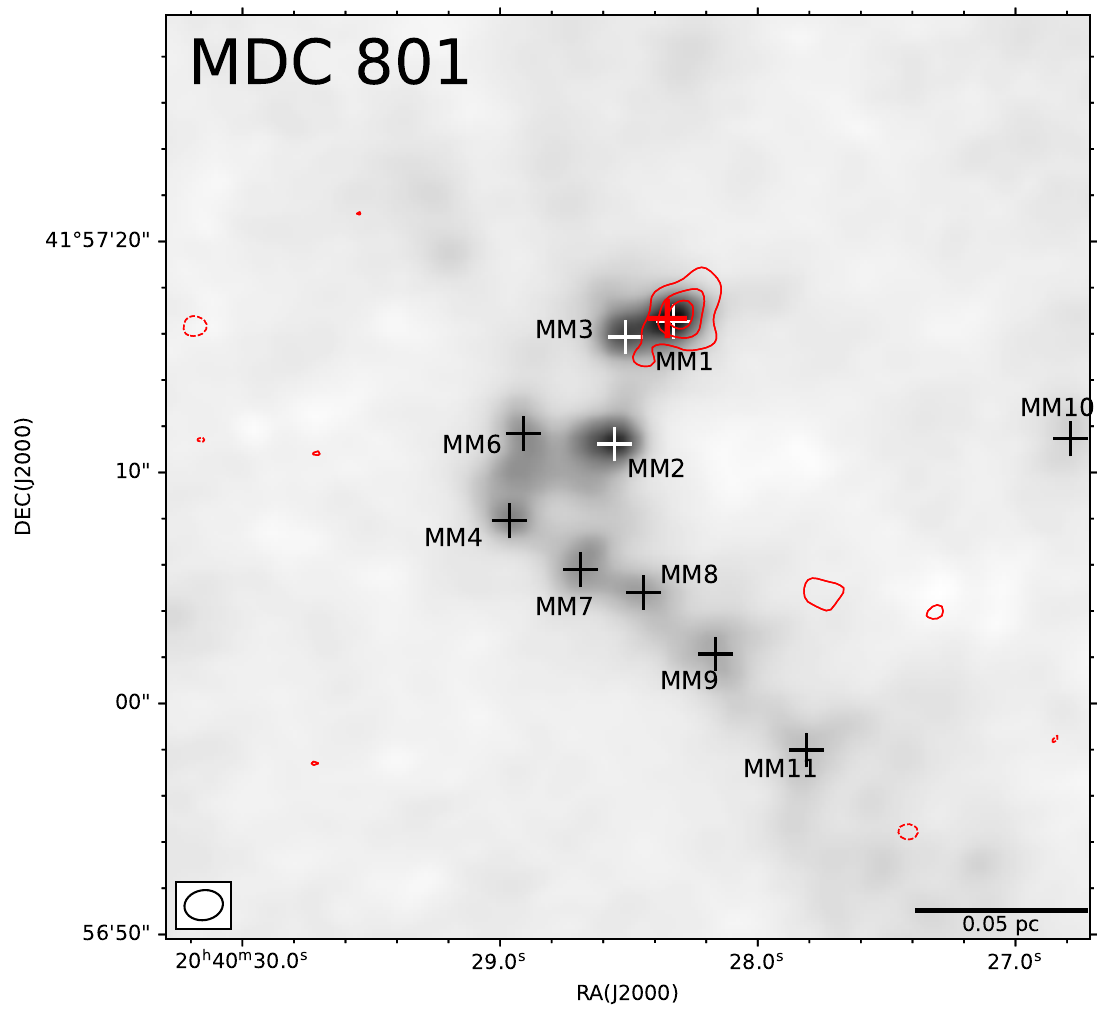} 
   
   \includegraphics[width=165pt]{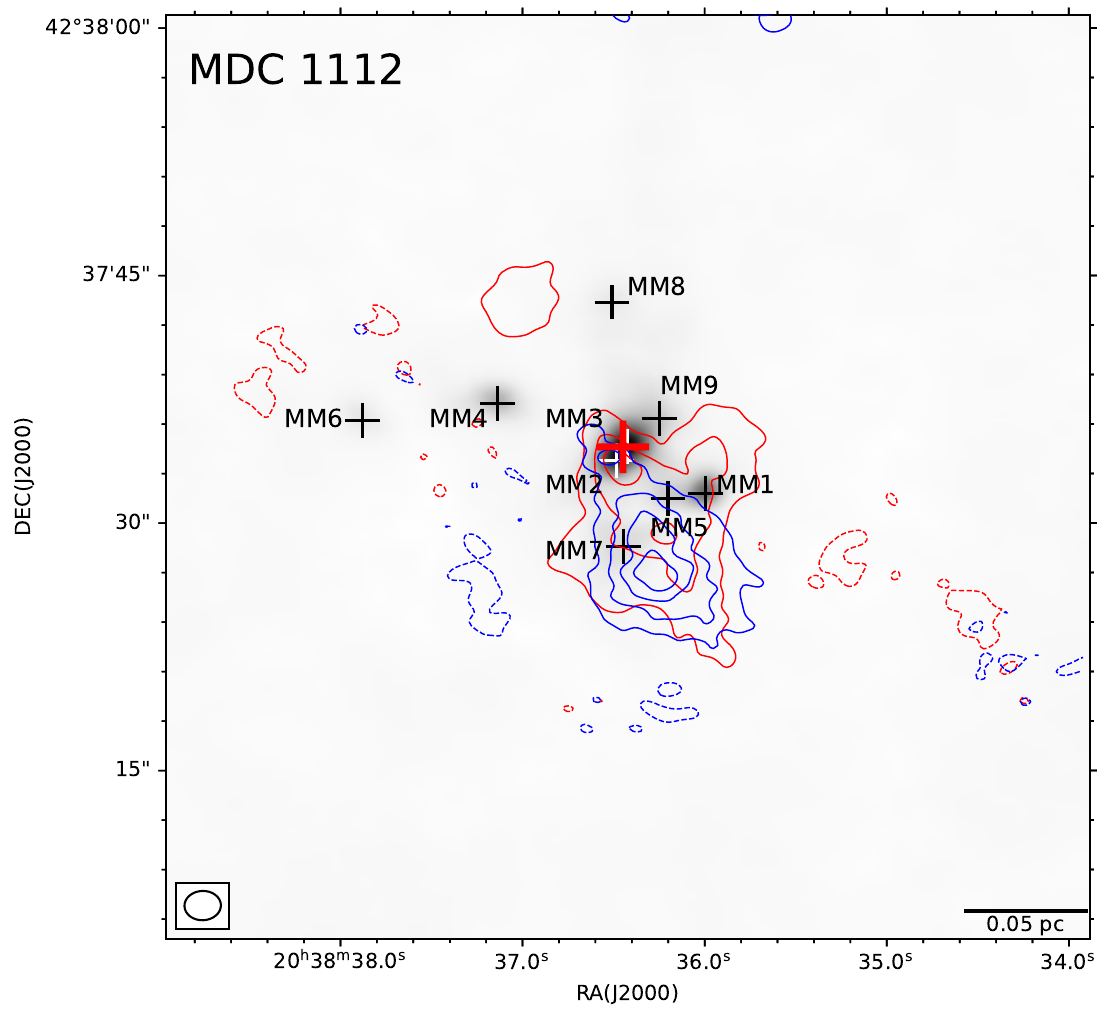} 
   \includegraphics[width=165pt]{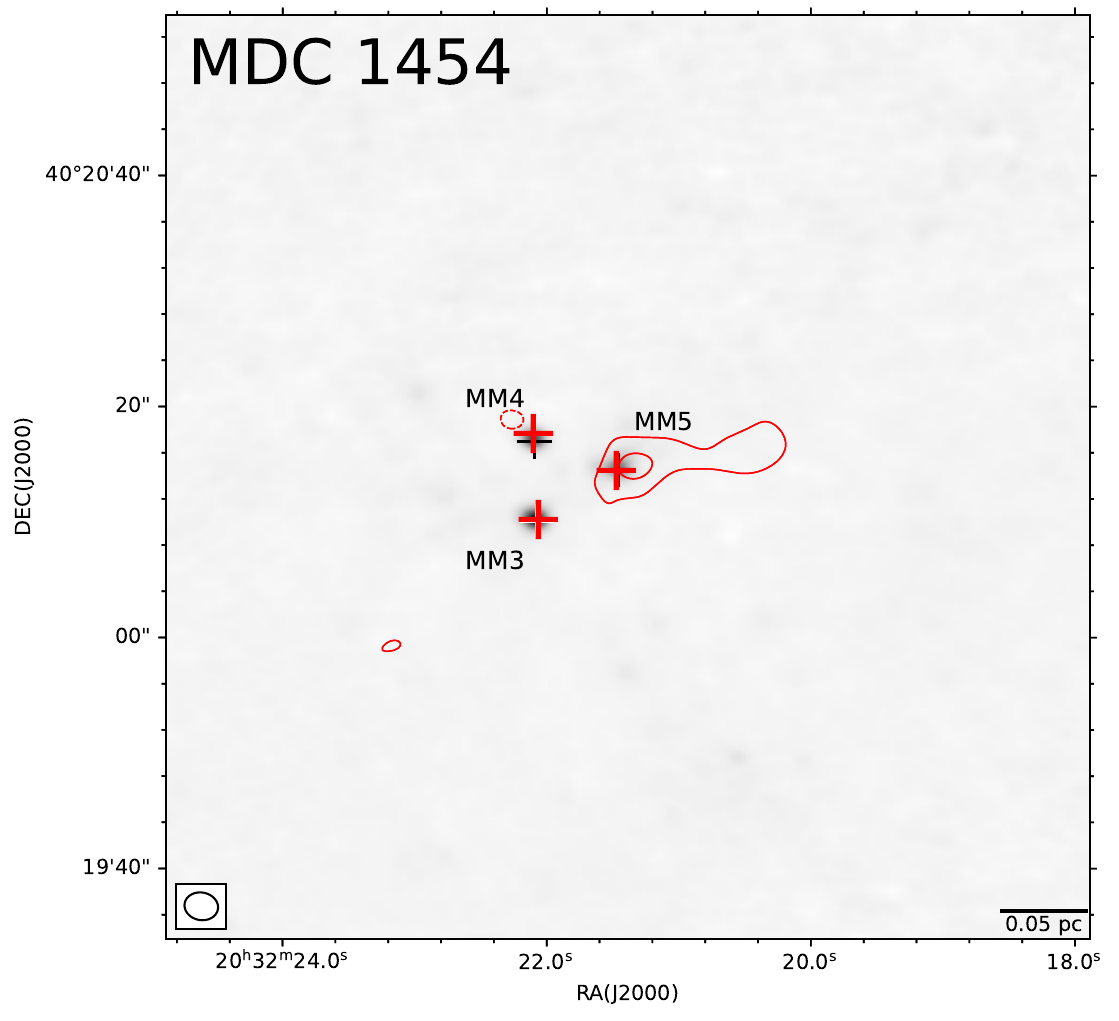} 
   \includegraphics[width=165pt]{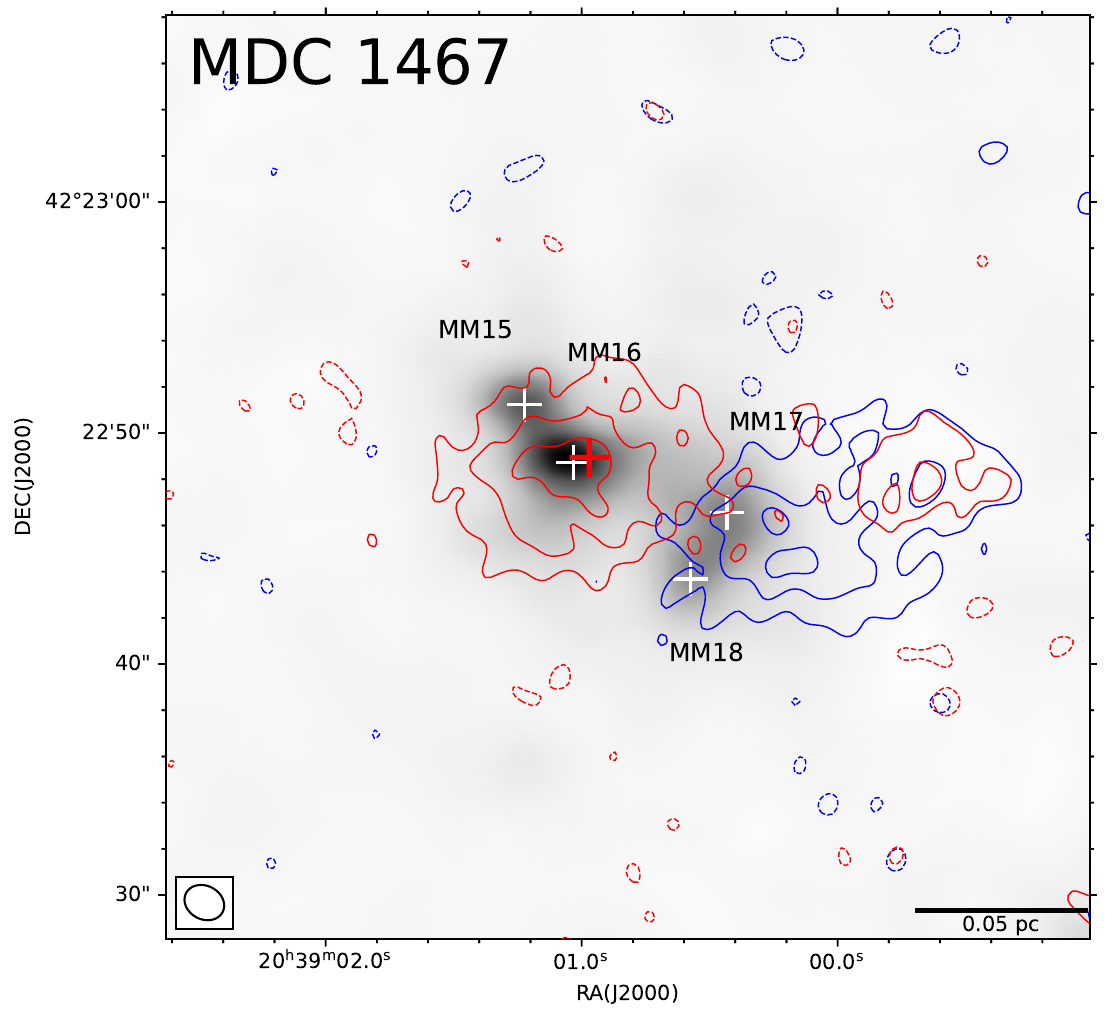} 
     \caption{ SiO emission versus radio emission associated with 15 continuum sources. The red and blue contours are the redshifted and blueshifted SiO (5$-$4) components with the same levels in Figures \ref{N03} $-$ \ref{N53}. The red crosses represent the positions and spatial scales of radio emissions. The grayscale maps represent 1.37 mm continuum emissions. The black and white crosses represent dust condensations identified by \citet{2021ApJ...918L...4C}. The MDC ID is marked in the top-left corner. The scale bar is presented at the bottom-right corner.}
  \label{fig:radio}
\end{figure*}
%-----------------------------------------------------------------------

\subsection{Properties of the SiO outflow central sources}

 \subsubsection{Associated radio continuum emissions} \label{rad}

%-----------------------------------------------------------------------
\begin{table*}
\caption{Properties of SiO (5$-$4) central sources.} \label{tab:ass_prop}
\begin{center}
\begin{tabular}{l ccc cccc}
\hline\hline
\noalign{\smallskip}
Source ID & M$_{\rm gas}$$^{~\rm a}$ & T$_{\rm NH_{3}}$ & T$_{\rm dust}$ & Radio emission$^{~\rm b}$ & IRAC$^{~\rm c}$ & MIPS$^{~\rm c}$ & Group$^{~\rm d}$ \\
& (M$_{\odot}$) & (K) & K &  & 3.6/5.8/8.0 $\mu$m & 24 $\mu$m & \\
\hline
MDC220-MM1 & 21.4 & $-$ & 18.1 & N & N & N & (1) \\
MDC220-MM2 & 7.64 & $-$ & 17.8 & N & Y & Y & (3) \\
MDC248-MM1 & 39.9 & 39.1 & 18.4 & K & Y & Y & (3) \\
MDC248-MM2 & 9.81 & 65.8 & 18.3 & K & Y & Y & (3) \\
MDC248-MM3 & 0.93 & 32.7 & 17.9 & N & N & N & (1) \\
MDC310-MM1 & 14.7 & 16.5 & 20.8 & K & Y & Y & (3) \\
MDC341-MM1 & 84.2 & $-$ & 17.4 & K & Y & Y & (3) \\
MDC351-MM2 & 10.3 & 17.0 & 18.0 & K & N & N & (1) \\
MDC351-MM3 & 3.15 & 19.5 & 18.1 & N & N & N & (1) \\
MDC507-MM3 & 8.54 & 35.0 & 17.9 & X/K & N & N & (1) \\
MDC507-MM5 & 7.16 & $-$ & 23.1 & C/X/K & Y & Y & (3) \\
MDC509-MM1 & 15.2 & 23.7 & 20.6 & K & N & Y & (2) \\
MDC509-MM4 & 1.57 & 16.5 & 17.3 & N & N & N & (1) \\
MDC509-MM7 & 1.48 & $-$ & 16.5 & K & N & Y & (2) \\
MDC684-MM1 & 4.20 & 29.2 & 17.8 & K & N & Y & (2) \\
MDC684-MM3 & 1.22 & $-$ & 17.4 & N & N & N & (1) \\
MDC698-MM1 & 9.11 & 21.9 & 17.8 & K & Y & Y & (3) \\
MDC698-MM3 & 0.61 & 18.1 & 16.8 & N & N & N & (1) \\
MDC753-MM4 & 10.3 & 21.6 & 17.0 & N & N & N & (1) \\
MDC753-MM8 & 2.62 & 23.6 & 20.0 & N & Y & Y & (3) \\
MDC801-MM1 & 3.45 & 17.9 & 18.1 & K & N & N & (1) \\
MDC1112-MM1 & 23.7 & $-$ & 27.8 & N & N & Y & (2) \\
MDC1112-MM2 & 10.9 & $-$ & 28.5 & C & Y & Y & (3) \\
MDC1112-MM3 & 19.8 & $-$ & 28.8 & C & Y & Y & (3) \\
MDC1454-MM5 & 8.33 & $-$ & 16.5 & K & N & N & (1) \\
MDC1467-MM17 & 43.1 & $-$ & 23.7 & N & N & N & (1) \\
MDC1467-MM18 & 27.4 & $-$ & 22.5 & N & N & N & (1) \\
MDC1599-MM2 & 31.5 & $-$ & 17.4 & N & N & Y & (2) \\
MDC1599-MM3 & 20.0 & $-$ & 17.2 & N & N & N & (1) \\
\hline\hline
\end{tabular}
\end{center}
Notes. \\
$^{~\rm a} $ Gas mass of the condensations obtained from \citet{2021ApJ...918L...4C}. \\
$^{~\rm b} $ Radio emissions \citep{2022ApJ...927..185W} at the position of SiO central sources: N for emission under 5$\sigma$, C/X/K for emission above 5$\sigma$ at C/X/K bands. \\
$^{~\rm c} $ Spitzer 3.6/5.8/8.0/24 $\mu$m emissions: N for emission under 10$\sigma$ after excluding emission from the background, Y for emission above 10$\sigma$.  \\
$^{~\rm d} $ Evolutionary group classification: (1) for sources undetected from 3.6 to 24 $\mu$m, (2) for sources detected at 24 $\mu$m but not or partly detected from 3.6 to 8.0 $\mu$m, (3) for sources detected at 3.6$-$24 $\mu$m.
\end{table*}
%-----------------------------------------------------------------------

\citet{2022ApJ...927..185W} made a comprehensive survey of the radio continuum emissions in MDCs in the Cygnus-X complex. All the central sources of the 32 SiO outflows identified here are covered in the radio continuum survey, and we find that 16 outflow sources are associated with the radio continuum emission (see Figure \ref{fig:radio} and Table \ref{tab:ass_prop}). For 12 SiO (5$-$4) outflows, radio continuum emissions are found at their central sources. For the MDC509-MM1 unipolar outflow and the MDC507-MM3 bipolar outflow, the radio emissions both arise from the central sources and extend along the same orientations as those of the redshifted lobes. For two outflows, we detect compact radio emissions very close to the central sources of the outflows, and the emissions overlaps with the blueshifted lobe of the MDC509-MM7 unipolar outflow and the redshifted lobe of the MDC684-MM1 bipolar outflow. For the radio continuum emission detected at the  central sources of the SiO outflows, we believe only the emission associated with MDC1112-MM2/MM3 to be arising from an H {\scriptsize II} region, and the radio emission in other sources is most likely attributed to ionized radio jets or winds \citep{2022ApJ...927..185W}.

The radio continuum detection could be an indicator of the evolutionary stage of the outflow central sources \citep{2021A&A...653A.117B}. At an early evolutionary stage, there is no radio emission associated with sources; as sources evolve, their associated radio emission originates from ionized jets or winds; sources with radio emission from H {\scriptsize II} regions are at a relatively late evolutionary stage.

 \subsubsection{Evolution of the central sources} \label{Herschel}

\citet{Zhang2023} performed VLA NH$_{3}$ ($J$,$K$) = (1,1) and (2,2) inversion line observations of the MDCs in Cygnus-X, and 15 SiO outflow central sources are covered 
here. The NH$_{3}$ lines are sensitive to temperatures of up to $\sim$ 30 K, and the temperatures derived from them for the 15 SiO outflow central sources studied here are listed in Table \ref{tab:ass_prop}. Temperature could be a tracer of evolutionary stage, where a lower temperature represents an earlier evolutionary stage.

We can also investigated the evolutionary stages of the SiO central sources by studying their infrared environments. We obtained the datasets in the archive, including Spitzer 3.6, 4.5, 8.0, and 24 $\mu$m and Herschel 70, 160, 250, 350, and 500 $\mu$m maps. For each source, we checked the infrared detection at the position and boundary identified by \citet{2021ApJ...918L...4C} to determine whether or not there is a counterpart. The infrared images for each MDC at wavelengths of 3.6, 4.5, 5.8, 8.0, and 24 $\mu$m are presented in Appendix \ref{Infrared}. Because of the poor spatial resolution (5.2$^{\prime\prime}$ at 70 $\mu$m, 12.0$^{\prime\prime}$ at 160 $\mu$m, 17.6$^{\prime\prime}$ at 250 $\mu$m, 23.9$^{\prime\prime}$ at 350 $\mu$m, and 35.2$^{\prime\prime}$ at 500 $\mu$m), we have not presented the infrared images with a wavelength larger than 70 $\mu$m. The detection at Spitzer 3.6, 5.8, 8.0, and 24 $\mu$m for the 29 SiO-central sources is summarized in Table \ref{tab:ass_prop}.

\begin{figure}
\centering
   \includegraphics[width=240pt]{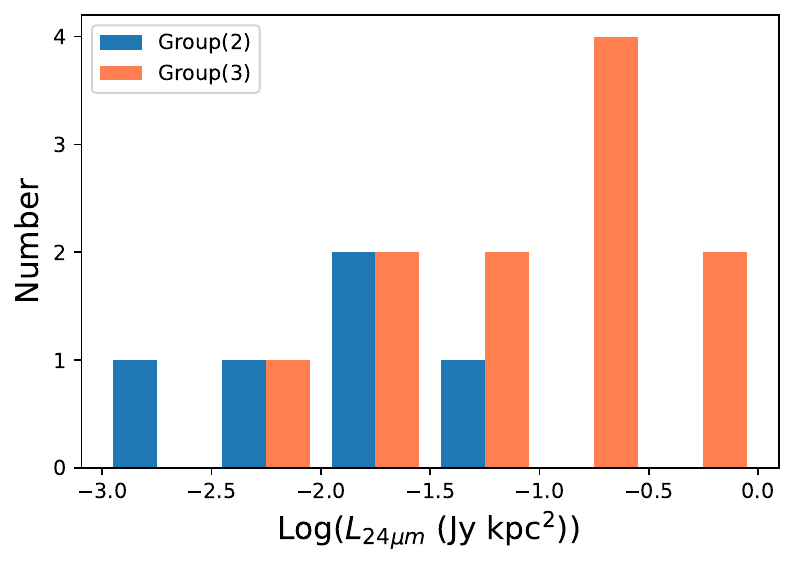}
     \caption{Distribution of the 24 $\mu$m luminosity between the SiO-central sources in Group(2) and Group(3).}
  \label{fig:seq_num}
\end{figure}

Based on the infrared (3.6$-$24 $\mu$m) environments, the 29 SiO central sources can be divided into three groups: (1) Fourteen sources are undetected in all infrared wavelengths, which suggests that they are at a relatively early evolutionary stage. Four sources are detected with radio emissions, while ten sources are not. In addition, eight of these sources have ammonia observations with an average T$_{\rm NH_{3}}$ of 22.3 K. (2) Five sources are detected at 24 $\mu$m but are dark or partly bright at other shorter wavelengths, indicating they are at a later evolutionary stage. Three of them are detected with radio emission. Also, two of these sources with ammonia observations have an average T$_{\rm NH_{3}}$ of 26.5 K. (3) The remaining ten sources are seen at 3.6$-$24 $\mu$m and are expected to be at the latest evolutionary stage. Most of them (8 out of 10) have been detected with radio emissions. Furthermore, five sources with T$_{\rm NH_{3}}$ have an average value of 28.9 K. Overall, although the ammonia kinetic temperature is not available for each source, the temperature increases as the star-forming region evolves. This is roughly consistent with the proposed picture of the evolutionary stages in Section \ref{rad}. The 24 $\mu$m luminosity is an indicator of the mass of the central sources; \citet{2019ApJS..241....1C} and \citet{2022ApJ...927..185W} classified MDCs in Cygnus-X based on the 24 $\mu$m luminosity. We show the distribution of the 24 $\mu$m luminosity at the condensation scale of $\sim$0.01 pc between the SiO-central sources in Group(2) and Group(3) in Figure \ref{fig:seq_num}. The Group(2) sources have a lower 24 $\mu$m luminosity median value ($\sim$0.03 Jy kpc$^{2}$) than that of the Group(3) sources ($\sim$0.2 Jy kpc$^{2}$), which indicates that the 24 $\mu$m luminosity, as well as the mass of the central condensation, appears to increase with the possible evolutionary stages.

%-----------------------------------------------------------------------
%-----------------------------------------------------------------------

The opening angle of CO outflows has been found to increase as the powering source evolves within high-mass star-forming regions (e.g., \citealt{2005ASSL..324..105B}; \citealt{2007prpl.conf..245A}; \citealt{2019ApJ...871..141Q}).  In our study, we find that among the 11 outflows powered by Group(3) sources, 8 exhibit a wide-angle morphology, as exemplified by the MDC341-MM1 outflows. Conversely, 9 out of the 15 outflows associated with Group(1) sources tend to display narrower opening angles, as seen in the MDC220-MM1 blueshifted outflow and the MDC1599-MM3 outflow. Consequently, similar to CO outflows, our SiO outflows also present an enlargement of their opening angles as the central sources grow.

%-----------------------------------------------------------------------

\section{Summary} \label{sec:sum}
As a part of the CENSUS project, we conducted a SiO (5$-$4) survey toward a sample of 48 MDCs in Cygnus-X with the SMA. Our results can be summarized as follows:

1. We detect SiO (5$-$4) emission in 16 out of 48 MDCs. 
We identify 14 bipolar and 18 unipolar SiO outflows with 29 central sources. 
Two bipolar SiO outflows share a common central source, two unipolar outflows have a joint central source, and there is no 1.37 continuum source associated with the bipolar outflow in MDC699.
In the Spitzer three-color maps, 11 bipolar and 13 unipolar outflows have shock-related excess 4.5 $\mu$m emission.

2. We find diffuse low-velocity ($\Delta{v}$ $\le$ 1.2 km s$^{-1}$) SiO (5$-$4) emission surrounding the continuum sources in MDC 220 and MDC 684, both of which have a different morphology from that of the high-velocity outflows. Further observations with higher velocity resolutions and higher sensitivities are needed to confirm whether this emission originates from decelerated outflow shocks or large-scale shocks induced by global cloud collapse.

3. With the assumption of the optically thin SiO (5$-$4) line in LTE and the SiO abundance (1 $\times$ 10$^{-8}$), we estimate the SiO column densities and different outflow parameters.
 
4. For all blueshifted and redshifted lobes of two SiO (5$-$4) bipolar outflows centered at MDC310-MM1 and MDC684-MM1, the outflow mass--velocity relation can be fitted with a broken power law with the indices steepening at higher velocities.

5. Radio continuum emission is found associated with 16 outflow-powering sources. The radio emission associated with MDC1112-MM2 and MM3 bipolar outflows is generated from H {\scriptsize II} region(s); the other areas of emission are likely to be ionized jets or winds. Based on the infrared environments (Spitzer 3.6, 5.8, 8.0, and 24 $\mu$m), we divide the 29 SiO central sources into three possible evolutionary stages. The average T$_{\rm NH_{3}}$ and 24 $\mu$m luminosity levels appear to increase with advancing evolutionary phase.

\begin{acknowledgements}
This work is supported by National Key R\&D Program of China No. 2022YFA1603100, No. 2017YFA0402604, and the National Natural Science Foundation of China (NSFC) grants U1731237 and 11590781. K.Q. acknowledges the science research grant from the China Manned Space Project. K.Y. acknowledges the supports from China Postdoctoral Science Foundation No. 2021M701669.
\end{acknowledgements}

%\bibliographystyle{aa} % style aa.bst
%\bibliography{bibliography} % your references

\begin{appendix}

%\onecolumn

\section{SiO (5$-$4) channel maps} \label{channel_map}

\begin{figure*}[!htb]
\centering
   \includegraphics[width=500pt]{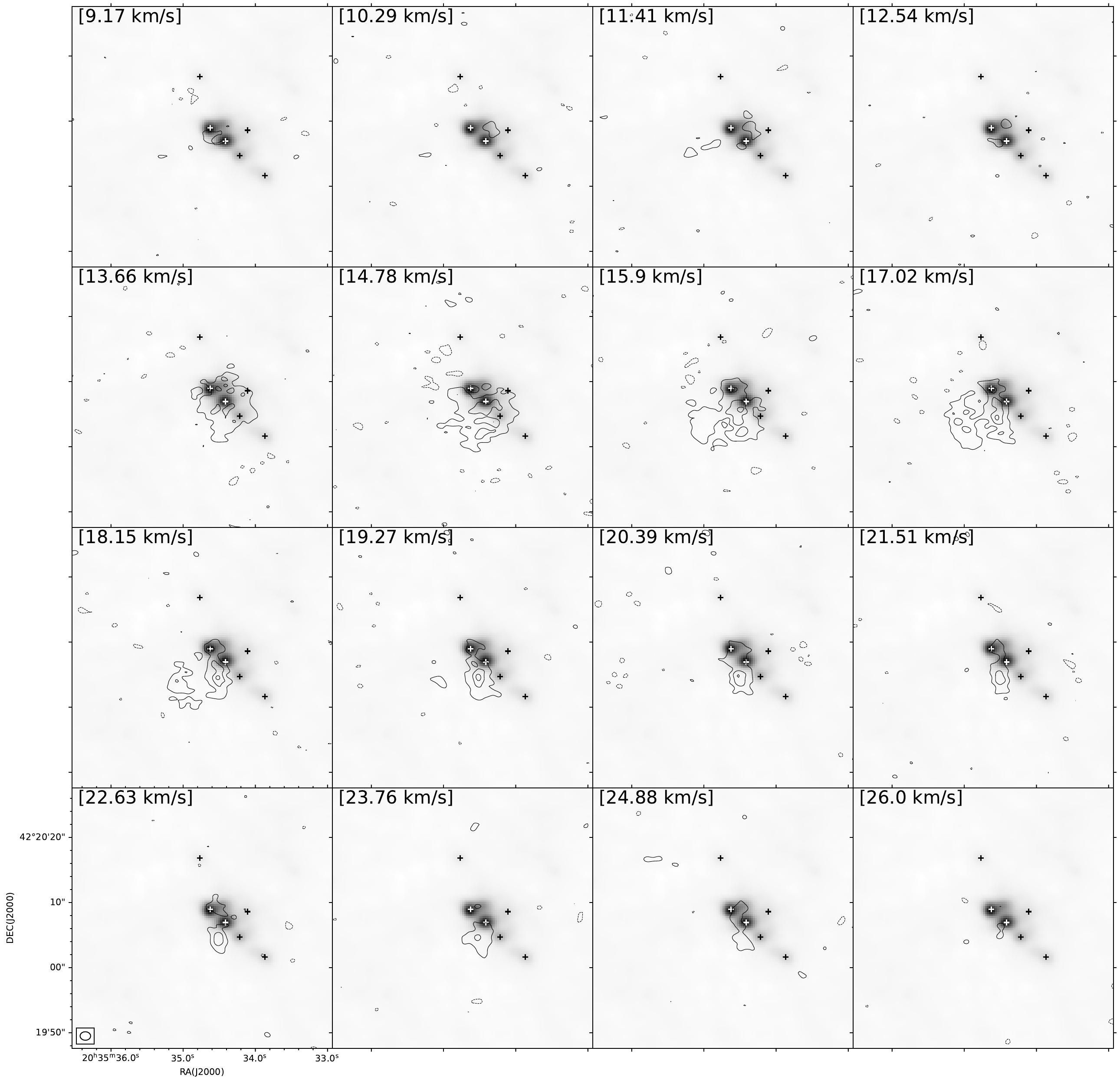}
     \caption{Channel maps of the SiO (5$-$4) emission overlaid on the 1.37 mm continuum emission for MDC 220. Solid contours represent emission starting at and continuing in steps of $\pm$ 3$\sigma$, where $\sigma$ = 0.055 Jy beam$^{-1}$ km s$^{-1}$. The black and white crosses represent dust condensations identified by \citet{2021ApJ...918L...4C}. The velocity of each panel is given in the upper left corner. The synthesized beam is shown in the bottom-left corner.}
  \label{fig:channel_map_N03}
\end{figure*}

\begin{figure*}
\centering
   \includegraphics[width=430pt]{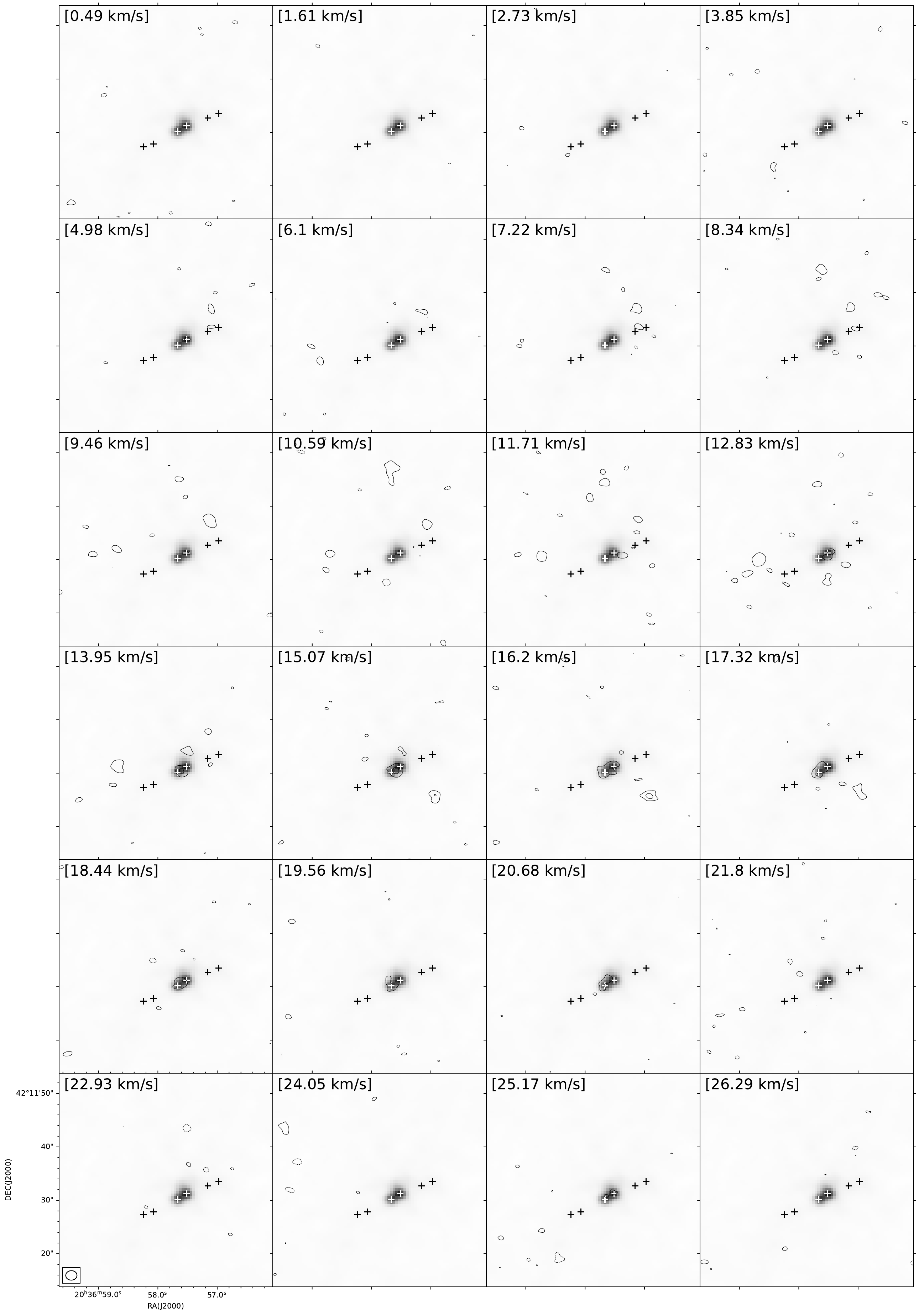}
     \caption{Same convention as Figure \ref{fig:channel_map_N03} but for MDC 248, with $\sigma$ = 0.065 Jy beam$^{-1}$ km s$^{-1}$.}
  \label{fig:channel_map_N12}
\end{figure*}

\begin{figure*}
\centering
   \includegraphics[width=430pt]{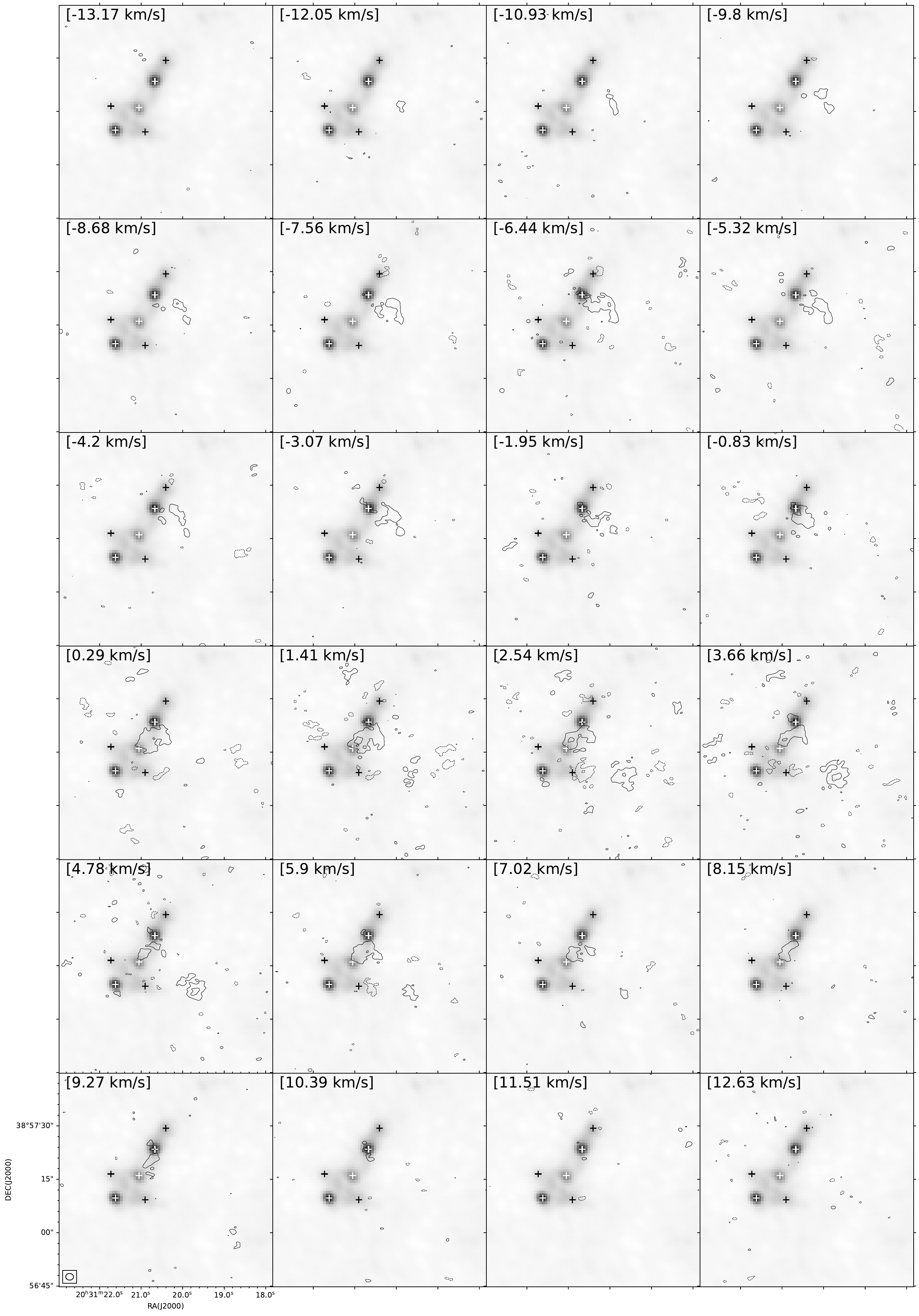}
     \caption{Same convention as Figure \ref{fig:channel_map_N03} but for MDC 351, with $\sigma$ = 0.080 Jy beam$^{-1}$ km s$^{-1}$.}
  \label{fig:channel_map_S32}
\end{figure*}

\begin{figure*}
\centering
   \includegraphics[width=500pt]{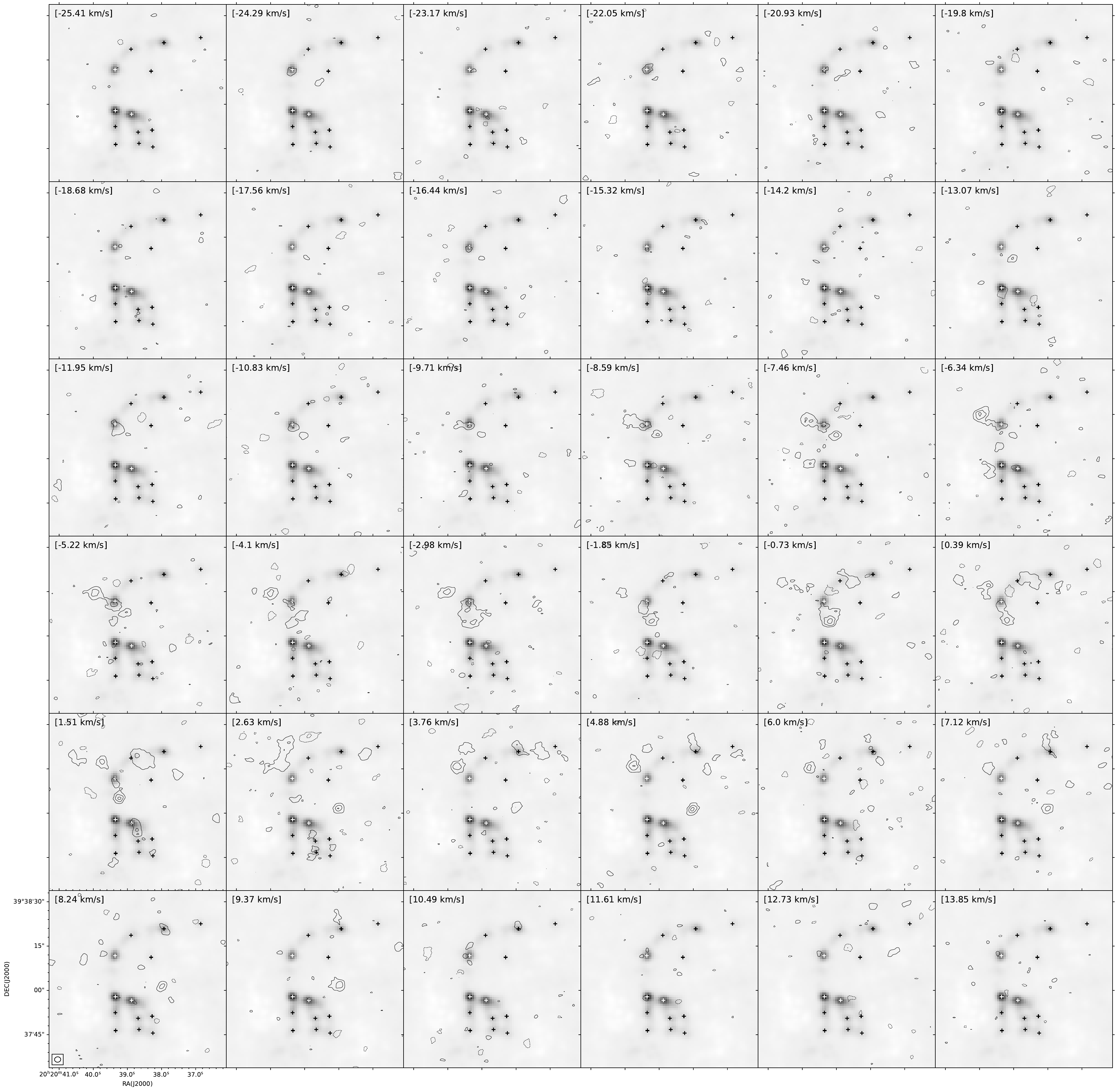}
     \caption{Same convention as Figure \ref{fig:channel_map_N03} but for MDC 507/753, with $\sigma$ = 0.055 Jy beam$^{-1}$ km s$^{-1}$.}
  \label{fig:channel_map_S07}
\end{figure*}

\begin{figure*}
\centering
   \includegraphics[width=500pt]{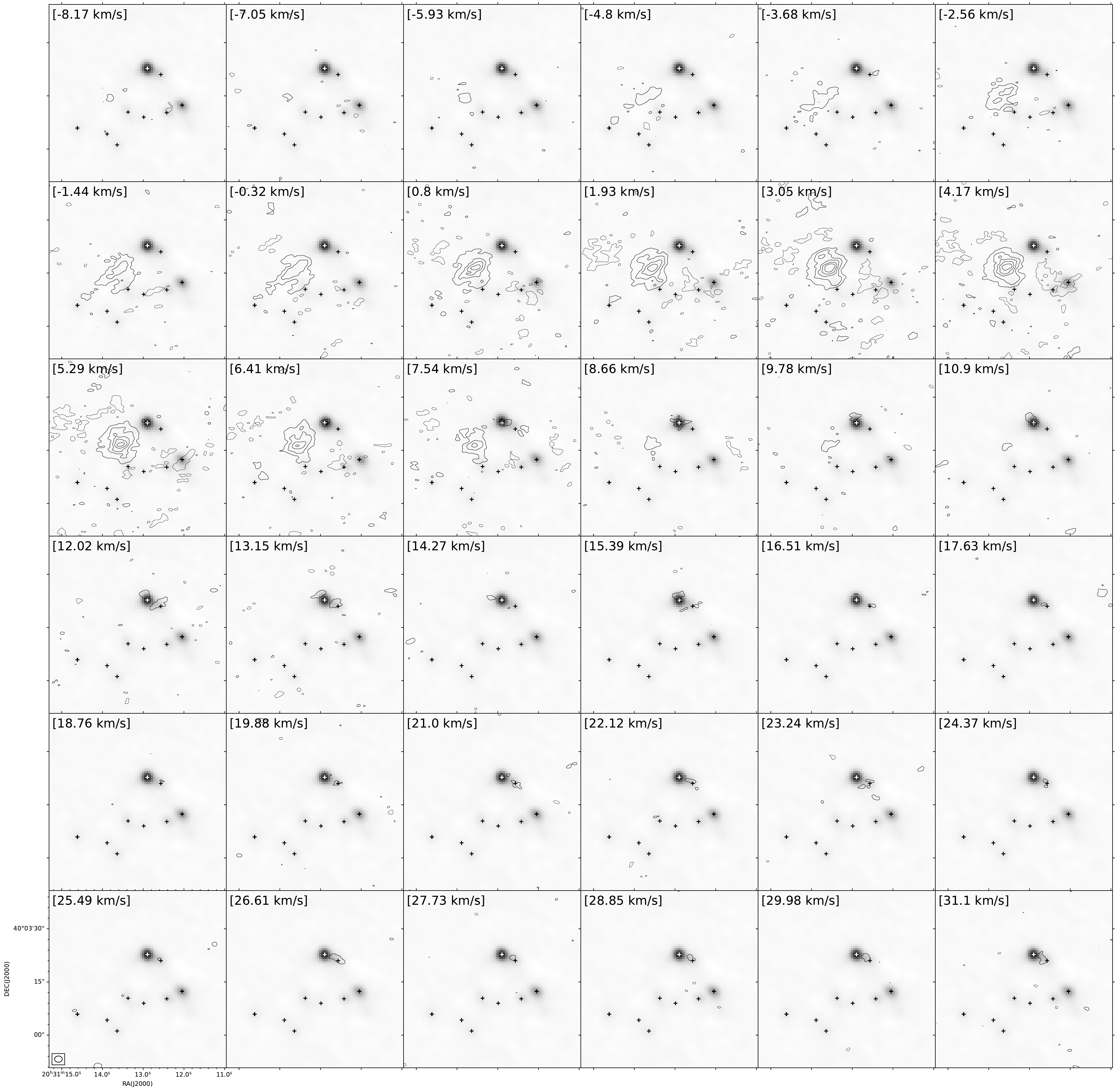}
     \caption{Same convention as Figure \ref{fig:channel_map_N03} but for MDC 509, with $\sigma$ = 0.095 Jy beam$^{-1}$ km s$^{-1}$.}
  \label{fig:channel_map_S30}
\end{figure*}

\begin{figure*}
\centering
   \includegraphics[width=350pt]{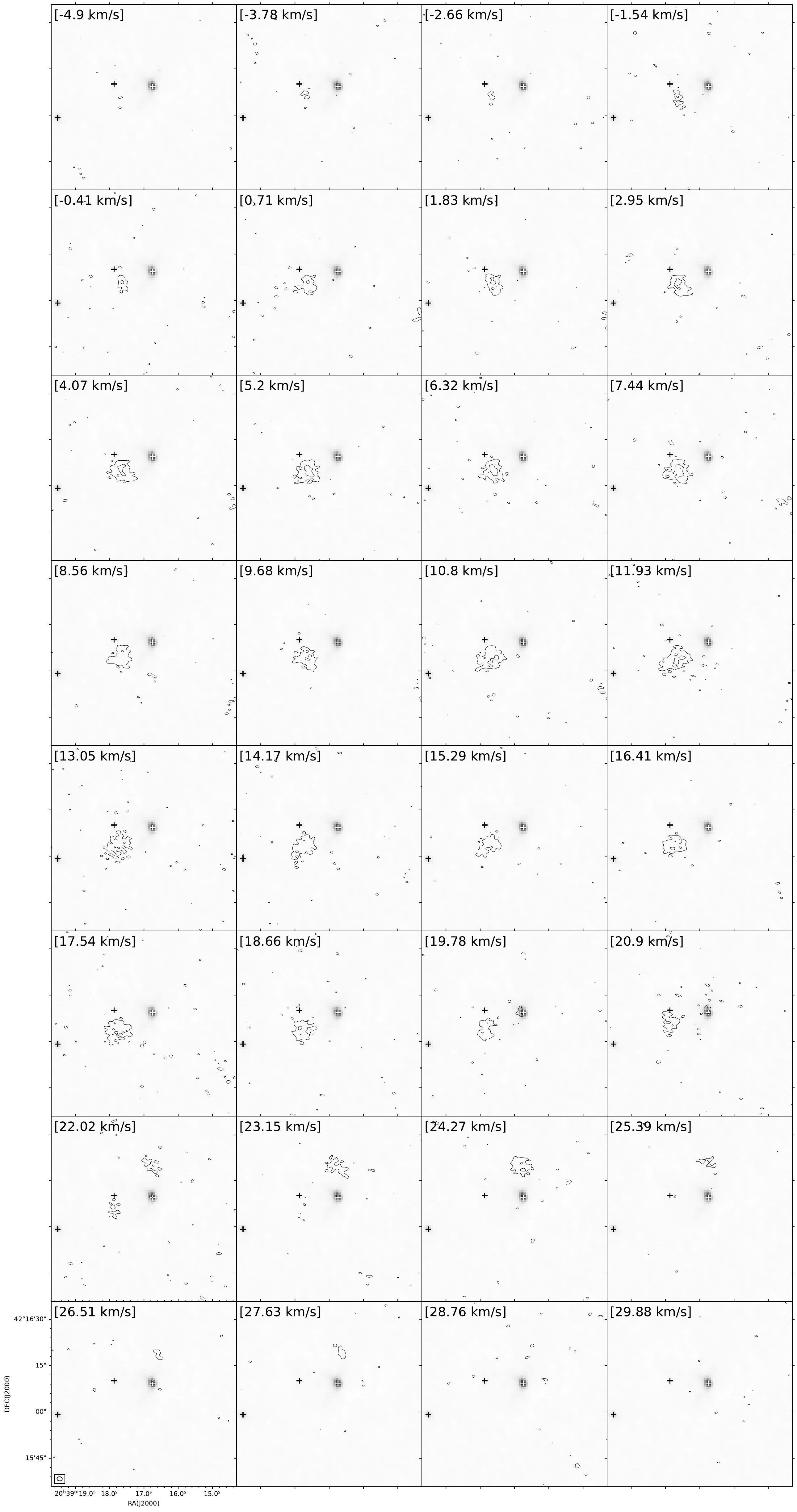}
     \caption{Same convention as Figure \ref{fig:channel_map_N03} but for MDC 698, with $\sigma$ = 0.060 Jy beam$^{-1}$ km s$^{-1}$.}
  \label{fig:channel_map_N56}
\end{figure*}

\begin{figure*}
\centering
   \includegraphics[width=450pt]{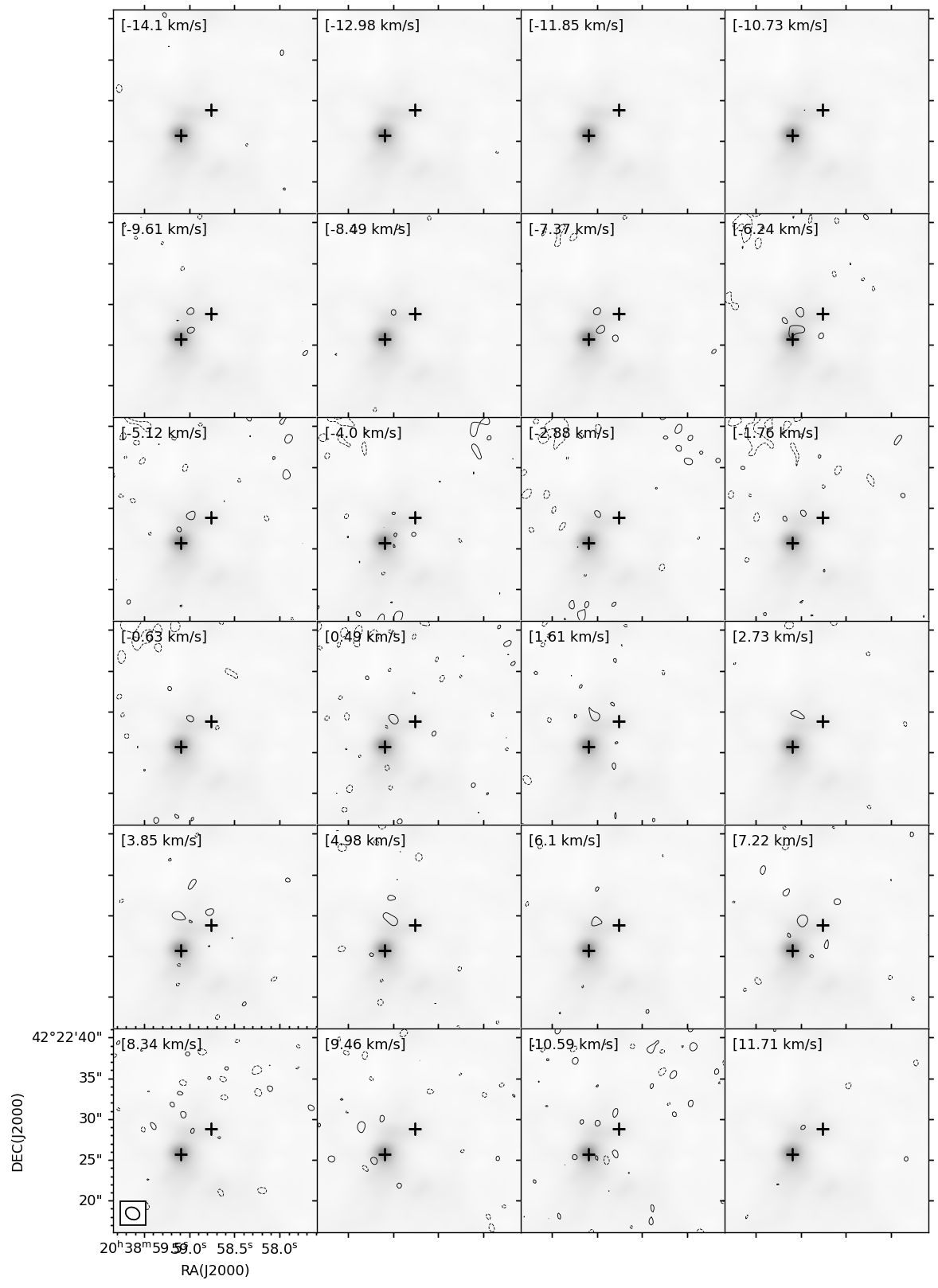}
     \caption{Same convention as Figure \ref{fig:channel_map_N03} but for MDC 699, with $\sigma$ = 0.140 Jy beam$^{-1}$ km s$^{-1}$.}
  \label{fig:channel_map_N38}
\end{figure*}

\begin{figure*}
\centering
   \includegraphics[width=500pt]{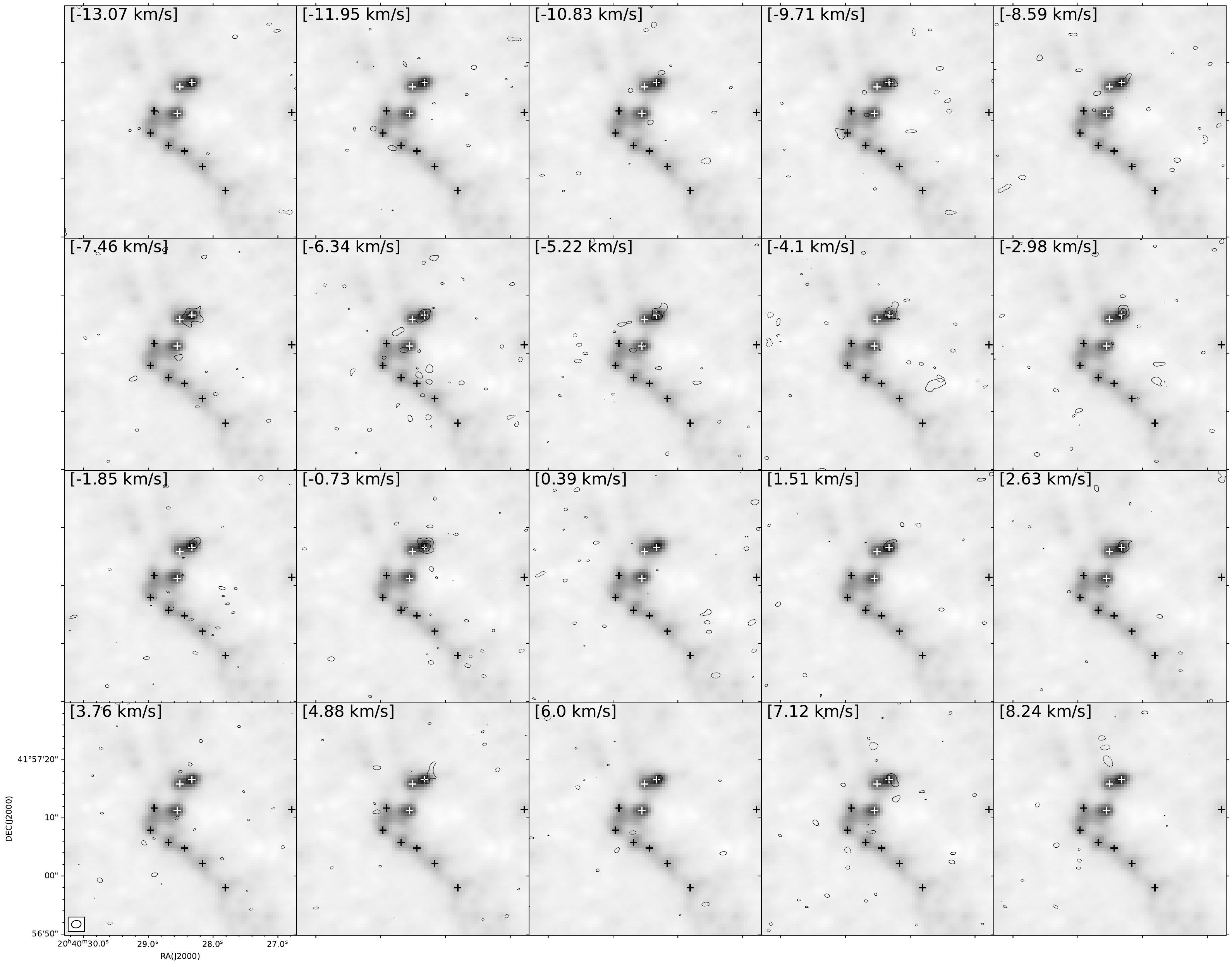}
     \caption{Same convention as Figure \ref{fig:channel_map_N03} but for MDC 801, with $\sigma$ = 0.055 Jy beam$^{-1}$ km s$^{-1}$.}
  \label{fig:channel_map_N65}
\end{figure*}

\begin{figure*}
\centering
   \includegraphics[width=450pt]{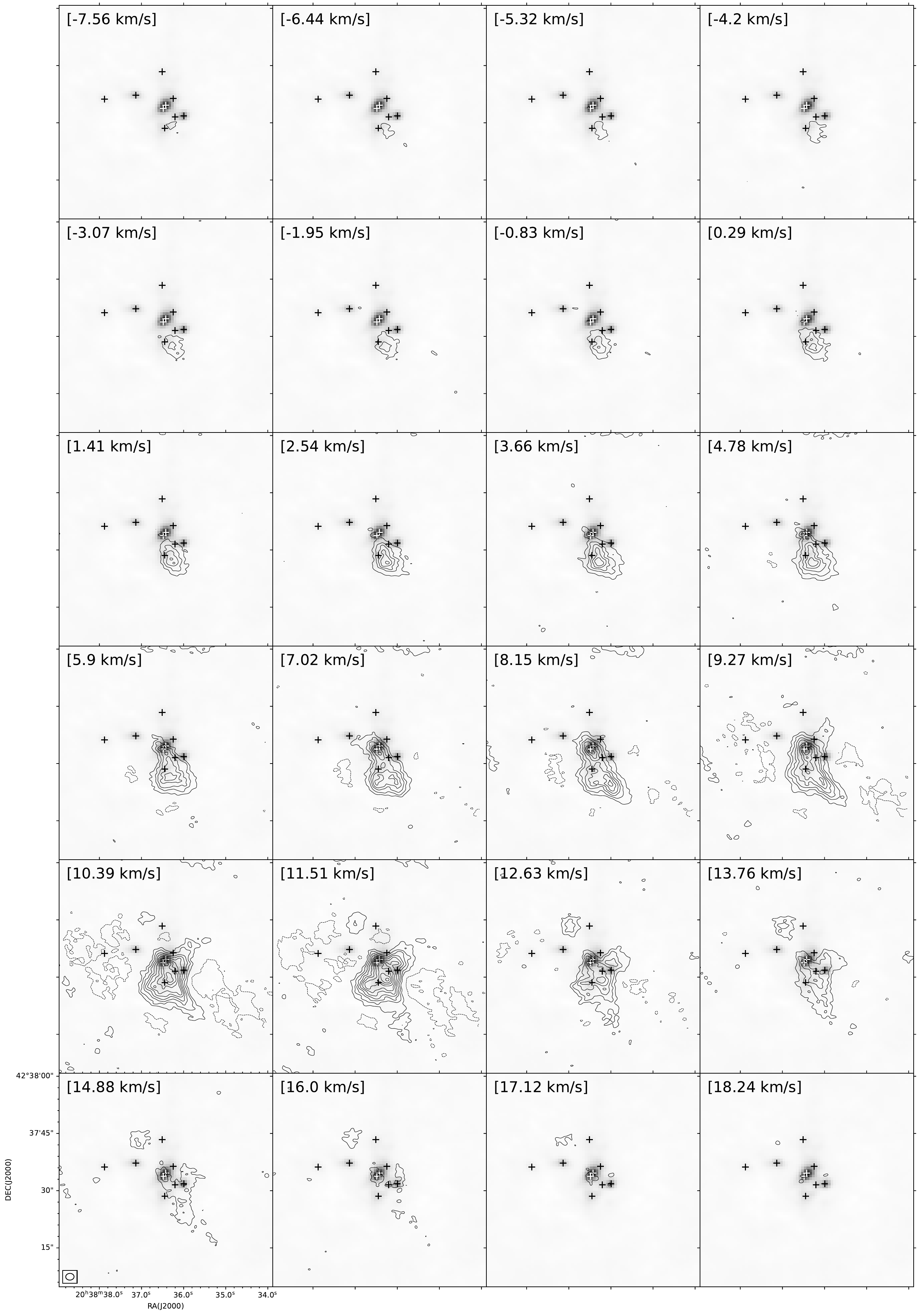}
     \caption{Same convention as Figure \ref{fig:channel_map_N03} but for MDC 1112, with $\sigma$ = 0.070 Jy beam$^{-1}$ km s$^{-1}$.}
  \label{fig:channel_map_N30}
\end{figure*}

\begin{figure*}
\centering
   \includegraphics[width=500pt]{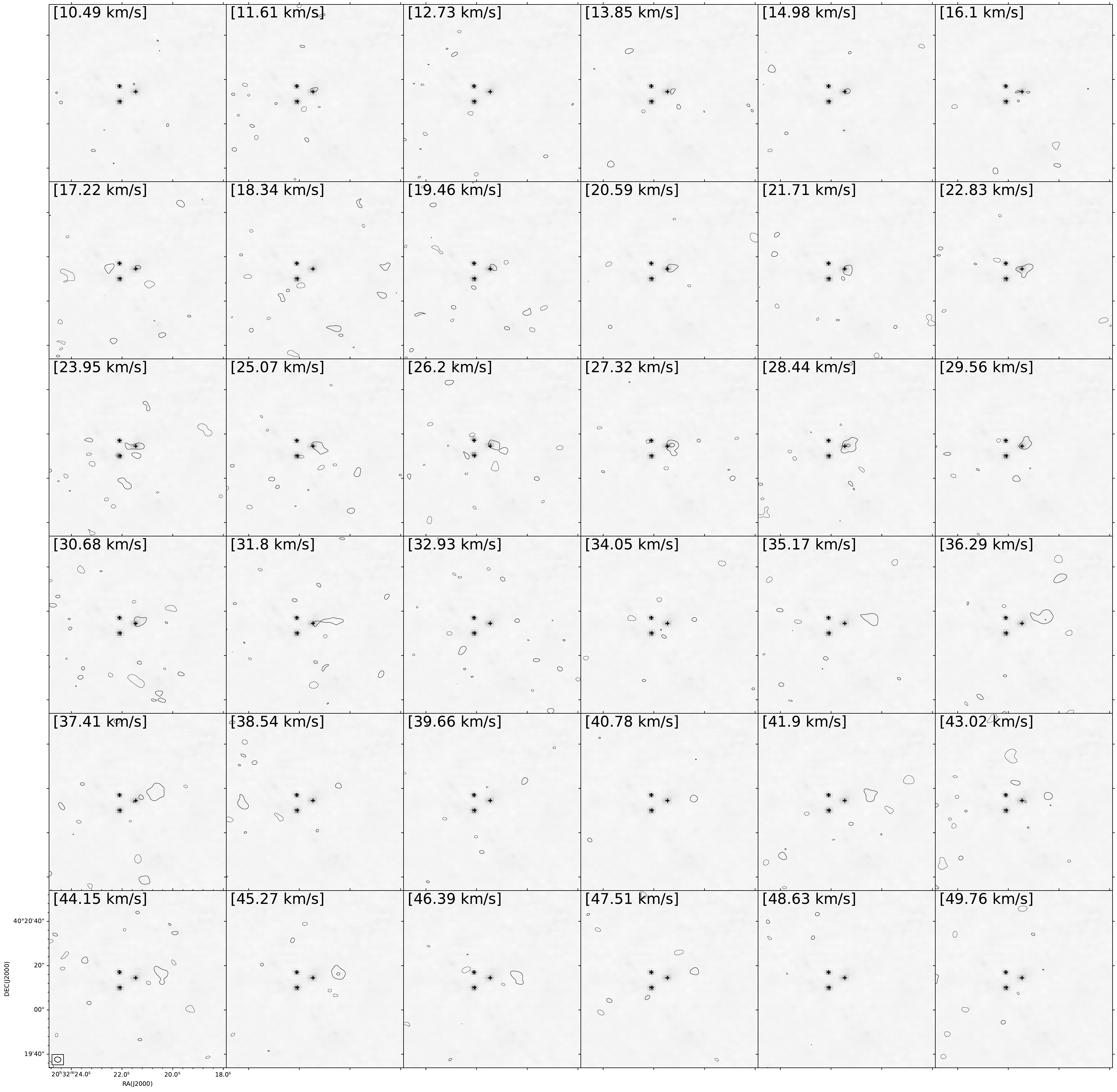}
     \caption{Same convention as Figure \ref{fig:channel_map_N03} but for MDC 1454, with $\sigma$ = 0.100 Jy beam$^{-1}$ km s$^{-1}$.}
  \label{fig:channel_map_DR15}
\end{figure*}

\begin{figure*}
\centering
   \includegraphics[width=500pt]{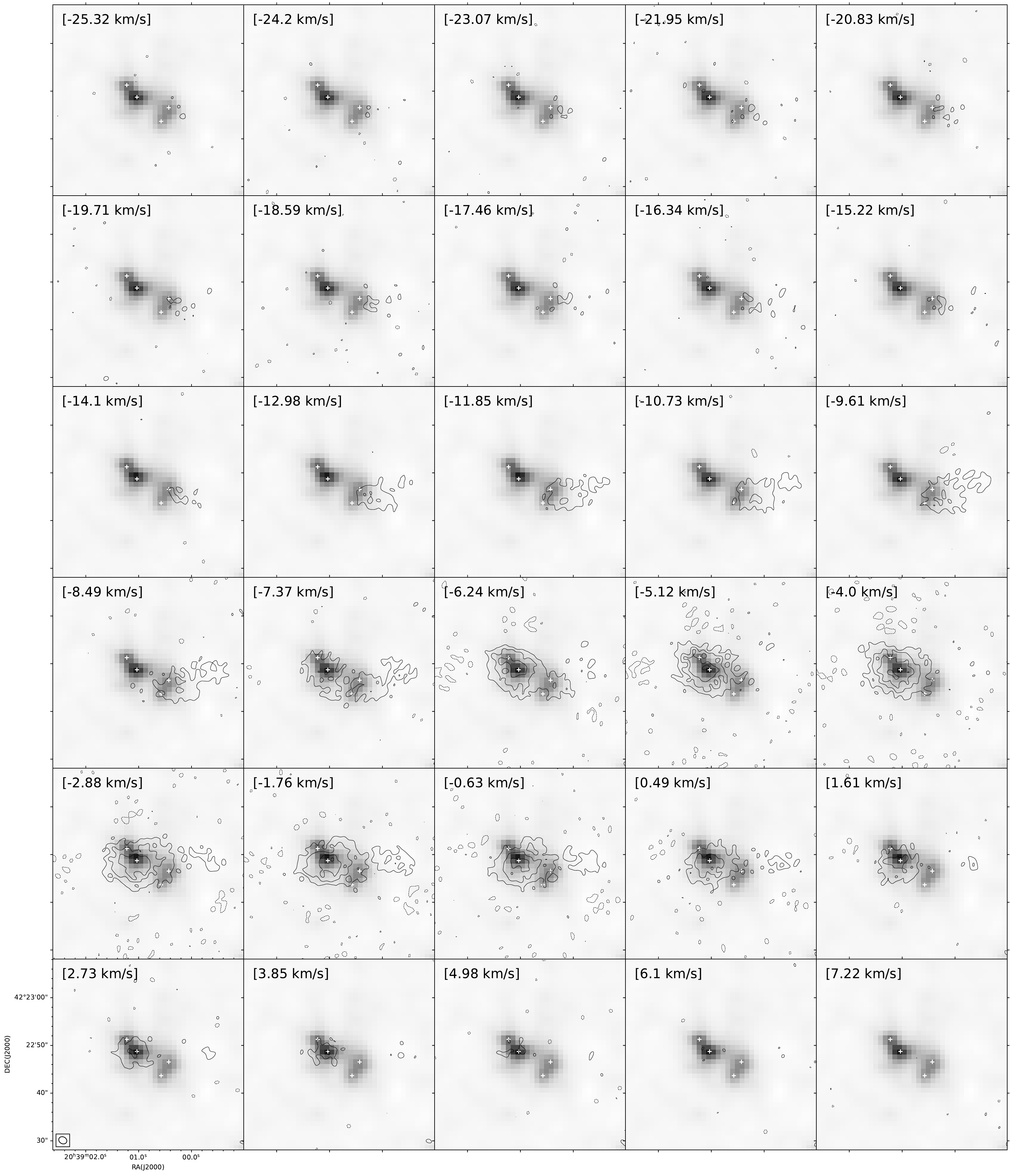}
     \caption{Same convention as Figure \ref{fig:channel_map_N03} but for MDC 1467, with $\sigma$ = 0.160 Jy beam$^{-1}$ km s$^{-1}$.}
  \label{fig:channel_map_N44}
\end{figure*}

\begin{figure*}
\centering
   \includegraphics[width=500pt]{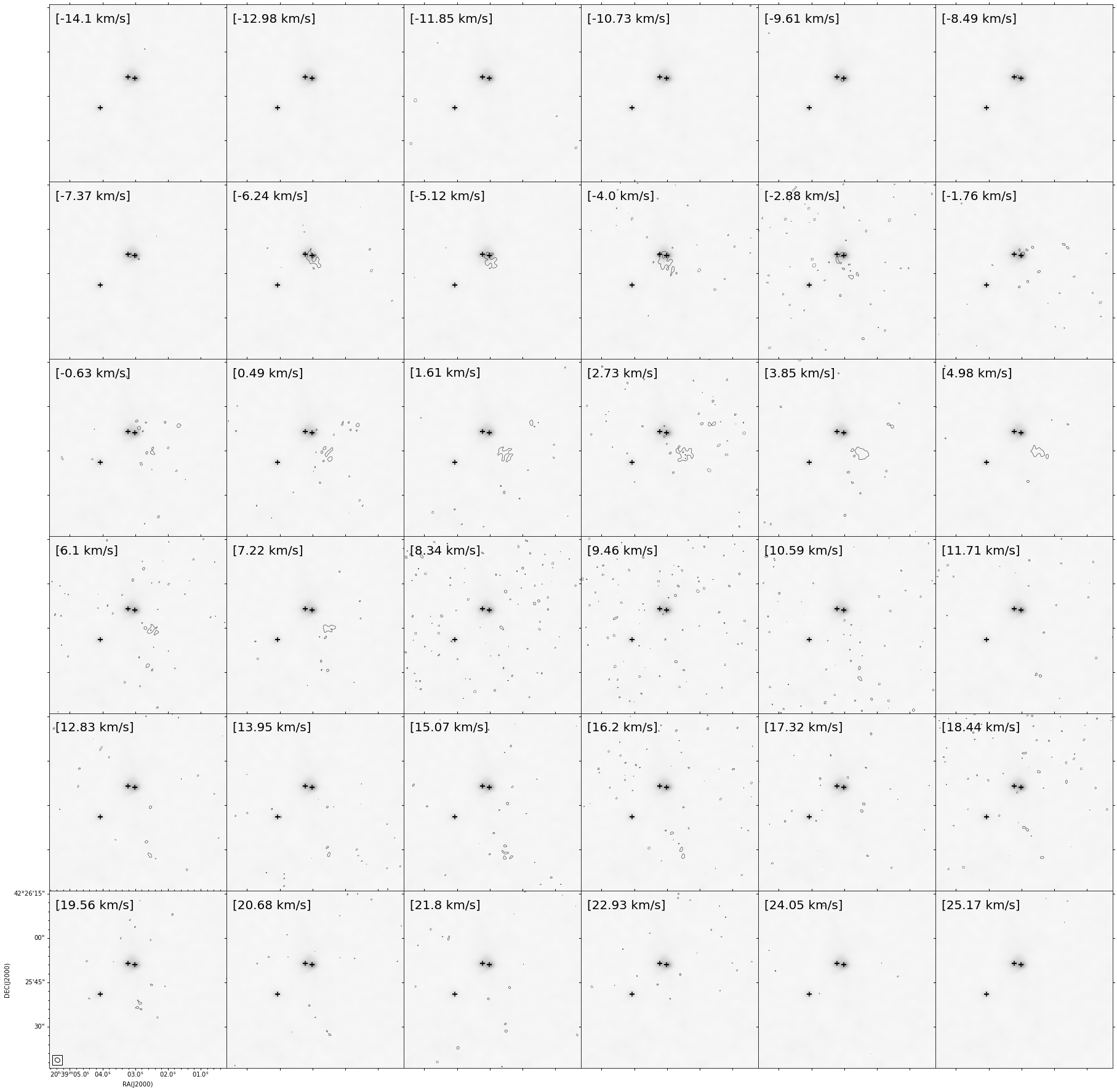}
     \caption{Same convention as Figure \ref{fig:channel_map_N03} but for MDC 1599, with $\sigma$ = 0.160 Jy beam$^{-1}$ km s$^{-1}$.}
  \label{fig:channel_map_N53}
\end{figure*}

In this section, we present channel maps of all the detected MDCs with SMA observations in Figure \ref{fig:channel_map_N03}$-$\ref{fig:channel_map_N53}.

%-----------------------------------------------------------------------

\clearpage

\begin{figure*}
\begin{center}
\begin{tabular}{l}
   \includegraphics[width=500pt]{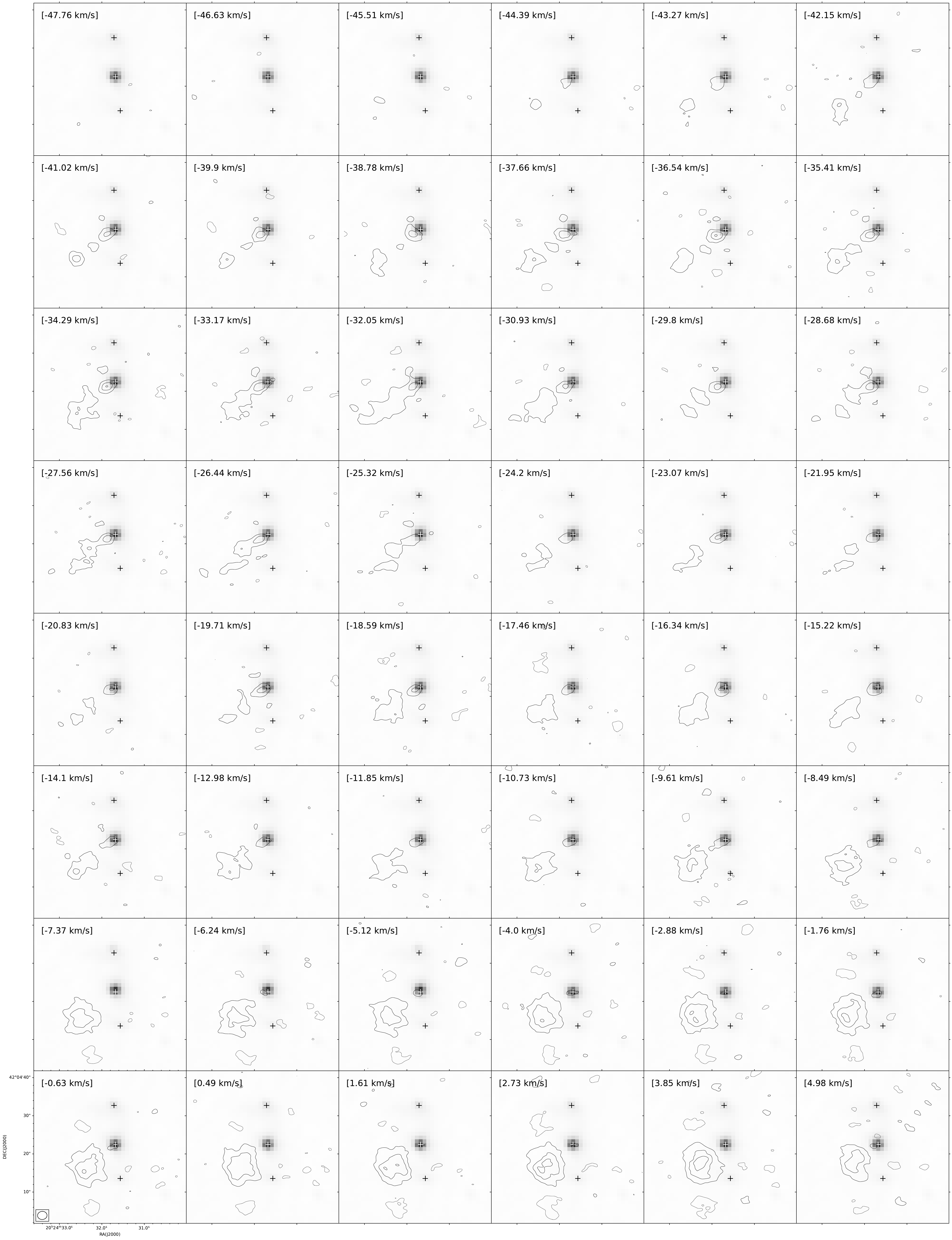}
\end{tabular}
\end{center}
\caption{Same convention as Figure \ref{fig:channel_map_N03} but for MDC 310, with $\sigma$ = 0.040 Jy beam$^{-1}$ km s$^{-1}$.}
\label{fig:channel_map_NW14}
\end{figure*}
\clearpage

\begin{figure*}
\addtocounter{figure}{-1}
\begin{center}
\begin{tabular}{l}
   \includegraphics[width=500pt]{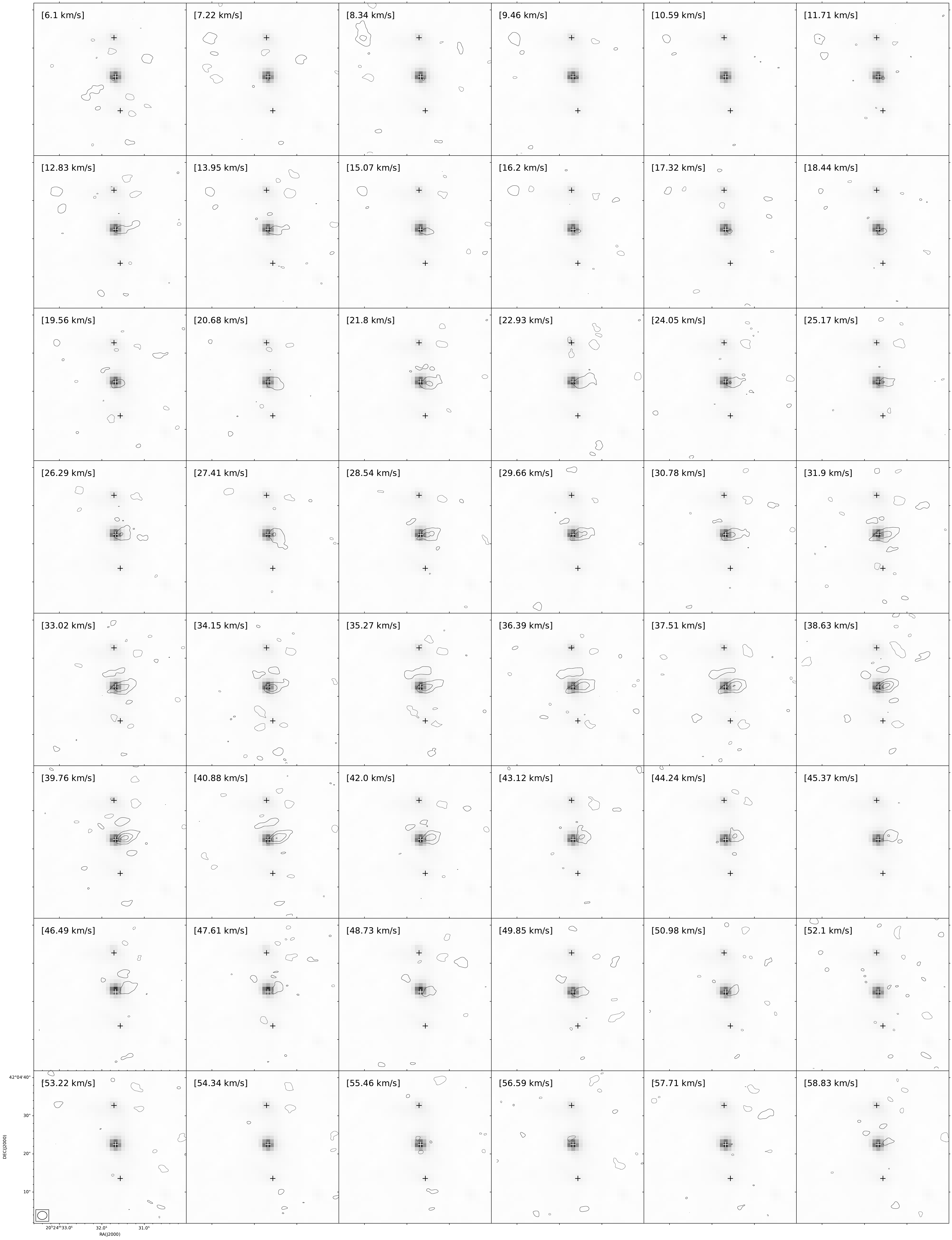}
\end{tabular}
\end{center}
\caption{(Continued.)}
\end{figure*}
\clearpage

\begin{figure*}
\begin{center}
\begin{tabular}{l}
   \includegraphics[width=500pt]{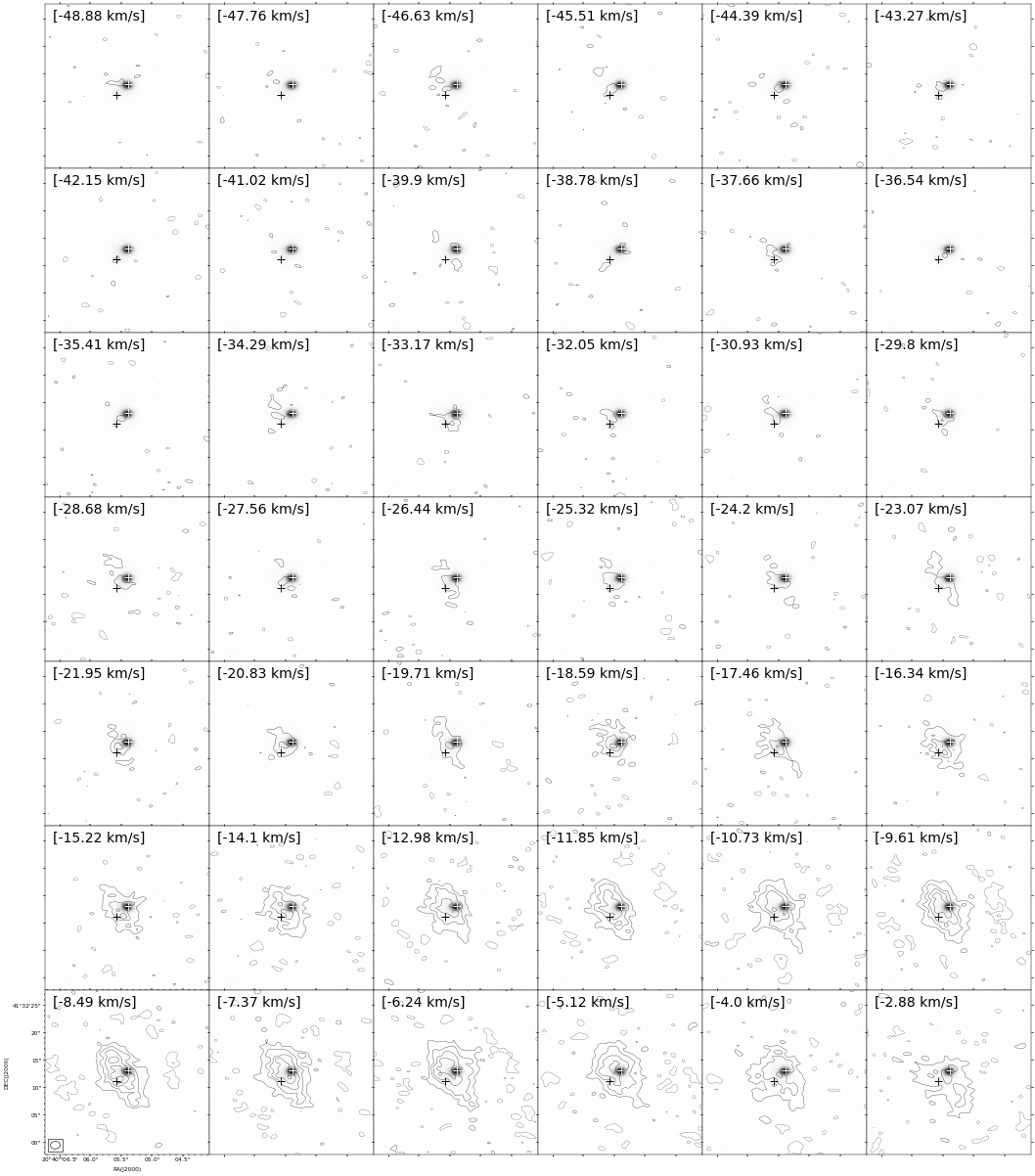}
\end{tabular}
\end{center}
\caption{Same convention as Figure \ref{fig:channel_map_N03} but for MDC 341, with $\sigma$ = 0.050 Jy beam$^{-1}$ km s$^{-1}$.}
\label{fig:channel_map_N63}
\end{figure*}
\clearpage

\begin{figure*}
\addtocounter{figure}{-1}
\begin{center}
\begin{tabular}{l}
   \includegraphics[width=500pt]{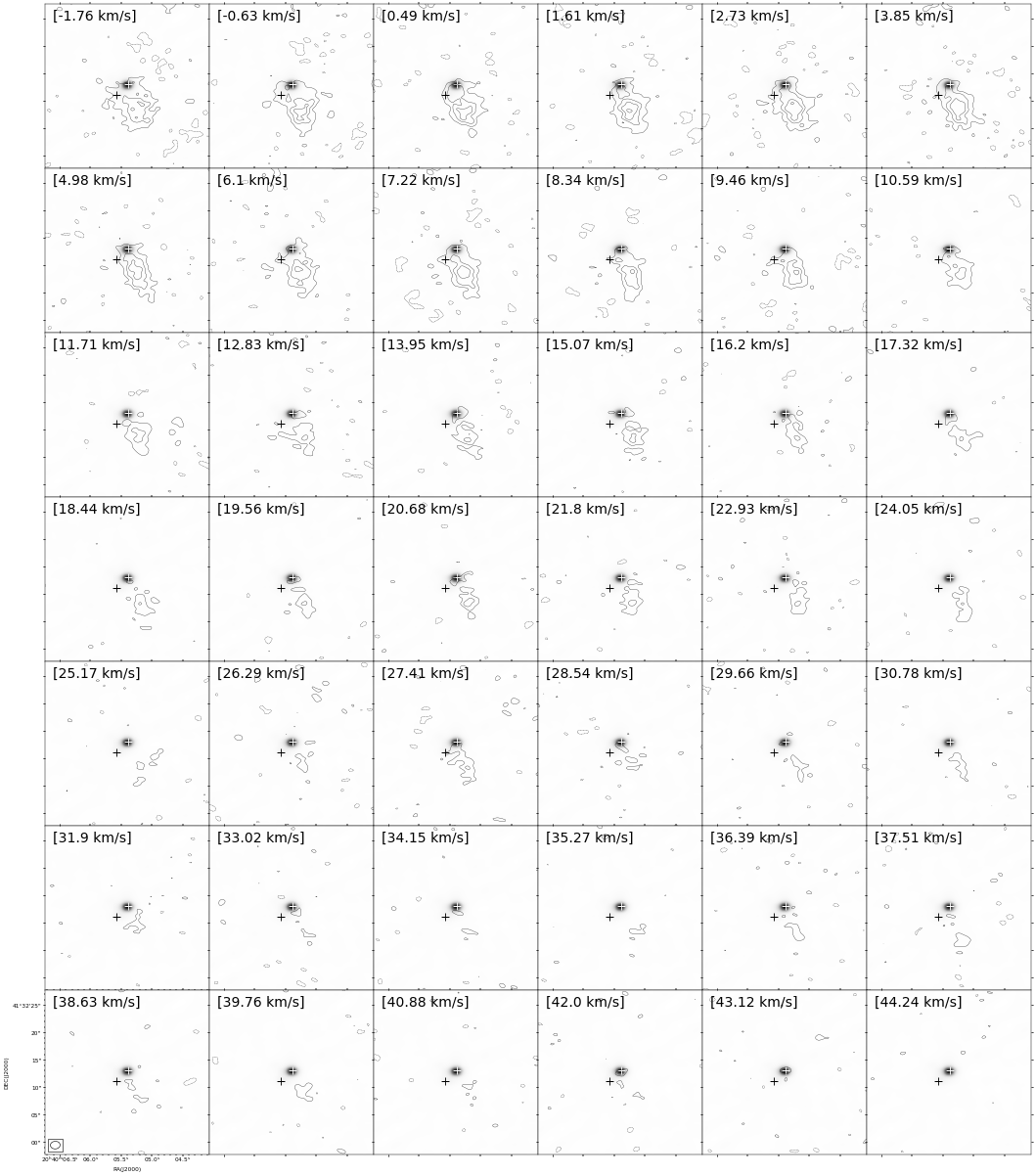}
\end{tabular}
\end{center}
\caption{(Continued.)}
\end{figure*}
\clearpage

\clearpage

\clearpage

\clearpage

\begin{figure*}
\begin{center}
\begin{tabular}{l}
   \includegraphics[width=500pt]{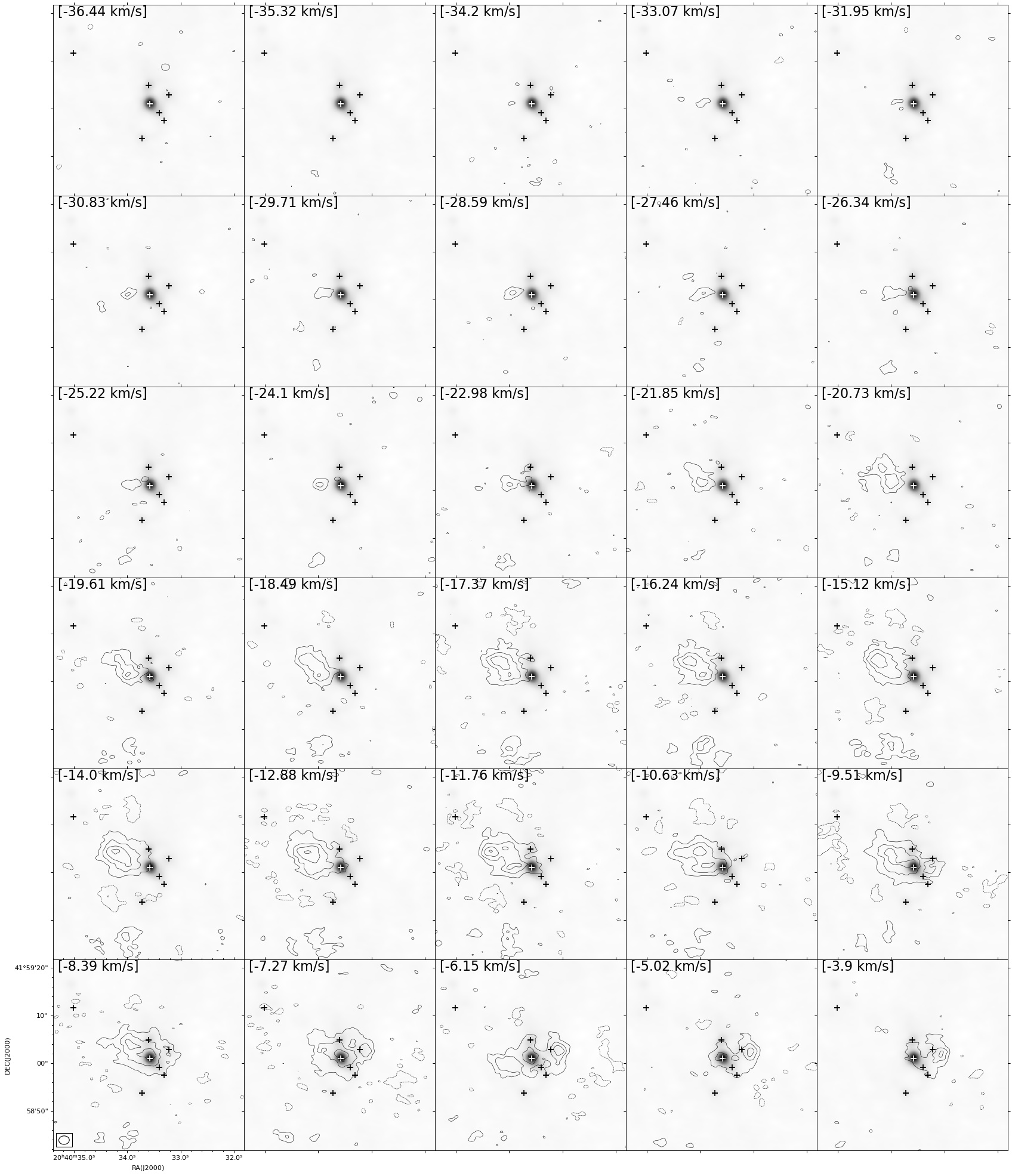}
\end{tabular}
\end{center}
\caption{Same convention as Figure \ref{fig:channel_map_N03} but for MDC 684, with $\sigma$ = 0.055 Jy beam$^{-1}$ km s$^{-1}$.}
\label{fig:channel_map_N68}
\end{figure*}
\clearpage

\begin{figure*}
\addtocounter{figure}{-1}
\begin{center}
\begin{tabular}{l}
   \includegraphics[width=500pt]{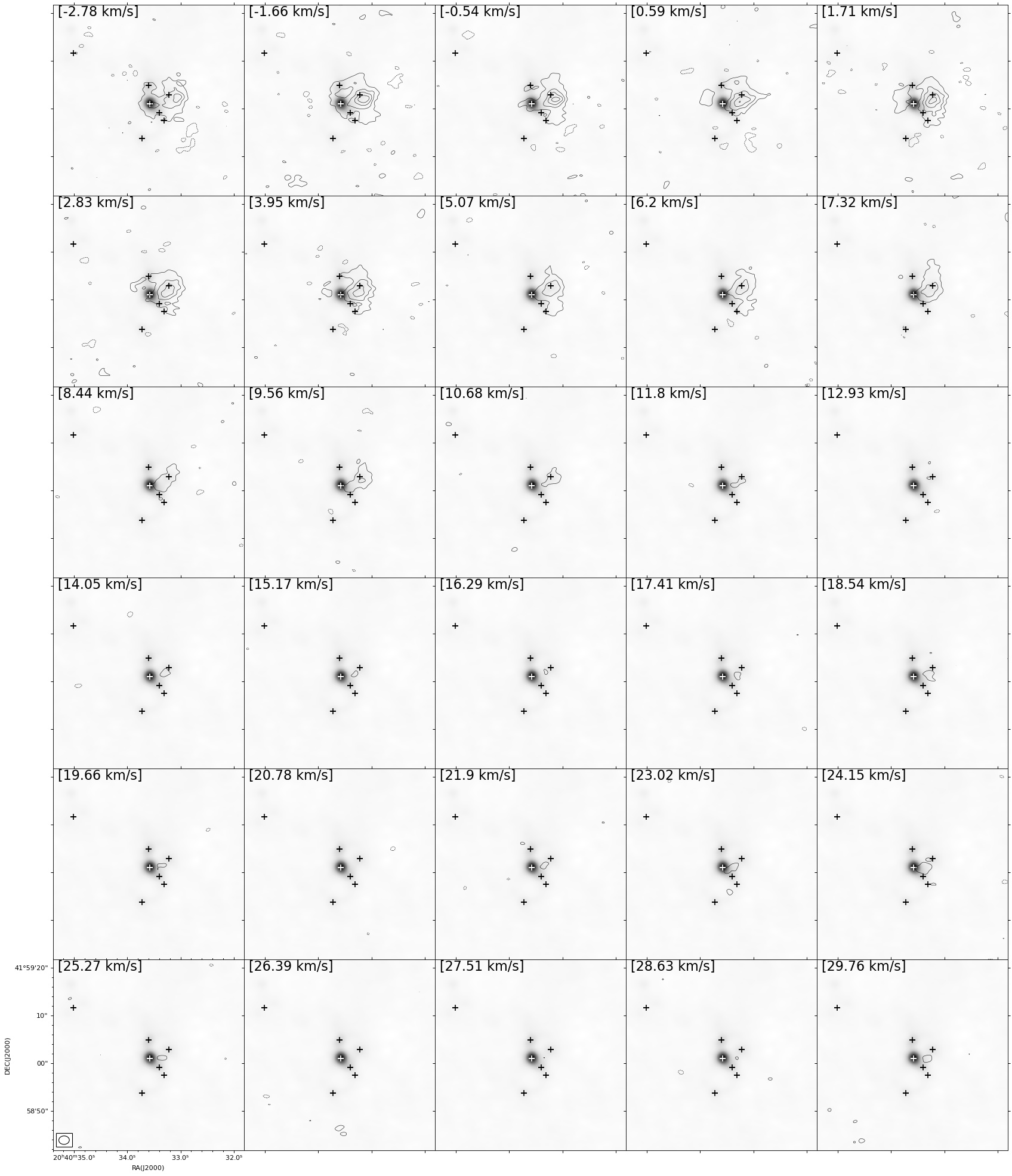}
\end{tabular}
\end{center}
\caption{(Continued.)}
\end{figure*}
\clearpage

\clearpage

\clearpage

\clearpage

\clearpage

\clearpage

\clearpage

%-----------------------------------------------------------------------

\clearpage

\section{Infrared environments} \label{Infrared}

In this section, we present the infrared images for each MDC at wavelengths of 3.6, 4.5, 5.8, 8.0, and 24 $\mu$m.

%-----------------------------------------------------------------------
\begin{figure*}[!htb]
\begin{center}
\begin{tabular}{l}
   \includegraphics[width=410pt]{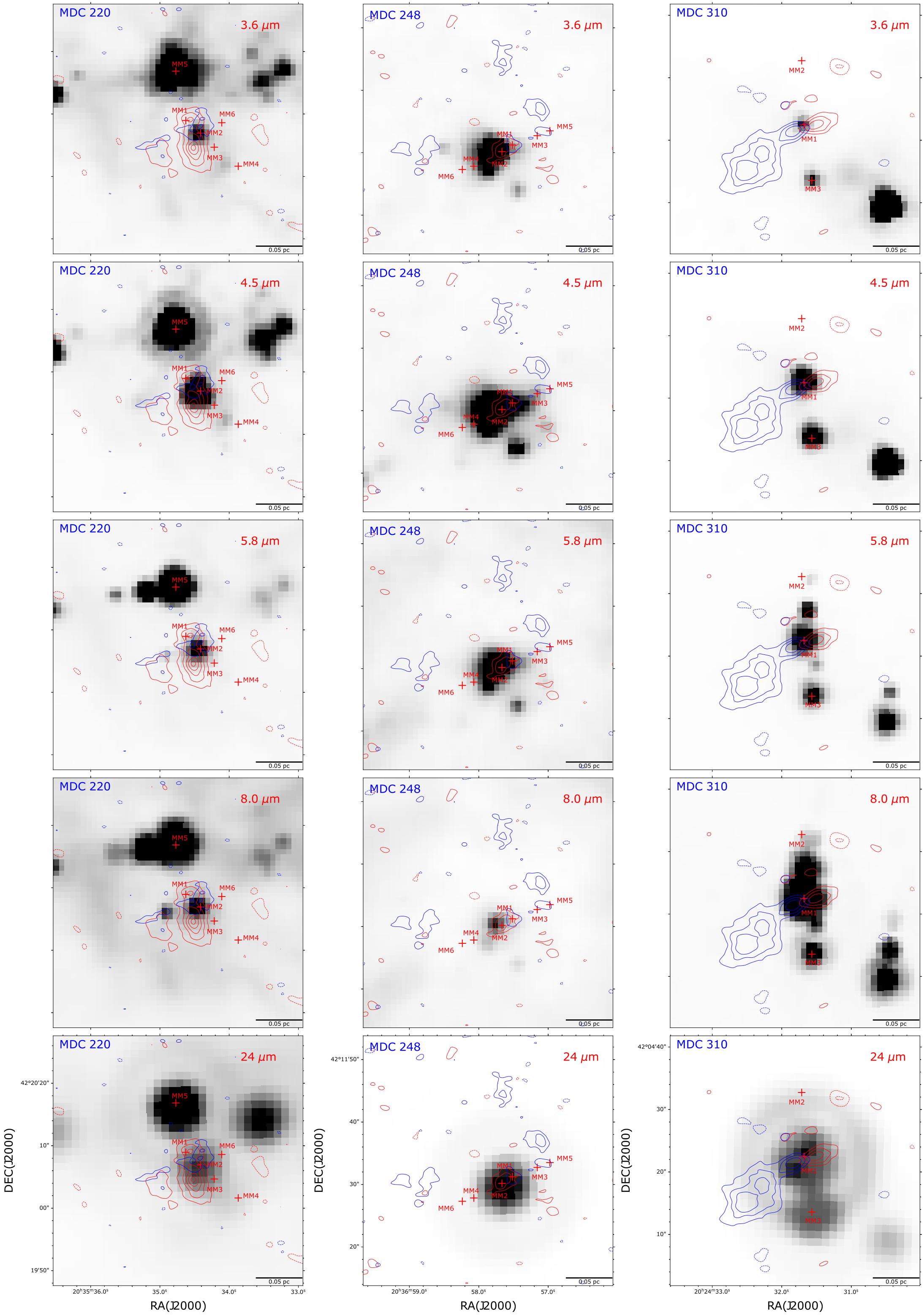}
\end{tabular}
\end{center}
\caption{Infrared images of the SiO (5$-$4)-detected MDCs. From top to bottom, we present the images of Spitzer 3.6/4.5/5.8/8.0/24 $\mu$m, and from left to right the images are in the region of different MDCs. The red and blue contours are the redshifted and blueshifted SiO (5$-$4) components with the same levels in Figure \ref{N03} $-$ \ref{N53}. The red crosses are the positions of identified 1.37 mm continuum sources. The wavelength is marked in the top-right corner. The scale bar is presented in the bottom-right corner.}
\label{fig:herschel}
\end{figure*}
\clearpage

\begin{figure*}
\addtocounter{figure}{-1}
\begin{center}
\begin{tabular}{l}
   \includegraphics[width=450pt]{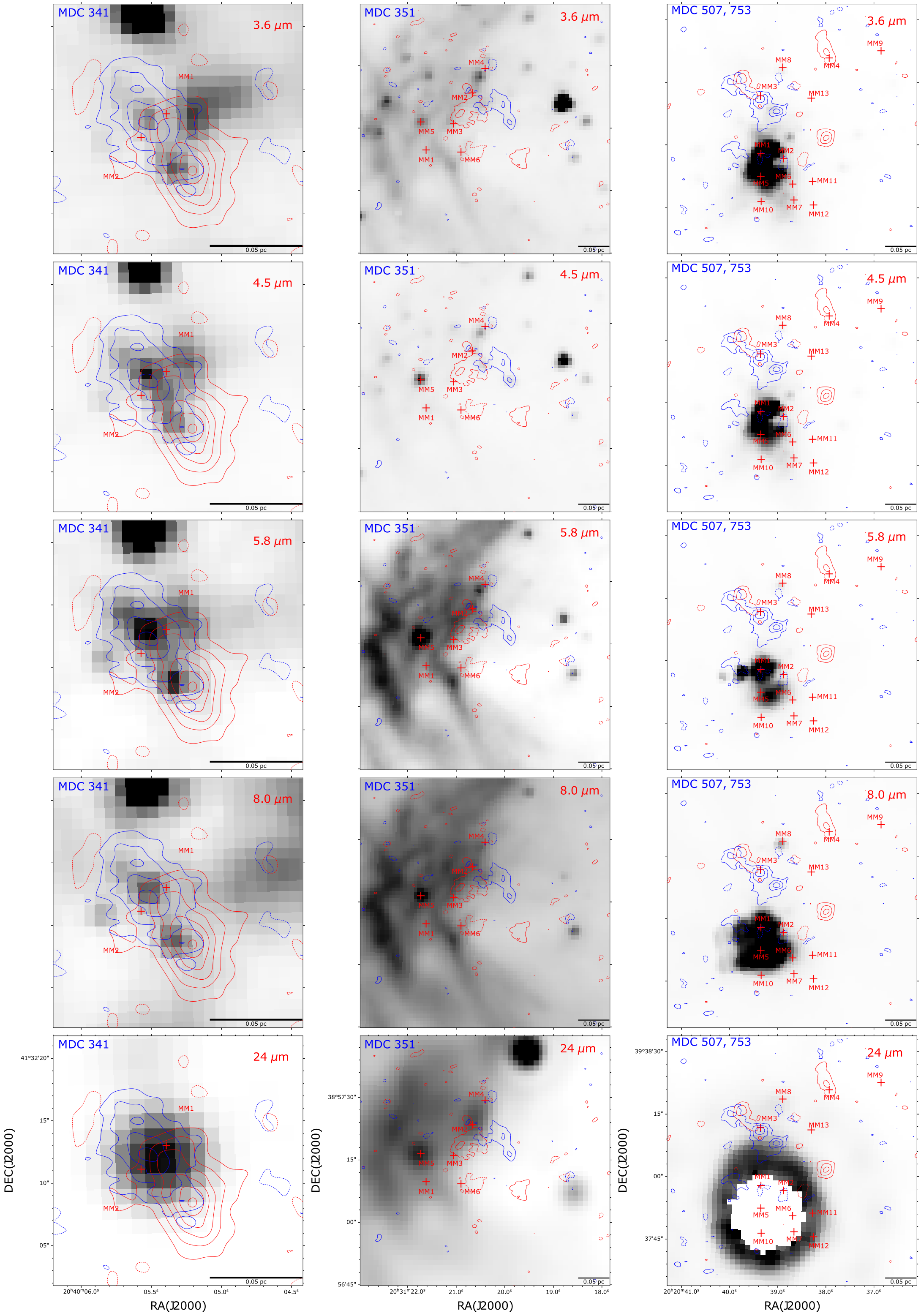}
\end{tabular}
\end{center}
\caption{(Continued.)}
\end{figure*}
\clearpage

\begin{figure*}
\addtocounter{figure}{-1}
\begin{center}
\begin{tabular}{l}
   \includegraphics[width=450pt]{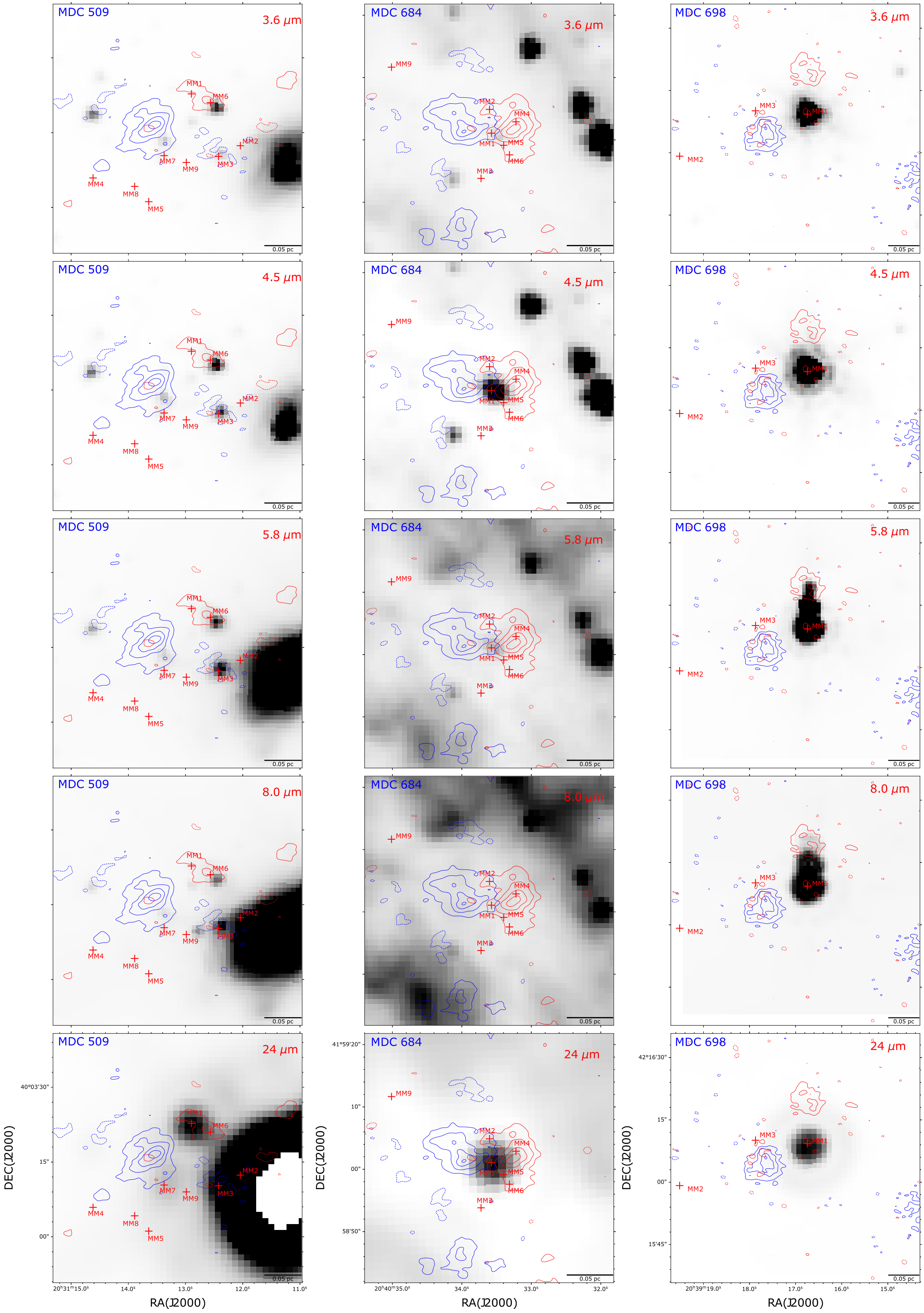}
\end{tabular}
\end{center}
\caption{(Continued.)}
\end{figure*}
\clearpage

\begin{figure*}
\addtocounter{figure}{-1}
\begin{center}
\begin{tabular}{l}
   \includegraphics[width=450pt]{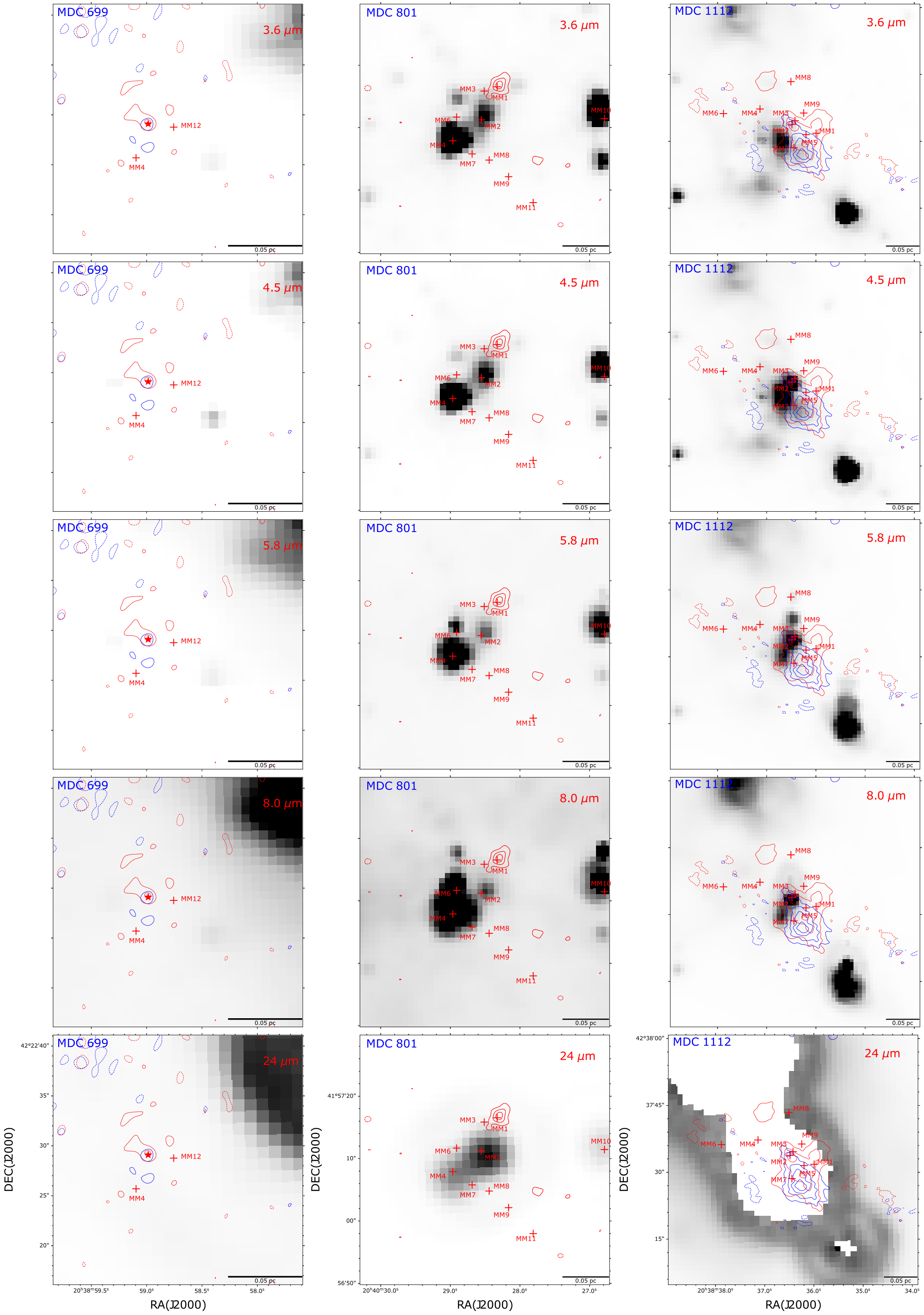}
\end{tabular}
\end{center}
\caption{(Continued.)}
\end{figure*}
\clearpage

\begin{figure*}
\addtocounter{figure}{-1}
\begin{center}
\begin{tabular}{l}
   \includegraphics[width=450pt]{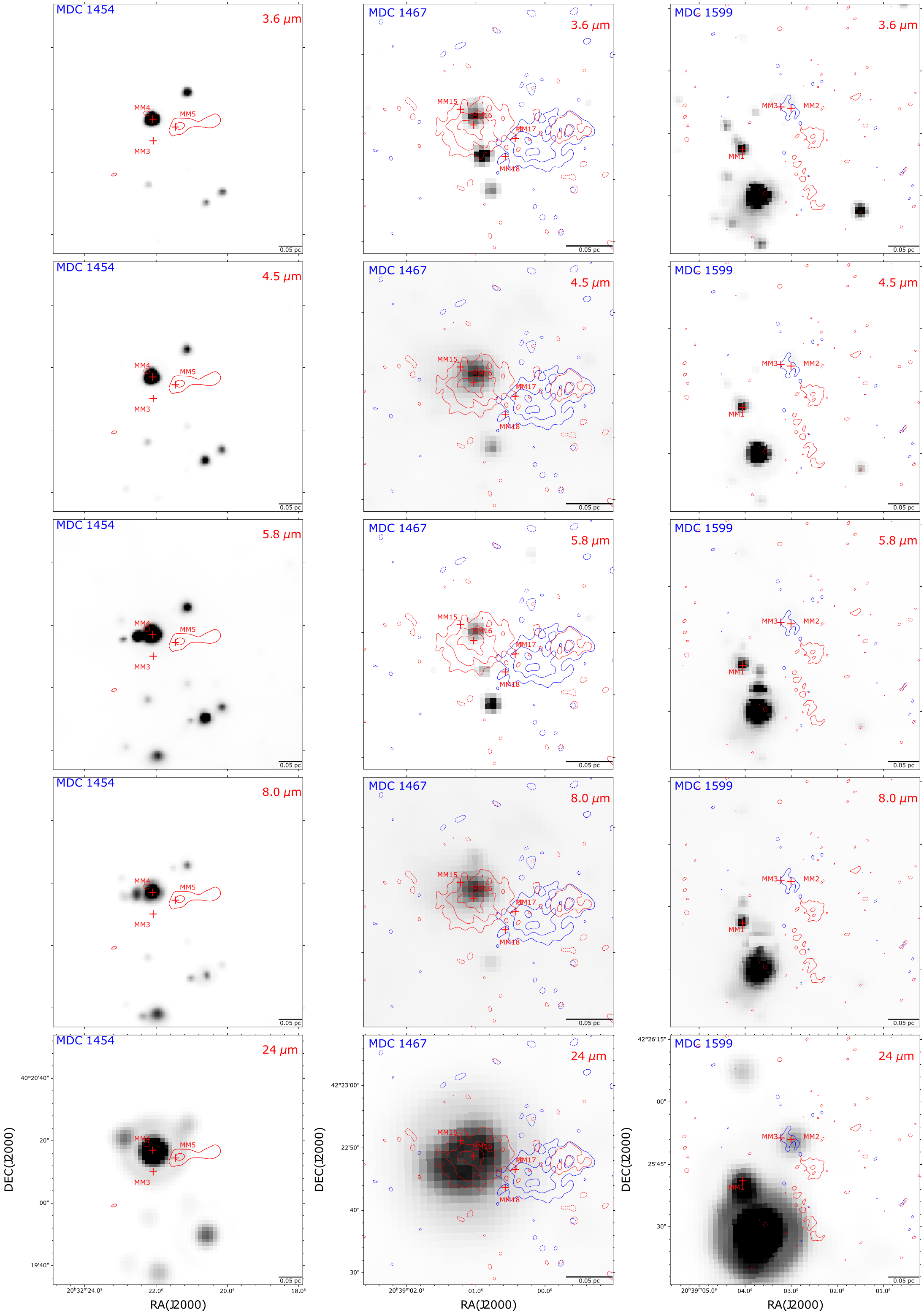}
\end{tabular}
\end{center}
\caption{(Continued.)}
\end{figure*}
\clearpage
%-----------------------------------------------------------------------

\end{appendix}

\end{document}